\documentclass[12pt]{article} 

\usepackage{ae,lmodern}
\usepackage[applemac]{inputenc} 
\usepackage[T1]{fontenc}

\usepackage{epsfig}
\usepackage{graphicx}
\usepackage{comment}
\usepackage{latexsym}
\usepackage{hyperref}
\usepackage{amsmath}
\usepackage{mathrsfs}
\usepackage[usenames, dvipsnames]{color}
\usepackage{amsbsy}
\usepackage{amssymb}
\usepackage{amsthm}
\usepackage{amsfonts}
\usepackage{cite}
\usepackage{enumitem}

\newcommand{\be}[0]{\begin{equation}}
\newcommand{\ee}[0]{\end{equation}}
\newcommand{\dis}{\displaystyle}
\renewcommand{\thefootnote}{\fnsymbol{footnote}}

\newcommand{\R}{\mathbb{R}}
\newcommand{\Z}{\mathbb{Z}}
\renewcommand{\natural}{\mathbb{N}}

\renewcommand{\O}{{\cal O}}

\newcommand{\ie}{{\em i.e.} }

\newcommand{\via}{{\it via} }

\newcommand{\where}{\mbox{where}}

\newcommand{\when}{\mbox{when}}
\renewcommand{\and}{\mbox{and}}

\newcommand{\esp}{\phantom{\!\!\overset{\displaystyle |}{|}}}
\newcommand{\espD}{\phantom{\!\!\underset{\displaystyle |}{\cdot}}}
\newcommand{\espDD}{\phantom{\!\!\underset{\displaystyle |}{|}}}

\newcommand{\bm}{\boldmath} 

\newcommand{\gray}{\color{Gray}}

\newcommand{\F}{{\cal F}}
\newcommand{\N}{{\cal N}}
\newcommand{\J}{{\cal J}}
\newcommand{\I}{{\cal I}}
\newcommand{\K}{{\cal K}}
\renewcommand{\P}{{\cal P}}
\newcommand{\G}{{\cal G}}
\newcommand{\A}{{\cal A}}
\renewcommand{\S}{{\cal S}}
\newcommand{\T}{{\cal T}}
\newcommand{\C}{{\cal C}}
\newcommand{\M}{{\cal M}}
\newcommand{\cZ}{{\cal Z}}

\newcommand{\Ms}{M_{\rm s}}

\newcommand{\nF}{n_{\rm F}}
\newcommand{\nB}{n_{\rm B}}
\newcommand{\Vone}{{\cal V}_{\mbox{\scriptsize 1-loop}}}

\newcommand{\Am}{\mathscr{A}}
\newcommand{\Gm}{\mathscr{G}}
\newcommand{\Cm}{\mathscr{C}}

\newcommand{\ti}{t_{\rm i}}
\newcommand{\tf}{t_{\rm f}}
\newcommand{\dyn}{\text{dyn}}

\catcode`\@=11
\def\marginnote#1{}
\newcount\hour
\newcount\minute
\newtoks\amorpm
\hour=\time\divide\hour by60 \minute=\time{\multiply\hour by60
\global\advance\minute by-\hour}
\edef\standardtime{{\ifnum\hour<12 \global\amorpm={am}%
        \else\global\amorpm={pm}\advance\hour by-12 \fi
        \ifnum\hour=0 \hour=12 \fi
        \number\hour:\ifnum\minute<10 0\fi\number\minute\the\amorpm}}
\edef\militarytime{\number\hour:\ifnum\minute<10 0\fi\number\minute}
\def\draftlabel#1{{\@bsphack\if@filesw {\let\thepage\relax
   \xdef\@gtempa{\write\@auxout{\string
      \newlabel{#1}{{\@currentlabel}{\thepage}}}}}\@gtempa
   \if@nobreak \ifvmode\nobreak\fi\fi\fi\@esphack}
        \gdef\@eqnlabel{#1}}
\def\@eqnlabel{}
\def\@vacuum{}
\def\draftmarginnote#1{\marginpar{\raggedright\scriptsize\tt#1}}
\def\draft{\oddsidemargin -.2truein
        \def\@oddfoot{\sl preliminary draft \hfil
        \rm\thepage\hfil\sl\today\quad\militarytime}
        \let\@evenfoot\@oddfoot \overfullrule 3pt
        \let\label=\draftlabel
        \let\marginnote=\draftmarginnote
   \def\@eqnnum{(\theequation)\rlap{\kern\marginparsep\tt\@eqnlabel}%
\global\let\@eqnlabel\@vacuum}  }
\def\thebibliography#1{
\vskip 0.5cm \centerline{\bf \Large References}
\list{
[\arabic{enumi}]}{\settowidth\labelwidth{[#1]}
\leftmargin\labelwidth
\advance\leftmargin\labelsep
\usecounter{enumi}}
\def\newblock{\hskip .11em plus .33em minus .07em}
\sloppy\clubpenalty4000\widowpenalty4000
\sfcode`\.=1000\relax}

\renewcommand{\theequation}{\arabic{section}.\arabic{equation}}
\renewcommand{\section}{\setcounter{equation}{0}\@startsection
{section}{1}{0mm}{-\baselineskip}{0.5\baselineskip} {\normalfont\Large\bfseries}}
\renewcommand{\subsection}{\@startsection
{subsection}{2}{0mm}{-\baselineskip}{0.5\baselineskip} {\normalfont\large\bfseries}}
\renewcommand{\subsubsection}{\@startsection
{subsubsection}{3}{0mm}{-\baselineskip}{0.5\baselineskip}
{\normalfont\normalsize\slshape}}

\begin{document}


\begin{titlepage}
\begin{flushright}
CPHT-RR009.022018, April  2018
\vspace{1.5cm}
\end{flushright}
\begin{centering}
{\bm\bf \Large Quantum no-scale regimes and moduli dynamics}

\vspace{5mm}

 {\bf Thibaut Coudarchet and Hervé Partouche\footnote{thibaut.coudarchet@ens-lyon.fr, herve.partouche@polytechnique.edu}}

 \vspace{1mm}

{Centre de Physique Théorique, Ecole Polytechnique,  CNRS\footnote{Unité  mixte du CNRS et de l'Ecole Polytechnique, UMR 7644.}, \\ Université Paris-Saclay, Route de Saclay, 91128 Palaiseau, France}

\end{centering}
\vspace{0.1cm}
$~$\\
\centerline{\bf\Large Abstract}\\

\begin{quote}

\hspace{.6cm} We analyze quantum no-scale regimes (QNSR) in perturbative heterotic string compactified on tori, with total spontaneous breaking of supersymmetry. We show that for marginal deformations initially at any point in moduli space, the dynamics of a flat, homogeneous and isotropic universe can always be attracted to a QNSR. This happens independently of the characteristics of the 1-loop effective potential $\Vone$, which can be initially positive, negative or vanishing, and maximal, minimal or at a saddle point. In all cases, the classical no-scale structure is restored at the quantum level, during the cosmological evolution. This is shown analytically by considering moduli evolutions entirely in the vicinity of their initial values. Global attractor mechanisms are analyzed numerically and depend drastically on the sign of $\Vone$. We find that all initially expanding cosmological evolutions along which $\Vone$ is positive are attracted to the QNSR describing a flat, ever-expanding universe. On the contrary, when~$\Vone$ can reach negative values, the expansion comes to a halt and the universe eventually collapses into a Big Crunch, unless the initial conditions are tuned in a tiny region of the phase space. This suggests that flat, ever-expanding universes with positive potentials are way more natural than their counterparts with negative potentials.
\end{quote}




\end{titlepage}
\newpage
\setcounter{footnote}{0}
\renewcommand{\thefootnote}{\arabic{footnote}}
 \setlength{\baselineskip}{.7cm} \setlength{\parskip}{.2cm}

\setcounter{section}{0}


\section{Introduction}

To account for an extremely small cosmological constant, a natural starting point in supergravity is the class of no-scale models~\cite{noscale}. The latter describe the spontaneous breaking of local supersymmetry at a scale $M$ that parameterizes a flat direction of a positive semi-definite potential. In perturbative string theory in $d$ dimensions, this setup can be realized at tree level by coordinate-dependent compactification~\cite{SSstring, Kounnas-Rostand}, which implements the Scherk-Schwarz mechanism~\cite{SS}. The magnitude of the supersymmetry breaking scale measured in $\sigma$-model frame,~$M_{(\sigma)}$, can be restricted to be lower than the string scale~$\Ms$, for Hagedorn-like instabilities~\cite{Hagedorn, Kounnas-Rostand} to be avoided. However, quantum effects lift in general the classical flat directions. At 1-loop, supposing for simplicity that there is no non-trivial mass scale lower than~$M$, a contribution of order~$(\nF-\nB) M^d$ to the effective potential is generated, where $\nF$ and $\nB$ are the numbers of massless fermionic and bosonic degrees of freedom. In this case, a mechanism responsible for the stabilization of $M$ would generically yield a large cosmological constant. For this reason, the theories satisfying $\nF=\nB$, which are sometimes referred as ``super no-scale models'', have attracted attention~\cite{Itoyama:1986ei, SNSM, SNSM2}, since their 1-loop effective potentials turn out to be exponentially suppressed. In some models, the potentials can even vanish exactly at 1-loop, at specific points in moduli space \cite{L=0}. However, even in these instances, the smallness of the potential happens to be invalidated once Higgs masses lower than $M$ are introduced \cite{SNSM, SNSM2}, and/or generic higher order loop corrections are taken into account \cite{L2}.

Alternatively, one may not assume the stabilization of the supersymmetry breaking scale. In this case, the motion of $M$ induced by the effective potential may be analyzed in a cosmological framework~\cite{CFP, attra}, and eventually at finite temperature~\cite{critical, attra, attraM/T,attT}. One of the main motivations of~\cite{CFP} was to find conditions (which we extend in the present paper) for flat, homogeneous and isotropic expanding universes to be allowed by the dynamics. In this reference, the analysis is done by taking into account a reduced set of fields, namely the volume~$vol$ of the torus involved in the Scherk-Schwarz supersymmetry breaking, the dilaton~$\phi$ and the scale factor~$a$ of the universe. For convenience, the degrees of freedom associated with $\ln(vol)$ and $\phi$  are implemented by two canonical fields $\Phi$ and $\phi_\bot$. They are orthogonal linear combinations, where $\Phi$ is the  ``no-scale modulus'' which satisfies~$\mbox{$M\equiv  e^{\alpha\Phi}\Ms$}$, with $\alpha$ a normalization factor. The history of   the universe described by a flat Friedmann-Lema\^itre-Robertson-Walker (FLRW) metric proves to depend drastically on the sign of the 1-loop effective potential\footnote{Technically, similar analyzes involving scalar fields with exponential potentials can be found in Ref.~\cite{solV}. They can be realized at tree level in string theory, with backgrounds involving compact hyperbolic internal spaces, S-branes or non-trivial fluxes \cite{accel}.}:

$\bullet$   For $\nF\ge \nB$, up to time reversal, the evolution is ever-expanding. At initial and late times, it  is driven by the kinetic energies of $\Phi$ and $\phi_\bot$, which dominate over the quantum effective potential. As a result, the  cosmological solution converges in both limits to classical ones, which are characterized by exact no-scale structures with free scalars $\Phi$ and $\phi_\bot$.\footnote{These limit solutions become exact trajectories in the super no-scale models \ie when  $\nF-\nB=0$.}   For this reason, the universe is said to be at early and late times in ``quantum no-scale regime'' (QNSR). It is only during an intermediate era  that connects both QNSRs that the effective potential is relevant. The latter may even induce a transient period of acceleration.

$\bullet$  For $\nF<\nB$, up to time reversal, three different histories can be encountered. In two of them, the universe starts with a Big Bang dominated by the total energy (kinetic plus potential) of $\Phi$. Then, it may forever expand by entering in QNSR, or it may reach a maximum size, before collapsing  into a Big Crunch again dominated by the total energy of the no-scale modulus. In the third kind of trajectories, the universe starts with a Big Bang in QNSR, reaches a maximal size and then collapses as before in a Big Crunch dominated by the total energy of $\Phi$. 

The goal of the present work is to improve the analysis of Ref.~\cite{CFP} by taking into account the dynamics of other moduli fields. To be specific, we consider the heterotic string compactified on a torus, where the Scherk-Schwarz spontaneous breaking of all supersymmetries  involves a single internal direction $X^d$. The latter is large, for $M_{(\sigma)}$ to be lower than $\Ms$. Due to the underlying maximally supersymmetric structure of the setup, all classical marginal deformations can  be interpreted as Wilson lines $y_{I\Upsilon}$, $I\in\{d,\dots,9\}$, $\Upsilon\in\{d,\dots,25\}$. In Sect.~\ref{V1l}, we first present the generic expression of the 1-loop effective potential obtained by switching on small deformations of any background (that has initially no non-trivial mass scale below~$M$). The Wilson lines associated with each gauge group factor can be massive, massless or tachyonic. Then, we focus on a specific configuration to be analyzed in great details, where the moduli of the internal directions $X^d$ and $X^{d+1}$ are allowed to vary, while all other deformations are frozen at extrema of the effective potential. 

Sect.~\ref{exact} is devoted to the derivation of exact results in the framework of the above simple model. Beside~$\Phi$,  the effective potential depends on $\phi_\bot$, which is no longer a free field, and on three  Wilson lines $y_{d,d+1}$, $y_{d+1,d}$ and  $y_{d+1,d+1}$. Physically,  $y_{d+1,d+1}$ parameterizes a Coulomb branch, and  $|y_{d+1,d+1}|\Ms$ is a contribution to the $\sigma$-model frame mass of the component fields belonging to supermultiplets charged under the gauge group. When $|y_{d+1,d+1}|\Ms \ll M_{(\sigma)}$,  supermultiplets are light and their Kaluza-Klein (KK) towers of states along the large direction~$X^d$ contribute effectively to the 1-loop potential. In this case, $y_{d,d+1}$ (or a combination of all three Wilson lines when $y_{d+1,d+1}$ is not exactly vanishing) plays the role of a phase, which determines whether it is the fermions or the bosons within these charged supermultiplets that acquire a mass by the supersymmetry breaking mechanism.  The exact kinetic terms of the model are also presented.  

QNSRs  compatible with weak string coupling are described in this setup in Sect.~\ref{QNSR}. They involve the scale factor~$a$, the scalars~$\Phi$,~$\phi_\bot$, and the Wilson lines $y_{d,d+1}$, $y_{d+1,d}$, $y_{d+1,d+1}$ which at this stage are restricted to be small perturbations of the initial background. Both types of regimes are considered, namely expanding eras $t\to +\infty$ or Big Bangs $t\underset{>}{\to} t_{\rm BB}$, where $t$ is cosmic time and $t_{\rm BB}$ a constant. Their existence is shown, regardless of the sign of $\nF-\nB$ and whether the small Wilson lines are massive, massless or tachyonic at 1-loop.  Compared to Ref.~\cite{CFP}, a novelty is that the moduli space metric is curved, which implies non-canonical kinetic terms. We find that this fact imposes a new condition for the universe to be in QNSR:  The scale $M_{(\sigma)}$ of supersymmetry breaking measured in sigma-model frame must increase as $t\to +\infty$ or $t\underset{>}{\to} t_{\rm BB}$. As a result, the regimes are valid until $M_{(\sigma)}$ reaches $\Ms$, when new stringy effects are expected to arise. Moreover, the new constraint reduces drastically the phase space where the system in QNSR can evolve. For instance, it reduces it by a factor of about 170 for $d=4$. However, this does not mean that the initial conditions that yield such regimes must be tuned within very narrow ranges, due to possible global attraction mechanisms. 

In Sect.~\ref{simu}, all results valid for small Wilson lines are checked by numerical simulations, in the case of the QNSR $t\to +\infty$.   Moreover, it turns out that the quadratic kinetic terms are exact for arbitrary $y_{d,d+1}$ and $y_{d+1,d}$, as long as $y_{d+1,d+1}$ is restricted to vanish. In Sect.~\ref{global}, we use this fact  to simulate large deformations of the initial background parameterized by the Wilson lines $y_{d,d+1}$, $y_{d+1,d}$. When the effective potential is positive, we find that for  arbitrary initial conditions (up to time reversal), the universe expands and is  attracted to the QNSR $t\to +\infty$. On the contrary, when the potential is negative, for the universe to be in QNSR $t\to+\infty$, its initial conditions must sit in the tiny phase space associated with this regime. Otherwise, the initially growing scale factor reaches a maximal size before collapsing. Altogether, these remarks suggest that in order to describe expanding universes in the framework we have considered, naturalness favors models having more fermions than bosons in their light spectra, $\nF-\nB\ge 0$. 

Our concluding remarks are given in Sect.~\ref{cl}, while technical derivations can be found in a long but self-content Appendix. The latter describes the implementation of continuous and discrete Wilson lines in an heterotic toroidal partition function, the spontaneous breaking of supersymmetry, and generic formulas for the effective potential. 


\section{1-loop effective potential}
\label{V1l}

The notion of QNSR in string theory was introduced in Ref.~\cite{CFP}, in the context of the heterotic string compactified on tori, where the total spontaneous breaking  
of supersymmetry is implemented by a stringy Scherk-Schwarz mechanism. However, beside the scale factor of the universe, only the dynamics of the dilaton and  that of the internal volume involved in the breaking of supersymmetry were taken into account. In order to remedy this fact, we consider in this section  features about the dependence of the 1-loop effective potential on all moduli fields. We restrict our analysis to the case where the breaking of supersymmetry is induced along a single internal direction. 


\vskip .2cm
\noindent {\large \em Marginal deformations}

\noindent  To be specific, we consider a  Minkowskian heterotic background in dimension $d\ge 3$,\footnote{The equations of motion and solutions we will derive in Sect.~\ref{QNSR} are formally valid for arbitrary real dimension $d>2$. Cosmological evolutions in 2 dimensions  could be find by taking the limit $d\to2_+$.} with internal space  $T^{10-d}$,
\be
\R^{0,d-1}\times T^{10-d}\, , 
\ee
where the total spontaneous breaking of supersymmetry is induced by a coordinate-dependent compactification along the direction $X^d$. The gauge symmetry group arising in this no-scale model from the Kac-Moody algebra realized on the right-moving bosonic side of the string is $\G_{26-d}$, where gauge groups of rank $r$ will in general be denoted $\G_r$.  At this stage, the background sits at a specific point of the Narain lattice moduli space \cite{Narain:1985jj}
\be
{SO(10-d,26-d)\over SO(10-d)\times SO(26-d)}\, ,
\ee
whose  real dimension is $(10-d)\times (26-d)$. This manifold can be parameterized by the internal metric $G_{IJ}$, the antisymmetric tensor $B_{IJ}$, $I,J\in\{d,\dots,9\}$, and the Wilson lines $Y_{I\J}$, $\J\in\{10,\dots, 25\}$. However, all of these $(10-d)\times (26-d)$ moduli fields can be interpreted from a KK point of view as the components along $T^{10-d}$ of 10-dimensional vector bosons in the Cartan subalgebra of $\G_{26-d}$. Thus, they can be viewed as Wilson lines, and it is natural to split the associated degrees of freedom into  initial background values $(G+B)^{(0)}_{IJ}$, $Y^{(0)}_{I\J}$ and arbitrary Wilson line deformations\footnote{Notations used in the core of the paper are slightly different from those used in the Appendix. The antisymmetric tensor $B$ stands for $B+\Delta B$ in Appendices~\ref{a4}--\ref{a6}. Moreover, in Eq.~(\ref{ba}), $Y^{(0)}_{d\J}$ is denoted $\eta^R_\J$, 
the arbitrary origin of the  fields $Y_{i\J}$ is chosen so that $Y_{i\J}^{(0)}=0$, and the continuous Wilson lines are denoted with upper indices~``$R$''.}
\begin{align}
&(G+B)_{IJ}=\left(\!\!\begin{array}{cc}(G+B)_{dd} & (G^{(0)}+B^{(0)})_{dj}+\sqrt{2}\, y_{dj}\\(G^{(0)}+B^{(0)})_{id}+\sqrt{2} \, y_{id}&(G^{(0)}+B^{(0)})_{ij}+\sqrt{2}\, y_{ij}\end{array}\!\!\right) , \nonumber  \\ 
&Y_{d\J}=Y^{(0)}_{d\J}+y_{d\J}\, , \quad Y_{i\J}=Y_{i\J}^{(0)}+y_{i\J}\, , \qquad  i,j\in\{d+1,\dots,9\} , \; \J\in\{10,\dots, 25\} . \esp 
\end{align}
In our conventions, $y_{I\Upsilon}$, $I\in\{d,\dots, 9\}$, $\Upsilon\in\{d,\dots, 25\}$, is the Wilson line along $X^I$ of the $U(1)$ Cartan generator arising from the right-moving bosonic coordinate $\Upsilon$. In particular, factors $\sqrt{2}$ are introduced in components of the matrix $(G+B)$ to account for the conventional length $\sqrt{2}$ of the roots of the simply laced Lie groups. In this setup, the scale of supersymmetry breaking measured in $\sigma$-model frame can be defined as the KK mass
\be
M_{(\sigma)}=\sqrt{G^{dd}}\, \Ms \, , 
\ee 
where $G^{IJ}\equiv (G^{-1})_{IJ}$. As long as $G_{dd}$ is at least slightly larger than 1, in which case $M_{(\sigma)}\simeq \Ms /\sqrt{G_{dd}}$, no scalar field can be tachyonic at tree level, \ie there is no possibility for a Hagedorn-like instability to take place \cite{Hagedorn}. Moreover, the gauge symmetry $\G_{26-d}$ is spontaneously broken to $U(1)\times \G_{25-d}$. 

Higgs instabilities may however occur at the quantum level. In fact, if the classical no-scale structure guaranties $M_{(\sigma)}$ and all  other marginal deformations $y$'s to be flat directions of a positive semi-definite tree-level potential \cite{noscale}, this is no longer the case when perturbative corrections are taken into account. As described extensively in the Appendix, a non-trivial  effective potential $\Vone^{(\sigma)}$ is already generated at 1-loop. Assuming that $M_{(\sigma)}$ is lower than the string scale $\Ms$, and that the spectrum in the initial background has no mass scale below~$M_{(\sigma)}$,\footnote{\label{tad}Relaxing this hypothesis amounts to shifting Wilson lines by small constant backgrounds, thus inducing tadpoles in Eq.~(\ref{Vgen}).} the generic form of $\Vone$ for small Wilson line deformations is \cite{SNSM, SNSM2}
\begin{align}
\Vone^{(\sigma)}= &\, (\nF-\nB)\,  v_{d}\,  M^d_{(\sigma)}\nonumber \\
&\, +  M^d_{(\sigma)} \, {v_{d-2}\over 2\pi} \sum_{\Upsilon=d+1}^{25} c_\Upsilon\Big[(d-1)y^2_{d\Upsilon}+{1\over G^{dd}}\sum_{i=d+1}^9 y^2_{i\Upsilon}\Big]\!+\cdots\nonumber \\
&\, +\O\big((c\Ms M_{(\sigma)})^{d\over 2}e^{-2\pi c{\Ms/M_{(\sigma)}}}\big)\, ,
\label{Vgen}
\end{align}
where the ellipses stand for higher order interactions in $y$'s.
In this expression, $\nB$ and $\nF$ are the numbers of massless bosonic and fermionic degrees of freedom in the undeformed background, while $v_d$ is a dressing coefficient that accounts for the towers of associated KK modes arising from the large supersymmetry breaking compact direction $X^d$,
\be
v_{d}={\Gamma({d+1\over 2})\, \zeta(d+1)\over 2^{d-1}\;\pi^{3d+1\over 2}}\, \Big(1-{1\over 2^{d+1}}\Big)\, .
\ee
In the last line, $c\Ms$ is the lowest mass scale  above $M_{(\sigma)}$. When the former is much larger than the latter, all states that are not in the above mentioned KK towers yield exponentially suppressed contributions. We see that the scalars $y_{id}$, $i\in\{d+1,\dots,9\}$,  are massless. Moreover, for all Cartan generator $\Upsilon\in\{d+1,\dots,25\}$, the coefficient $c_\Upsilon$ determines whether the Wilson lines $y_{I\Upsilon}$, $I\in\{d,\dots,9\}$,  are massive, massless or tachyonic at 1-loop.\footnote{Strictly speaking, the notion of ``mass'' here is a misnomer when $M_{(\sigma)}$ is treated as a dynamical field, in which case all terms in Eq.~(\ref{Vgen}) are  interactions.} Actually, decomposing $\G_{25-d}$ into simple Lie groups and $U(1)$ factors as follows,
\be
\G_{25-d}=\prod_\lambda \G^{(\lambda)}_{r_\lambda}\qquad \where \qquad \sum_\lambda r_\lambda=25-d\, ,
\ee
the Wilson lines $y_{I\Upsilon}$, $I\in\{d,\dots, 9\}$,  where $\Upsilon$ takes values corresponding to the Cartan generators of $\G_{r_\lambda}^{(\lambda)}$, share a common coefficient $c_\Upsilon\equiv c_{\G_{r_\lambda}^{(\lambda)}}$. The latter is related to the quadratic charges of the representations ${\cal R}^{(\lambda)}_{\rm B}$ and ${\cal R}^{(\lambda)}_{\rm F}$ of the massless bosons and fermions charged under $\G^{(\lambda)}_{r_\lambda}$ in the initial background (see Eq.~(\ref{CR})),
\be
c_{\G_{r_\lambda}^{(\lambda)}}=8\big(C_{{\cal R}^{(\lambda)}_{\rm B}}-C_{{\cal R}^{(\lambda)}_{\rm F}}\big)\, .
\ee
By switching on small $y$-deformations of the background we started with, some charged and initially massless states acquire Higgs masses lower than $M_{(\sigma)}$. This reduces the dimension (but not the rank) of the gauge symmetry, which enters a Coulomb branch. 


\vskip .2cm
\noindent {\large \em Example 1}

\noindent To illustrate the above generalities, let us consider the supersymmetric $E_8\times E'_8$ heterotic string compactified on $T^{10-d}$. Taking $G_{dd}\gg 1$, the gauge symmetry arising from the right-moving sector is $\G_{26-d}=U(1)\times \G_{9-d}\times E_8\times E_8'$. When we sit at a point in moduli space where $\G_{9-d}$ is maximally enhanced \ie contains no $U(1)$ factor, the model presents only two scales, namely the KK mass $M_{(\sigma)}$ and the much greater string scale $\Ms$. As reviewed in Appendices~\ref{a1}--\ref{a3}, denoting $a\in\Z_2$ the fermionic number, the simplest choice of implementation of the Scherk-Schwarz breaking of supersymmetry along the large compact direction $X^d$ induces KK masses\footnote{The representative of $a\in\Z_2$ in $\{0,1\}$ is understood.}  ${1\over 2}aM_{(\sigma)}$ to all initially massless degrees of freedom. As a result, the no-scale model has no massless fermions, $\nF=0$, and  its  mass coefficients are positive, $c_{\G_{r_\lambda}^{(\lambda)}}=8C_{{\cal R}^{(\lambda)}_{\rm B}}$. For instance, one obtains for $\G_{9-d}=SU(2)^{9-d}$ 
\begin{align}
&\nB=8\, \big[d-2+\dim\!\big(U(1)\times SU(2)^{9-d}\times E_8\times E_8'\big)\big] \!= 8\, (522-2d)\, , \nonumber \\
& c_{SU(2)}=8\times C_{[3]_{SU(2)}}=8\times 2=16\qquad \mbox{for the $9-d$ $SU(2)$ factors} \,  ,\nonumber \\
& c_{E_8}\,\;\; \,\,=8\times C_{[248]_{E_8}}\: =8\times 30=240\quad \mbox{for $E_8$ and $E_8'$} \,  .
\end{align}
This yields a negative effective potential and no Higgs instabilities for the $y$-fields at the quantum level.


\vskip .2cm
\noindent {\large \em Example 2}

\noindent For massless fermions to be present in the no-scale model, a more sophisticated choice of  supersymmetry breaking must be considered\footnote{In the notations of Appendix~\ref{a6}, this can be done by switching on discrete Wilson lines, as in Eq.~(\ref{desdef}),~(\ref{etabreak}).}. For instance, one can define a charge $\gamma\in\Z_2$ in terms of which the affine character of $E_8$ can be divided into $SO(16)$ ones, 
\be
{1\over 2}\sum_{\gamma,\delta\in\Z_2}\left({\bar \theta[^\gamma_\delta]\over \bar \eta}\right)^8 = \bar O_{16}+\bar S_{16}\, .
\ee
In this relation, $\gamma=0$ leads to the unit character $\bar O_{16}$, while $\gamma=1$ corresponds to the spinorial one $\bar S_{16}$. At the massless level, this amounts to splitting the adjoint representation of $E_8$ into adjoint and spinorial representations of $SO(16)$, $[248]_{E_8}=[120]_{SO(16)}\oplus [128]_{SO(16)}$. As seen in Appendix~\ref{a4}, it is possible to implement a Scherk-Schwarz supersymmetry breaking that induces KK masses\footnote{The representative of $a+\gamma+\gamma'\in\Z_2$ in $\{0,1\}$ is understood.}  ${1\over 2}(a+\gamma+\gamma')M_{(\sigma)}$ to all initially massless degrees of freedom, where $\gamma'\in\Z_2$ is the charge similar to $\gamma$ but associated with $E_8'$. As in Example~1, the fermions acquire masses if they belong to supermultiplets with  $\gamma+\gamma'$ even. The situation is however reversed when $\gamma+\gamma'$ is odd, since it is the bosons which become massive. As a result, the mechanism breaks spontaneously all supersymmetries as well as the gauge symmetry $E_8\times E_8'\to SO(16)\times SO(16)'$.  The  gauge group $\G_{9-d}$ can be chosen as before to be $SU(2)^{9-d}$. However, it is  instructive to also consider Coulomb branches $\G_{9-d}= SU(2)^{9-d-s}\times U(1)^s$, $s\in\{0,\dots,9-d\}$, when the masses of the non-Cartan gauge bosons are greater than $M_{(\sigma)}$, for Eq.~(\ref{Vgen}) to be valid. In this case, we obtain

\begin{align}
&\nB=8\, \big[d-2+\dim\!\big(U(1)\times SU(2)^{9-d-s}\times U(1)^s\big)+ 120+ 120\big] = 8\, (266-2d-2s)\, , \nonumber \\
&\nF=8\, (128+ 128) = 8\times256\, , \nonumber \\
& c_{SU(2)}=8\times C_{[3]_{SU(2)}}=8\times 2=16\qquad \:\qquad \mbox{for the $9-d-s$ $SU(2)$ factors} \,  ,\nonumber \\
& c_{U(1)}\,\,=0\;\;\;\qquad\qquad\qquad\qquad \qquad \qquad\qquad \mbox{for the $s$ $U(1)$ factors of $\G_{9-d}$} \,  ,\nonumber \\
& c_{SO(16)} =8\times (C_{[120]_{SO(16)}}-C_{[128]_{SO(16)}})=8\,(14-16)=-16\; \;\mbox{for $SO(16)$ and $SO(16)'$} \,  .
\end{align}

We see that $\nF-\nB=8(2d+2s-10)$, which is greater or equal to 0 for $d\ge 5$ and can be positive, negative or null for $d=3$ and 4. 
However, the Wilson lines of $SO(16)\times SO(16)'$ are all tachyonic at 1-loop and Higgs instabilities may arise. 


\vskip .2cm
\noindent {\large \em Example 3}

\noindent  We are naturally invited to reconsider Example~2 after Higgs transition, in the Coulomb branch where $SO(16)\times SO(16)'\to U(1)^{16}$. Again, we assume the masses of the $2\times 112$ non-Cartan gauge bosons to be greater than $M_{(\sigma)}$, for Eq.~(\ref{Vgen}) to be applicable. Since all states in the spinorial representations $[128]_{SO(16)}$ and $[128]_{SO(16)'}$ are also massive, we automatically obtain vanishing mass coefficients $c_{U(1)}$'s for the 16 Cartan $U(1)$'s. Moreover, we are back to a configuration where $\nF=0$, and the effective potential is necessarily negative. In fact, we could have reached the same point in moduli space by considering the Coulomb branch in Example~1, where $E_8\times E_8'\to U(1)^{16}$. 

The above 3 simple examples illustrate the fact that at the quantum level, local  stability (eventually marginal) of the Wilson lines ($C_{{\cal R}^{(\lambda)}_{\rm B}}\ge C_{{\cal R}^{(\lambda)}_{\rm F}}$, for all $\lambda$) and non-negativity of the effective potential ($\nF\ge \nB$) are conditions that are easily in contradiction. Actually, it would be interesting to clarify whether they may be compatible. However, insofar as in the present paper we are interested in flat FLRW cosmological evolutions where the effective potential is dominated by the kinetic energies of moduli fields, it happens that the sign of $\nF-\nB$ as well as those of the mass coefficients $c_{\G_{r_\lambda}^{(\lambda)}}$'s do not play significant roles in the existence of QNSRs. Before showing this  in Sect.~\ref{QNSR} in an heterotic context we will now describe, we signal that global attractor mechanisms are nonetheless sensitive to the signs, as will be seen in the numerical simulations of Sect.~\ref{global}. 


\vskip .2cm
\noindent {\large \em A specific setup}

\noindent  Before $y$-deformation, we consider from now on a background to be studied in great details, 
\be
\R^{0,d-1}\times S^1_{\rm SS}(R_d)\times S^1(R_{d+1}=1)\times T^{8-d}\, ,
\label{Bi}
\ee
where the index $\rm SS$ signals a supersymmetry breaking coordinate-dependent compactification along the circle of radius $R_d\equiv \sqrt{G_{dd}}\gg 1$. Note that the block-diagonal form of the internal space metric does not say anything about the antisymmetric tensor, so that we may choose
\be
\label{backg}
2 B^{(0)}_{d,d+1}=\eta^R_{d+1}\in \Z\, , \quad G^{(0)}_{d,d+1}=0\, , \quad G^{(0)}_{d+1,d+1}=1\, . 
\ee
Altogether, these data imply the gauge symmetry group generated by the right-moving sector to be factorized as $\G_{26-d}=U(1)\times \G_1\times \G_{8-d}\times \G_{16}$.\footnote{This is clear for $B_{d,d+1}^{(0)}=0$ mod 2 but  remains true for arbitrary real $B_{d,d+1}^{(0)}$, as will be described in details in Sect.~\ref{exact}.} In the  decompactification limit $\mbox{$R_d\to +\infty$}$, where supersymmetry is restored, our choice of radius $R_{d+1}\equiv \sqrt{G^{(0)}_{d+1,d+1}}=1$ for the second circle  implies an $SU(2)$ enhancement of the gauge symmetry (in 5 dimensions). However, as explained at the end of Appendix~\ref{a4}, two cases can arise at finite $R_d$, depending on the parity of the ``discrete Wilson line'' $\eta^R_{d+1}\equiv 2(G^{(0)}+B^{(0)})_{d,d+1}\in \Z$ (see Eq.~(\ref{ydis})). When $\eta^R_{d+1}$ is even, all fermionic degrees of freedom of the supermultiplets in the adjoint representation of $SU(2)$ acquire a mass ${1\over 2}M_{(\sigma)}$. In this case, the enhancement of the gauge symmetry is preserved, and $\G_1=SU(2)$. On the contrary, when $\eta^R_{d+1}$ is odd, the spontaneous breaking operates simultaneously on supersymmetry and on the $SU(2)$ gauge symmetry. In practice, the bosonic degrees of freedom of the $SU(2)$ non-Cartan supermultiplets acquire a mass ${1\over 2}M_{(\sigma)}$. As a result, only the Cartan gauge symmetry is preserved, $\G_1=U(1)$, and the latter is coupled to the massless fermions belonging to the non-Cartan supermultiplets of charges~$\pm\sqrt{2}$.  In all instances, the mass coefficients are given by 
\begin{align}
&\eta^R_{d+1}=0 \mbox{ even} \quad \Longrightarrow\quad \G_1=SU(2)\, , \;\;  c_{\G_1}=8\times C_{[3]_{SU(2)}}=8\times 2 =16\, , \nonumber\\
&\eta^R_{d+1}=0 \mbox{ odd} \,\,\quad \Longrightarrow\quad \G_1=U(1)\, , \;\:  \;\;\,c_{\G_1} =-8\times C_{\pm\sqrt{2}}=-8\times 2=-16\, .
\end{align}

Our goal being  to switch on moduli fields in order to study their dynamics later on, we will  make further assumptions for the sake of simplicity. We suppose that the undeformed background~(\ref{Bi}) does not introduce mass scales below $M_{(\sigma)}$. As already mentioned in Footnote~\ref{tad}, this  ensures that the 1-loop potential does not induce tadpoles for the $\mbox{$y$-fields}$. This can be realized by considering maximally enhanced gauge groups $\G_{8-d}\times \G_{16}$ or points in their Coulomb branches where the non-Cartan generators have masses above $M_{(\sigma)}$. Under these conditions, it is consistent to freeze to 0 the Wilson lines of $\G_{8-d}\times \G_{16}$ along $S_{\rm SS}^1(R_d)\times S^1(R_{d+1}=1)\times T^{8-d}$, as well  as those of $U(1)\times \G_1$ along $T^{8-d}$. In fact, the configuration 
\be
y_{I,d+2}\equiv \dots\equiv y_{I,25}\equiv 0, \; I\in\{d,\dots,9\}\, , \qquad y_{id}\equiv y_{i,d+1}\equiv 0, \; i\in\{d+2,\dots,9\}\, , 
\label{rest}
\ee
solves trivially the equations of motion of the associated degrees of freedom, even when the $\mbox{1-loop}$  potential is included in the effective supergravity. Given these restrictions, we are left with non-trivial Wilson lines $y_{d,d+1}$,  $y_{d+1,d}$, $y_{d+1,d+1}$, which are those of $U(1)\times \G_1$ along the compact directions $X^d$ and $X^{d+1}$. For small deformations, the effective potential then becomes:
\begin{align}
\Vone^{(\sigma)}= &\, (\nF-\nB)\,  v_{d}\,  M^d_{(\sigma)}\nonumber \\
&\, +  M^d_{(\sigma)} \, {v_{d-2}\over 2\pi} \,  c_{\G_1}\Big[(d-1)y^2_{d,d+1}+{y^2_{d+1,d+1}\over G^{dd}}\Big]\!+\cdots\nonumber\espD \\
&\, +\O\big((c\Ms M_{(\sigma)})^{d\over 2}e^{-2\pi c{\Ms/M_{(\sigma)}}}\big)\, .
\label{Vap}
\end{align}
Of course, if $M_{(\sigma)}$ acquired a vacuum expectation value, it would be very artificial to impose Eq.~(\ref{rest}) and expand $\Vone^{(\sigma)}$ at $y_{d,d+1}=y_{d+1,d}=y_{d+1,d+1}=0$, when $c_{\G^{(\lambda)}_{r_\lambda}}<0$ for some $\lambda$'s. However, as already announced, the situation happens to be  drastically different in a QNSR. Before observing  this fact for $c_{\G_1}=-16$ in Sect.~\ref{QNSR}, we find instructive to make explicit the ellipses in Eq.~(\ref{Vap}), by presenting the exact expression of $\Vone^{(\sigma)}$ valid for arbitrary Wilson lines $y_{d,d+1}$,  $y_{d+1,d}$, $y_{d+1,d+1}$. 


\section{Exact formulas}
\label{exact}

In this section, we would like to have a better idea of the global structure of the ``reduced'' moduli space parameterized by the continuous Wilson lines $y_{d,d+1}$,  $y_{d+1,d}$, $y_{d+1,d+1}$. In particular, we will describe periodicity properties of the effective potential, as well as the exact kinetic terms. 


\vskip .2cm
\noindent {\large \em Effective potential}

\noindent  For arbitrary deformations, the expression of  the 1-loop effective potential $\Vone^{(\sigma)}$ is obtained by applying the generic formula Eq.~(\ref{Lf}), derived in Appendix~\ref{a5}. Up to the $\O\big((c\Ms M_{(\sigma)})^{d\over 2}e^{-2\pi c{\Ms/M_{(\sigma)}}}\big)$ exponentially suppressed terms, $\Vone^{(\sigma)}$ can be written as a sum over a finite number of  KK towers of states associated with the large compact direction $X^d$ (along which they have vanishing winding numbers, $n_d=0$). These KK towers are those characterized by mass scales denoted $\M'_{L0}$ that are lower than the KK \ie supersymmetry breaking scale $M_{(\sigma)}$. They always appear in groups of 8, due to the degeneracy arising from the left-moving supersymmetric side of the string. For the initial background satisfying Eqs~(\ref{Bi}), ~(\ref{backg}), they can be listed as follows:

$(i)$ The 8 KK towers at right-moving oscillator level $\ell_R=0$, whose right-moving quantum numbers are a given root of  $SU(2)$, and that are neutral under $\G_{8-d}\times \G_{16}$. For each root $\epsilon \sqrt{2}$, $\mbox{$\epsilon\in\{-1,1\}$}$, the momentum and winding numbers along $T^{8-d}$ are
\be
m_{d+1}=-n_{d+1}=-\epsilon\, , \qquad m_i=n_i=0\, , \quad i\in\{d+2,\dots,9\}\, . 
\ee
These towers arise in the Neveu-Schwarz sector of the 32 extra right-moving worldsheet fermions, $\vec a^R=\vec 0$. 
   
$(ii)$ The 8 KK towers at oscillator level $\ell_R=0$, whose right-moving quantum numbers are a given root or weight\footnote{In the notations of Eq.~(\ref{ydis}), non-adjoint representations of $\G_{8-d}$ exist when some of the discrete Wilson lines $\eta^R_j\in\Z$, $\mbox{$j\in\{d+2,\dots,9\}$}$, are odd. In this case, $\G_{8-d}$ may contain $U(1)$ factors coupled to fermions, with non-trivial charges we still refer as weight vectors' components.} vector (of length equal to $\sqrt{2}$) of a representation of $\G_{8-d}$, and that are neutral under $\G_1\times \G_{16}$. They have non-trivial momentum and winding numbers along $T^{8-d}$ and arise in sector $\vec a^R=\vec 0$.

$(iii)$ The 8 KK towers at oscillator level $\ell_R=0$, whose right-moving quantum numbers are a given root or weight vector (of length equal to $\sqrt{2}$) of a representation of $\G_{16}$, and that are neutral under $\G_1\times \G_{8-d}$. They have trivial momentum and winding numbers along $T^{8-d}$ and arise in any sector $\vec a^R$. 

$(iv)$ The $8\times 24$ KK towers at oscillator level $\ell_R=1$ that are neutral under $\G_1\times \G_{8-d}\times \G_{16}$. They have trivial momentum and winding numbers along $T^{8-d}$ and arise in sector $\vec a^R=\vec 0$. 

Due to our restriction on the allowed non-trivial Wilson line deformations, Eq.~(\ref{rest}), all KK towers $(ii)$--$(iv)$ have characteristic masses $\M'_{L0}=0$ (see Eq.~(\ref{Mgene})) and ``phases'' $\zeta^d=0$. Defined in Eq.~(\ref{zgene}), $\zeta^d$ actually determines the relative weights of the bosonic and fermionic modes within a given KK tower.  The non-trivial Wilson lines $y_{d,d+1}$,  $y_{d+1,d}$, $y_{d+1,d+1}$ however impact the characteristic masses and phases of the $8\times 2$ KK towers~$(i)$. In total, the 1-loop effective potential given in Eq.~(\ref{Lf}) takes the specific form 
\begin{align}
\Vone^{(\sigma)}= &\, \big(\nF-\nB+(-1)^{\eta^R_{d+1}}\, 8\times 2\big)\,  v_{d}\,  M^d_{(\sigma)}\nonumber \\
&\, -(-1)^{\eta^R_{d+1}}\, 8\times 2\; {2M^d_{(\sigma)}\over (2\pi)^{3d+1\over 2}} \,  \sum_{\tilde m_d}\, {\cos \!\big(2\pi( 2\tilde m_d+1) z\big)\over |2\tilde m_d+1|^{d+1}} \, \nonumber F\Big(2\pi|2\tilde m_d+1| {\M\over M_{(\sigma)}}\Big)\espDD\\
&\, +\O\big((c\Ms M_{(\sigma)})^{d\over 2}e^{-2\pi c{\Ms/M_{(\sigma)}}}\big)\, .
\label{Vexact}
\end{align}
In this formula, the definition of the function $F$ can be found in Eq.~(\ref{Ffunc}), we have introduced $z$ instead of $\zeta^d$ for notational convenience, and the non-trivial characteristic mass is denoted $\M$,  
\be
z=\sqrt{2}\, \bigg(y_{d,d+1}-{y_{d,d+1}+y_{d+1,d}\over \sqrt{2}\left(1+\sqrt{2}\, y_{d+1,d+1}\right)}\: y_{d+1,d+1}\bigg) , \qquad \M={\sqrt{2}\, |y_{d+1,d+1}|\over \sqrt{1+\sqrt{2}\,y_{d+1,d+1}}}\, .
\label{zM}
\ee
If it is physically natural to use  $M_{(\sigma)}$ and $y_{d,d+1}$,  $y_{d+1,d}$, $y_{d+1,d+1}$ to parameterize the classical moduli space, it is however a matter of convention. Another choice may be to consider the   ``volume'' $G_{dd}$ as the remaining degree of freedom independent of $y_{d,d+1}$,  $y_{d+1,d}$, $y_{d+1,d+1}$,  in terms of which the supersymmetry breaking scale satisfies
\be
M_{(\sigma)}^2\equiv G^{dd}\Ms^2={\Ms^2\over G_{dd}\left(1-{\displaystyle (y_{d,d+1}+y_{d+1,d})^2\over \displaystyle2G_{dd}\, (1+\sqrt{2}\, y_{d+1,d+1})}\right)}\, .
\ee

Some remarks about Eq.(\ref{Vexact}) are in order:

$\bullet$ The dependence in Wilson lines of $\Vone^{(\sigma)}$ involves only two combinations of fields,~$z$ and~$\M$. Thus, a flat direction exists at 1-loop. 

$\bullet$  The expansions of the cosine and function $F$  for small arguments contain exclusively even powers. However, depending on $d$, only a finite number of  monomials can be summed term by term. At order $z^2$ and $\M^2$, summing over $\tilde m_d$ and restricting to the quadratic terms in Wilson lines, one obtains the approximate result~(\ref{Vap}). 

$\bullet$   Due to the factor $|2\tilde m_d+1|^{d+1}$ in the denominator, as well as the exponential suppression of the function $F$ for large argument, the discrete sum in Eq.~(\ref{Vexact}) is numerically very close to  that restricted to $\tilde m_d=0$ and $-1$. The error introduced this way in the sum is about 1\% or (much) less.

$\bullet$  The potential is 1-periodic in $z$. A half-period shift $z\to z+{1\over 2}$ flips the sign of the second line in Eq.~(\ref{Vexact}).

$\bullet$  The mass $\M$, which  characterizes as a whole each KK tower $(i)$, depends only on $y_{d+1,d+1}$, due to an exact cancellation of the contributions of  $y_{d,d+1}$ and $y_{d+1,d}$ in the general expression~(\ref{Mgene}). This is remarkable, since the mass of each KK mode (see Eq.~(\ref{MKKgene})) does depend on the three Wilson lines. 

$\bullet$  For instance in the case $\eta^R_{d+1}$ even, when $\M=0$ \ie $y_{d+1,d+1}=0$, the lightest KK masses and $z$ (for example in the range  $[-{1\over 2},{1\over 2}]$) satisfy 
\be
\label{mz}
m^2=\left({a\over 2}-|z|\right)^2 M_{(\sigma)}^2\, , \qquad z=\sqrt{2}\, y_{d,d+1}\, .
\ee
For $z=0$, the associated states are the massless $SU(2)$ non-Cartan gauge and scalar bosons ($a=0$) and their fermionic superpartners ($a=1$) of masses ${1\over 2}M_{(\sigma)}$. As a result, the second line of Eq~(\ref{Vexact}) cancels the contribution $8\times 2$ in the first line. The situation is reversed for $z=\pm{1\over 2}$, for which the fermions are massless and the bosons massive, so that the role of the second line of Eq~(\ref{Vexact}) is to shift $\nF\to \nF+8\times 2$ in the first line. When $z$ varies between these two extreme cases, the KK towers do not contain massless states. Their  absolute contributions are lower  and actually vanish for $z=\pm{1\over 4}$. In fact, when $|z| \in(0, {1\over 2}]$, the gauge symmetry is in the Coulomb branch, $SU(2)\to U(1)$. On the contrary, when $\eta^R_{d+1}$ is odd, $a$ is replaced with $1-a$ in the mass formula of Eq.~(\ref{mz}) and the roles of bosons and fermions are reversed. In particular, for $z=\pm{1\over 2}$, the second line of Eq~(\ref{Vexact}) simply shifts  $\nB\to \nB+8\times 2$.

$\bullet$  When $y_{d+1,d+1}$ is switched on, the dependence of $z$ and thus $\Vone$ on $y_{d+1,d}$ becomes non-trivial. For instance, in the neighborhood of the undeformed background,  $y_{d+1,d}$ appears at lowest order  in Eq.~(\ref{Vexact}) in the interaction term
\be
\label{3pt}
-M^d_{(\sigma)} \, {v_{d-2}\over 2\pi} \,  c_{\G_1}\, (d-1)\, \sqrt{2}\, y_{d+1,d}\, y_{d,d+1}\, y_{d+1,d+1}\, .
\ee  
Thus, even if it is still massless, it is not identified anymore with the flat direction of the 1-loop potential.

$\bullet$  The function $F$ is even, positive, shaped like a bell centered at the origin, and exponentially suppressed for large a argument. As a result, when $\M$ is non-vanishing but still smaller than $M_{(\sigma)}$, the magnitude (at fixed phase $z$) of the contributions of the $8\times 2$ KK towers $(i)$ is lowered. In fact, $y_{d+1,d+1}$ induces a small Higgs mass, so that the towers do not contain massless modes, even for $z=0$ or ${1\over 2}$ mod 1, and the gauge theory always sits in the Coulomb branch $SU(2)\to U(1)$. When $\M$ is greater than  $M_{(\sigma)}$, the Higgsing is large and we are free to omit the second line of Eq.~(\ref{Vexact}).\footnote{At the transition, \ie when $\M$ is slightly greater than $M_{(\sigma)}$, omitting the second line should be accompanied by fixing $c=\M$ in the last one.} In this case, the Wilson lines $y_{d,d+1}$,  $y_{d+1,d}$, $y_{d+1,d+1}$ are flat directions, up to exponentially suppressed terms. 


\vskip .2cm
\noindent {\large \em Kinetic terms}

\noindent  At tree level, imposing the restriction~(\ref{rest}), the  massless degrees of freedom allowed to have non-trivial homogeneous and isotropic backgrounds\footnote{In dimension $d=4$, we also impose the axion field dual to the spacetime antisymmetric tensor to be constant.}  are the graviton, the dilaton and the complex moduli
\be
\label{TU}
{\cal T}=B_{d,d+1}+i\sqrt{G_{dd}G_{d+1,d+1}-G_{d,d+1}^2} \, , \qquad {\cal U}={G_{d+1,d}+i\sqrt{G_{dd}G_{d+1,d+1}-G_{d,d+1}^2}\over G_{dd}}\, .
\ee
Splitting the dilaton into a constant plus a dynamical field, $\phi_{\rm dil}\equiv \langle \phi_{\rm dil}\rangle+\phi$, the Einstein frame metric is defined as $g_{\mu\nu}=e^{-{4\over d-2}\phi}G_{\mu\nu}$ and the classical effective action of the above degrees of freedom reduces to their kinetic terms, 
\be
\label{STU}
S_{\rm tree}={1\over \kappa^2}\int d^dx\, \sqrt{-g}\, g^{\mu\nu}\bigg[{{\cal R}_{\mu\nu}\over 2}-{2\over d-2}\,\partial_\mu\phi\partial_\nu\phi +{\partial_\mu {\cal T}\partial_\nu \bar {\cal T}\over ({\cal T}-\bar {\cal T})^2}+{\partial_\mu {\cal U}\partial_\nu \bar {\cal U}\over ({\cal U}-\bar {\cal U})^2}\bigg]\, .
\ee
In our conventions, the signature of the metric is $(-,+,\cdots,+)$, ${\cal R_{\mu\nu}}$ is the Ricci tensor  and $\kappa^2=e^{2\langle \phi_{\rm dil}\rangle}/\Ms^{d-2}$ is Einstein's constant. To make contact with the arbitrary Wilson lines $y_{d,d+1}$,  $y_{d+1,d}$, $y_{d+1,d+1}$, the following dictionary can be used, 
\begin{align}
&G_{d+1,d+1}=1+\sqrt{2}\, y_{d+1,d+1}\, , \nonumber\\
&G_{d,d+1}\;\;\:\,={1\over \sqrt{2}}\, (y_{d,d+1}+y_{d+1,d})\equiv h_{d,d+1}\, , \nonumber \\
& B_{d,d+1}\;\;\;\,={\eta^R_{d+1}\over 2}+{1\over \sqrt{2}}\, (y_{d,d+1}-y_{d+1,d})\, .
\end{align}
Moreover, the supersymmetry breaking scale measured in Einstein frame is dressed with a dilaton factor and can be redefined in terms of the so-called ``no-scale modulus'' $\Phi$, 
\be
M\equiv e^{{2\over d-2}\phi}\, M_{(\sigma)}\equiv e^{\alpha\Phi}\Ms \, , \quad \where \quad \alpha\Phi={2\over d-2}\, \phi+\ln\sqrt{G^{dd}}\, , \quad \alpha=\sqrt{d-1\over d-2}\, .
\ee
Noticing that the kinetic terms of $T$ and $U$ yield, among other things, a contribution $-{1\over 2}(\partial\ln \sqrt{G_{dd}})^2$, it is natural to relate the latter to $-{1\over 2}(\partial\ln \sqrt{G^{dd}})^2$ by using the identity
\be
G^{dd}G_{dd}={1+\sqrt{2}\, y_{d+1,d+1}\over 1+f}\, , \quad \where \quad  f=\sqrt{2}\, y_{d+1,d+1}-{h_{d,d+1}^2\over G_{dd}}\, .
\ee
In this way, the kinetic terms of $\phi$ and $\ln\sqrt{G^{dd}}$ can be combined into those of $\Phi$ and an ``orthogonal'' combination $\phi_\perp$,
\be
\sqrt{d-1}\, \phi_\perp=2\phi-\ln\sqrt{G^{dd}}\, . 
\ee  
In total, we ultimately find
\begin{align}
\label{kinexact}
S_{\rm tree}={1\over \kappa^2}\int d^dx\, &\sqrt{-g}\, \bigg[{{\cal R}\over 2}-{1\over 2}(\partial\Phi)^2-{1\over 2} (\partial\phi_\bot)^2\nonumber \\
&-{\Omega_1\over 4}\, G^{dd} \Big((\partial y_{d,d+1})^2+(\partial y_{d+1,d})^2\Big)-{\Omega_2\over 4}\, (\partial y_{d+1,d+1})^2+\Omega_3\bigg]\,  ,
\end{align}
where we have defined 
\begin{align}
&\Omega_1= {1\over 1+\sqrt{2}\, y_{d+1,d+1}}\, , \qquad \Omega_2={1-{h^4_{d,d+1}\over G^2_{dd}}\over (1+f)^2}\, , \nonumber \espD \\
&\Omega_3={\partial y_{d+1,d+1}\over 2\sqrt{2}}\bigg(\partial \Big({h^2_{d,d+1}\over G_{dd}}\Big)\, {1-{h^2_{d,d+1}\over G_{dd}}\over (1+f)^2}+{\partial G_{dd}\over G_{dd}}\, {h^2_{d,d+1}\over G_{dd}}\, {1\over (1+\sqrt{2}\, y_{d+1,d+1})(1+f)}\bigg)\, .
\end{align}


\newpage 
\noindent {\large \em Local marginal deformations}

\noindent With exact formulas for the potential and kinetic terms at hand, we can make precise the notion of ``small Wilson lines deformations'' used in Sect.~\ref{V1l}, for the backgrounds satisfying Eqs~(\ref{Bi}),~(\ref{backg}) and (\ref{rest}): 
\be
\label{small}
|y_{d+1,d+1}| \ll \sqrt{G^{dd}}\ll 1\, , \quad |y_{d,d+1}| \ll 1\, , \quad  |y_{d+1,d}| \ll 1\, .
\ee
Our goal being to study the dynamics of moduli fields, the restriction on $y_{d+1,d+1}$ implies $\M\ll M_{(\sigma)}$ so that the three Wilson lines are not flat directions. The conditions on $y_{d,d+1}$ and $y_{d+1,d}$ imply $|h_{d,d+1}| \ll 1$. Noticing that 
\be
\Omega_3={G^{dd}\over 2\sqrt{2}}\, \partial y_{d+1,d+1}\, \partial h_{d,d+1}^2+\cdots\, , 
\label{O3}
\ee
where the ellipses stand for at least quartic terms in Wilson lines, it is then consistent at leading order to set $(\Omega_1,\Omega_2,\Omega_3)= (1,1,0)$  in the kinetic terms. Moreover, the cubic interaction~(\ref{3pt}) and higher order ones in the potential can also be neglected, compared to the quadratic mass terms in Eq.~(\ref{Vap}). 


\section{Quantum no-scale regimes}
\label{QNSR}

Our goal in this section is to show that QNSRs do exist when the dynamics of marginal deformations of the internal space are taken into account. Indeed, we will find conditions under which such regimes can be reached in the setup described at the end of Sect.~\ref{V1l}.

When the assumptions~(\ref{small}) are fulfilled, the 1-loop effective  action in Einstein frame  can be written  as  
\begin{align}
S_{\mbox{\scriptsize 1-loop}}=&\, {1\over \kappa^2}\int d^dx\, \sqrt{-g}\, \bigg[{{\cal R}\over 2}- {1\over 2}(\partial\Phi)^2-{1\over 2} (\partial\phi_\bot)^2 \nonumber\espD \\
&-{G^{dd}\over 4}(\partial y_{d,d+1})^2-{G^{dd}\over 4}(\partial y_{d+1,d})^2-{1\over 4} (\partial y_{d+1,d+1})^2+\cdots-\kappa^2\Vone\bigg] \, ,
\label{S1}
\end{align}
where the potential is given by, 
\be
\Vone=  e^{d\alpha\Phi} \Ms^d \Big[(\nF-\nB)\,  v_{d}+ {v_{d-2}\over 2\pi} \,  c_{\G_1}\Big(\!(d-1)y^2_{d,d+1}+{y^2_{d+1,d+1}\over G^{dd}}\Big)\Big]\!+\cdots\, .
\label{VEap}
\ee
In the kinetic terms, the ellipses correspond to 2-derivatives, cubic and higher order terms in Wilson lines $y_{d,d+1}$,  $y_{d+1,d}$, $y_{d+1,d+1}$, while in $\Vone$ they stand for cubic and higher order interactions, or exponentially suppressed corrections when $c/\sqrt{G^{dd}}\gg 1$. In the following, we will neglect all of these subdominant contributions. 


\vskip .2cm
\noindent {\large \em Equation for $a$}

\noindent  Focusing on homogeneous and isotropic cosmological evolutions in flat space, we consider a metric and scalar field ansatz 
\begin{align}
&ds^2=-N(x^0)^2(dx^0)^2+a(x^0)^2\Big((dx^1)^2+\cdots+(dx^{d-1})^2\Big)\, , \nonumber \espD\\
&\Phi(x^0)\, , \; \phi_\bot(x^0)\, , \; y_{d,d+1}(x^0)\, , \; y_{d+1,d}(x^0)\, , \; y_{d+1,d+1}(x^0)\, . 
\end{align}
The equations of motion for the lapse function $N$ and scale factor $a$ take the following forms, in the gauge $N\equiv 1$ which defines cosmic time $x^0\equiv t$,  
\begin{align}
{1\over 2}\, (d-1)(d-2) H^2&=\K +\kappa^2 \Vone\, , 
\label{fri}\espD \\
(d-2) \dot H+{1\over 2}\, (d-1)(d-2)H^2&=-\K+\kappa^2\Vone\, , \label{e2}
\end{align}
where $H\equiv \dot a/a$ and the kinetic terms are
\be
\K={1\over2}\, \dot\Phi^2+{1\over 2}\, \dot\phi_\bot^2+{G^{dd}\over 4}\, \dot y_{d,d+1}^2+{G^{dd}\over 4}\, \dot y_{d+1,d}^2+{1\over 4}\, \dot y_{d+1,d+1}^2\, .
\ee
Interested in QNSRs, we eliminate $\K$ between Eqs~(\ref{fri}) and~(\ref{e2}), 
\be
{1\over d-1}\, {(a^{d-1})^{\dis \cdot\cdot}\over a^{d-1}}\equiv \dot H+(d-1)H^2={2\over d-2}\, \kappa^2\Vone\, ,
\label{vseul}
\ee
and look for cosmological evolutions satisfying either 
\begin{align}
a(t)\underset{t-t_+\to +\infty}{\longrightarrow} +\infty \qquad \mbox{or} \qquad a(t)\underset{t-t_-\to 0_+}{\longrightarrow} 0\, , 
\end{align}
for some constants $t_\pm$, with the effective potential dominated by $H^2$. To be specific, we assume the solutions to satisfy 
\be
\kappa^2 \Ms^d\, e^{d\alpha \Phi} =\O\!\left({H^2\over a^{K_\pm}}\right)
\label{h2a}    
\ee
in the above limits, where $\pm K_\pm>0$ are constants to be determined. The $t-t_+\to +\infty$ asymptotic regime describes an ever-expanding universe, while $t-t_-\to 0_+$ corresponds to a Big Bang arising at $t=t_-$. Of course, contracting evolutions in QNSR may also be found by time reversal.  
Under these hypotheses, and supposing a power law behavior of the scale factor, Eq.~(\ref{vseul}) can be integrated once, 
\be
C_\pm-{1\over H}=-(d-1)(t-t_\pm)\!\left(1+\O\!\left({1\over a^{K_\pm}}\right)\right) .
\ee
Without loss of generality, the constant $C_+$ can be absorbed in a redefinition of $t_+$, while $C_-$ has to vanish for $a(t)$ to vanish at $t_-$.  Integrating a second time, one obtains
\be
\label{a0}
a=\Am (t-t_\pm)^{1\over d-1} \left(1+\O\!\left({1\over a^{K_\pm}}\right)\right) ,
\ee
where $\Am>0$ is a not yet specified constant. Up to the subdominant term $\O(1/a^{K_\pm})$, the time-dependence of the scale factor is by no way surprising since a negligible potential energy implies the evolution of the  universe to be driven by the moduli kinetic energies, \ie a cosmic fluid of energy density $\rho$ and pressure $P$ satisfying $\rho\sim P$.


\vskip .2cm
\noindent {\large \em Equation for $y_{d+1,d}$}

\noindent At quadratic order in Wilson line deformations, $y_{d+1,d}$ has a vanishing potential but a non-canonical kinetic term. Thus, its equation of motion is that of a free field, with non-conventional friction term, 
\be
\ddot y_{d+1,d}+\big[(d-1)H+(\ln G^{dd})^{\dis \cdot}\big]\, \dot y_{d+1,d}=0\, , 
\label{er1}
\ee
which yields
\be
\dot y_{d+1,d}= {2\, c_{d+1,d}\over a^{d-1}G^{dd}}\, ,
\label{r1}
\ee 
where $c_{d+1,d}$ is an integration constant.  
A consequence of Eq.~(\ref{a0}) is that the l.h.s. of Friedmann equation (\ref{fri}) is 
\be
{1\over 2}\, (d-1)(d-2)H^2={d-2\over 2(d-1)}\, {\Am^{2(d-1)}\over a^{2(d-1)}} \left(1+\O\!\left({1\over a^{K_\pm}}\right)\right) , 
\ee
while the kinetic and potential terms in the r.h.s. satisfy
\be
\K\ge {c_{d+1,d}^2\over a^{2(d-1)}G^{dd}}\, ,  \qquad |\kappa^2 \Vone|\ll H^2\, . 
\ee
For these facts to be consistent, we proceed by assuming a power law behavior 
\be
G^{dd}\sim\Gm(t-t_\pm)^{J_\pm}\, , 
\label{G}
\ee
for some coefficient $\pm J_{\pm} > 0$ to be determined,  and a constant $\Gm>0$. In this case, the kinetic term of $y_{d+1,d}$ is subdominant in $\K$,
\be
H^2\O_1\equiv \O\!\left(G^{dd}\dot y_{d+1,d}^2\right)\!=\O\!\left({c_{d+1,d}^2\over \Am^{2(d-1)}}{H^2\over G^{dd}}\right)\!\ll \K=\O(H^2)\, .
\ee 
In the end, we obtain
\be
\label{y21}
 y_{d+1,d}\simeq  y_{d+1,d}^{(0)}- {\Cm_{d+1,d}\over J_{\pm} (t-t_\pm)^{J_{\pm}}}\, ,\quad \where \quad \Cm_{d+1,d}={2\, c_{d+1,d}\over \Am^{d-1}\G}\, ,
\ee
and the second integration constant satisfies $|y_{d+1,d}^{(0)}| \ll 1$. 

Notice that  in the QNSR $t-t_+\to +\infty$, the initial hypothesis~(\ref{h2a}) implies $M$ to drop. This is also the case for the QNSR $t-t_-\to 0_+$, if ${|K_-|\over d-1}>2$. 
On the contrary, Eq.~(\ref{G})  implies the supersymmetry breaking scale measured in $\sigma$-model frame, $M_{(\sigma)}$, to rise and formally tend to infinity, when $t-t_+\to +\infty$ (or $t-t_-\to 0_+$). This means that in the QNSRs, $t$ should not exceed some maximal value $\tf$ (or reach values below $\tf$) such that $G^{dd}(\tf)=c^2$. After (or before) $\tf$, the exponential terms in the effective potential~(\ref{Vap}) are no more suppressed.\footnote{When $c=\O(1)$, Hagedorn-like transitions may even occur when $G^{dd}=\O(1)$.}  


\vskip .2cm
\noindent {\large \em Equation for $y_{d+1,d+1}$}

\noindent In order to determine $y_{d+1,d+1}$ in the QNSRs, one can insert in its equation of motion,
\be
\label{y}
\ddot y_{d+1,d+1}+(d-1) H\dot y_{d+1,d+1}+{2v_{d-2}\over \pi} \, c_{\G_1}  \kappa^2\Ms^d\, {e^{d\alpha\Phi}\over G^{dd}}\, y_{d+1,d+1}=0\, , 
\ee
the behaviors of $(d-1)H\sim 1/(t-t_\pm)$, $G^{dd}\sim \Gm (t-t_\pm)^{J_\pm}$ and  $e^{d\alpha\Phi}\sim \# H^2/a^{K_\pm}$. For $c_{\G_1}>0$, the generic solution of the differential equation can be expressed in terms of Bessel functions of the first kind, $J_0$, and second kind, $Y_0$,
\be
y_{d+1,d+1}=C \, J_0\!\left({L\over (t-t_\pm)^{{1\over 2}(J_\pm+{K_\pm\over d-1})}}\right)\!+C' \, Y_0\!\left({L\over (t-t_\pm)^{{1\over 2}(J_\pm+{K_\pm\over d-1})}}\right),
\ee
where $L>0$ and the arbitrary $C$, $C'$ are constants.  For $c_{\G_1}<0$, the Bessel functions are ``modified'' into $I_0$ and $K_0$. In both cases, the value of $L$ is irrelevant when taking the limit $t-t_+\to +\infty$ or $t-t_-\to 0_+$, and we obtain
\be
\label{ysol}
y_{d+1,d+1}\simeq \Cm_{d+1,d+1}\ln{t-t_\pm\over t_0-t_\pm}\, ,
\ee
where $\Cm_{d+1,d+1}$ and $t_0$ are constants. 
Notice that this logarithmic behavior  is not in contradiction with the smallness of $y_{d+1,d+1}$ we have assumed in Eq.~(\ref{small}). This follows from the fact that in a QNSR, $|y_{d+1,d+1}|/  \sqrt{G^{dd}}$ decreases, due to the power-dependence of $G^{dd}$ in time. Physically, the supersymmetry breaking scale in $\sigma$-model frame $M_{(\sigma)}$ grows faster than the Higgs mass $|y_{d+1,d+1}| \Ms$. 

Before proceeding, it is instructive to use  
\be
\dot y_{d+1,d+1}\sim {\Cm_{d+1,d+1}\over t-t_\pm}\, ,
\ee
in order to evaluate the mass term, 
\be
\label{O2}
\O\!\left(\kappa^2\Ms^d\,{e^{d\alpha\Phi}\over G^{dd}}\, y_{d+1,d+1}\right)\!=\O\!\left(H  \dot y_{d+1,d+1}{1\over a^{K_{\pm}}G^{dd}}\ln{t-t_\pm\over t_0-t_\pm}\right)\!\equiv H\dot y_{d+1,d+1}\O_2\, .
\ee
With this result, Eq.~(\ref{y}) becomes
\be
\ddot y_{d+1,d+1}+(d-1) H \dot y_{d+1,d+1}\big(1+\O_2\big)\!=0\, ,
\ee
which can be  integrated once to yield the more accurate result
\be
\label{Cc}
\dot y_{d+1,d+1}={2\, c_{d+1,d+1}\over a^{d-1}}\big(1+\O_2\big)\, , \quad\; \where\; \quad\Cm_{d+1,d+1}={2\, c_{d+1,d+1}\over \Am^{d-1}}\, .
\ee


\vskip .2cm
\noindent {\large \em Equation for $y_{d,d+1}$}

\noindent As before, one can solve the equation of motion of $y_{d,d+1}$, 
\be
\ddot y_{d,d+1}+\big[(d-1)H+(\ln G^{dd})^{\dis \cdot} \big]\, \dot y_{d,d+1}+{2v_{d-2}\over \pi} \, (d-1)\,c_{\G_1} \kappa^2\Ms^d\, {e^{d\alpha\Phi}\over G^{dd}}\, y_{d,d+1} =0\, , 
\label{x}
\ee
after substituting $H$, $G^{dd}$ and  $e^{d\alpha\Phi}$ with their limit behaviors. For $c_{\G_1}>0$, the generic solution turns out to be expressed in terms of Bessel functions of the first kind, 
\be
\label{y12}
y_{d,d+1}={C \over (t-t_\pm)^{J_\pm\over2}}\, J_k\!\left({L\over (t-t_\pm)^{{1\over 2}(J_\pm+{K_\pm\over d-1})}}\right)\!+{C' \over(t-t_\pm)^{J_\pm\over2}}\,J_{-k}\!\left({L\over (t-t_\pm)^{{1\over 2}(J_\pm+{K_\pm\over d-1})}}\right),
\ee
where $k=J_{\pm}/(J_\pm+{K_\pm\over d-1})$. For $c_{\G_1}<0$, the Bessel functions are ``modified'', $J_k,J_{-k}\to I_k,I_{-k}$.  In the limit $t-t_+\to +\infty$ or $t-t_-\to 0_+$ we are interested in, this leads to 
\be
\label{ydd+1}
y_{d,d+1}\simeq y_{d,d+1}^{(0)}-{\Cm_{d,d+1}\over J_\pm(t-t_\pm)^{J_\pm}}\equiv y_{d,d+1}^{(0)}(1+\tilde \O_1)\, , 
\ee
where $y_{d,d+1}^{(0)}$ and $\Cm_{d,d+1}$ are integration constants, with $|y_{d,d+1}^{(0)}| \ll1$.
Alternatively, one can write
\be
\label{r2}
\dot y_{d,d+1}\sim {2\, c_{d,d+1}\over a^{d-1}G^{dd}}\, , \qquad \where \qquad\Cm_{d,d+1}={2\, c_{d,d+1}\over \Am^{d-1}\G}\, .
\ee
The kinetic energies of $y_{d,d+1}$ and $y_{d+1,d}$ are thus of same order, 
\be
H^2\O_1\equiv \O\!\left(G^{dd}\dot y_{d,d+1}^2\right)\!=\O\!\left({c_{d,d+1}^2\over \Am^{2(d-1)}}{H^2\over G^{dd}}\right)\!\ll \K=\O(H^2)\, .
\ee 
 

\vskip .2cm
\noindent {\large \em Equation for $\phi_\bot$}

\noindent Once Wilson lines are taken into account, the scalar $\phi_\bot$ is no longer a free field. Due to the fact that 
\be
\label{gdd}
G^{dd}=e^{{2\over\alpha} \Phi}\, e^{-{2\over \sqrt{d-1}}\phi_\bot}\, , 
\ee
$\phi_\bot$ couples non-trivially to kinetic and mass terms, and  its equation of motion is highly non-linear,
\be
\label{ebot}
\ddot\phi_\bot+(d-1) H \dot \phi_\bot=-{G^{dd}\over 2\sqrt{d-1}}\big(\dot y_{d,d+1}^2+\dot y_{d+1,d}^2\big)- {v_{d-2}\over \pi\sqrt{d-1}} \,c_{\G_1} \kappa^2\Ms^d\, {e^{d\alpha\Phi}\over  G^{dd}}\, y_{d+1,d+1}^2\,.
\ee 
However,  up to a numerical factor, the two terms in the r.h.s. show up respectively in $\K$ and $\kappa^2\Vone$. We have already seen that the former is of order $H^2\O_1$, while the second is of order 
\be
\label{o3}
H^2\O_3\equiv \O\!\left(\kappa^2\Ms^d\, {e^{d\alpha\Phi}\over  G^{dd}}\, y_{d+1,d+1}^2\right)\!= \O\!\left({H^2\over a^{K_\pm}}{y_{d+1,d+1}^2\over G^{dd}}\right)\ll  H^2\, .
\ee
For reasons that will become clearer later, it is useful to explain the term $\O_1$. 
Assuming $\dot \phi_\bot=\O(H)$, we define the constant $C_\bot$ such that 
\be
\label{Cor}
{1\over a^{2(d-1)}}\sim C_\bot (d-1)H\dot \phi_\bot\, , 
\ee
and write Eq.~(\ref{ebot}) in the following form,
\be
\ddot\phi_\bot+(d-1)H \dot \phi_\bot \left(1+{2C_\bot\over \sqrt{d-1}}\, {c_{d,d+1}^2+c_{d+1,d}^2\over G^{dd}}+\cdots+\O_3\right)=0\, 
\ee
where the ellipses stand for subdominant contributions in the $\O_1$ term.  
Integrating once, we obtain 
\be
\dot \phi_\bot=\sqrt{2}\, {c_\bot\over a^{d-1}} \left(1+{2C_\bot\over \sqrt{d-1}}{c_{d,d+1}^2+c_{d+1,d}^2\over J_\pm\,  G^{dd}}+\cdots +\O_3\right), 
\label{solbot}
\ee
where $c_\bot$ is an arbitrary  constant. Using the above result, 
Eq.~(\ref{Cor}) is consistent and we can identify 
\be
C_\bot={1\over \sqrt{2} \, c_\bot \, \Am ^{d-1}}\, .
\ee


\vskip .2cm
\noindent {\large \em Equation for $\Phi$}

\noindent The treatment of the no-scale modulus $\Phi$ can be similar. Its equation of motion, 
\be
\alpha\ddot\Phi+(d-1) H \alpha \dot \Phi=-d\alpha^2\kappa^2\Ms^d \Vone+{G^{dd}\over 2}\big(\dot y_{d,d+1}^2+\dot y_{d+1,d}^2\big)\!+ {v_{d-2}\over \pi} \,c_{\G_1}\kappa^2\Ms^d\, {e^{d\alpha\Phi}\over  G^{dd}}\, y_{d+1,d+1}^2\,,
\ee 
can be linearly combined with Eq.~(\ref{vseul}) to eliminate the term proportional to $\Vone$. One obtains
\begin{align}
\Big(\alpha\dot\Phi+{\alpha^2\over 2}\, d(d-2)H\Big)^{\displaystyle\cdot} &+(d-1)\, H\Big(\alpha \dot\Phi+{\alpha^2\over 2}\, d(d-2)H\Big)=\nonumber \\
&{G^{dd}\over 2}\big(\dot y_{d,d+1}^2+\dot y_{d+1,d}^2\big)\!+ {v_{d-2}\over \pi} \,c_{\G_1}\kappa^2\Ms^d\, {e^{d\alpha\Phi}\over  G^{dd}}\, y_{d+1,d+1}^2\, , \esp
\label{phieq}
\end{align}
which is an equation whose form is identical to that of $\phi_\bot$. Thus, assuming $\dot\Phi=\O(H)$, we can proceed in a similar way to obtain 
\be
\alpha \dot\Phi+{\alpha^2\over 2}\, d(d-2)H={c_\Phi\over a^{d-1}}\left(1-{2C_\Phi}{c_{d,d+1}^2+c_{d+1,d}^2\over J_\pm\,  G^{dd}}+\cdots+\O_3\right) ,
\label{dphi}
\ee
where $c_\Phi$ is an arbitrary constant and 
\be
C_\Phi={1\over c_\Phi \, \Am^{d-1}}\, .
\ee


\vskip .2cm
\noindent {\large \em Friedmann constraint}

\noindent We started our discussion by solving Eq.~(\ref{vseul}) for the scale factor $a(t)$, which introduced two integration constants $\Am$ and $t_\pm$ in the solution~(\ref{a0}). However, Friedmann differential equation~(\ref{fri}) being only first-order, it can be used to fix $\Am$ in terms of the other parameters. 

To reach this goal, we first collect all results found for the scalar fields to write  the total kinetic energy as
\begin{align}
\label{Kfor}
\K=&\; {1\over 8}\, d^2(d-2)^2\alpha^2H^2-{1\over 2}\,d(d-2)H{c_\Phi\over a^{d-1}}+{{c^2_\Phi\over 2\alpha^2}+c_\bot^2+c_{d+1,d+1}^2\over a^{2(d-1)}}\nonumber \espD \\
&+{c_{d,d+1}^2+c_{d+1,d}^2\over a^{2(d-1)}G^{dd}}\, (C_\K+\cdots)+H^2\O_3+{c^2_{d+1,d+1}\over a^{2(d-1)}}\, \O_2\, ,
\end{align}
where  
\be
C_\K= 1+{1\over J_\pm}\!\left({\sqrt{2\over d-1}}\, {2c_\bot\over \Am^{d-1}}-{2\over \alpha^2}\, {c_\Phi\over \Am^{d-1}}+{d\over \alpha^2}\right).
\ee
In the expression of $\K$, all terms in the first line are $\O(H^2)$. In the second line,  the contribution proportional to $1/(a^{2(d-1)}G^{dd})$ and $H^2\O_3$ arise from the kinetic terms of $y_{d,d+1}$ and $y_{d+1,d}$, as well as the subdominant contributions of those associated with  $\phi_\bot$ and $\Phi$. Moreover, the last term, which is the subdominant part of the kinetic energy of $y_{d+1,d+1}$, can be compared to the other contributions by noticing that 
\be
{c^2_{d+1,d+1}\over a^{2(d-1)}}\, \O_2=\O\big(H^2\Cm_{d+1,d+1}^2\big) \O_2=\O\!\left({H^2\over a^{K_\pm}}{y_{d+1,d+1}^2\over G^{dd}}{1\over \ln\big({t-t_\pm\over t_0-t_\pm}\big)}\right)=H^2{\O_3\over \ln\big({t-t_\pm\over t_0-t_\pm}\big)}\, .
\ee
The latter being dominated by $H^2\O_3$, it can be omitted in Eq.~(\ref{Kfor}). 
In a similar spirit, the 1-loop effective potential can be written as
\be
\Vone=e^{d\alpha\Phi} \Ms^d \Big[(\nF-\nB)\,  v_{d}+ {v_{d-2}\over 2\pi} \,  c_{\G_1}(d-1){{y^{(0)}}^2}_{\!\!\!\!\!\!\!\!\!\!\!\!d,d+1}(1+\tilde\O_1)\Big] \!+H^2\O_3\, .
\ee
Finally, we may follow Ref.~\cite{CFP} by defining   
\be
\label{tau}
\tau\equiv {(d^2-4)(d-1)\over 2dc_\Phi} \, H a^{d-1}={(d^2-4)\over 2dc_\Phi} \, (a^{d-1})^{\displaystyle\cdot} \, , 
\ee
in terms of which the l.h.s. of Friedmann equation~(\ref{fri}) and the dominant terms~$\O(H^2)$ of~$\K$ combine into a suitable form. The result is
\begin{align}
\label{friast}
-{d^2c_\Phi^2\over 2(d-1)(d+2)}\, {\P(\tau)\over a^{2(d-1)}}= &\; \kappa^2\Ms^d\, e^{d\alpha\Phi}  \Big[(\nF-\nB)\,  v_{d}+ {v_{d-2}\over 2\pi} \,  c_{\G_1}(d-1){{y^{(0)}}^2}_{\!\!\!\!\!\!\!\!\!\!\!\!d,d+1}(1+\tilde\O_1)\Big]\nonumber \\
&+{c_{d,d+1}^2+c_{d+1,d}^2\over a^{2(d-1)}G^{dd}}\, (C_\K+\cdots)+H^2\O_3\, , 
\end{align} 
where $\P$ is a quadratic polynomial, 
\be
\P(\tau)=\tau^2-2\tau+\!\left(1-{4\over d^2}\right)\!\!\Big(1+2\alpha^2\, {c_\bot^2+c_{d+1,d+1}^2\over c_\Phi^2}\Big)\, . 
\ee

In the limits we are interested in, the behavior~(\ref{a0}) of the scale factor implies $\tau$ to converge to a constant, 
\be
\label{taulim}
\tau=\tau_0\left(1+\O\!\left({1\over a^{K_\pm}}\right)\right) , \qquad \where \qquad \tau_0={d^2-4\over 2d} \, {\Am^{d-1}\over c_\Phi} \, ,
\ee
and $\P(\tau)$ to satisfy
\be
\P(\tau)=\P(\tau_0)+\O\!\left({1\over a^{K_\pm}}\right).
\ee
Thus, for Eq.~(\ref{friast}) to be consistent, two conditions must be fulfilled:

($i$) $\tau_0$ must be a root of $\P$. In this instance only, instead of being $\O(H^2)$, the l.h.s. of Eq.(\ref{friast}) satisfies 
\be
\label{frP}
-{d^2c_\Phi^2\over 2(d-1)(d+2)}\, {\P(\tau)\over a^{2(d-1)}}=\O\!\left({H^2\over a^{K_\pm}}\right).
\ee

($ii$) We must have
\be
\label{cii}
\O\!\left({H^2\over a^{K_\pm}}\right) \gg {c_{d,d+1}^2+c_{d+1,d}^2\over a^{2(d-1)}G^{dd}}\, (C_\K+\cdots)\, , \qquad   \O\!\left({H^2\over a^{K_\pm}}\right)\gg H^2\O_3 \, ,
\ee
for our initial defining assumption of a QNSR to be true, Eq.~(\ref{h2a}). 

Condition $(i)$ requires the discriminant of $\P$ to be positive, which amounts to having
\be
\Big({c_\bot\over \gamma_{\rm c} c_\Phi}\Big)^2+\Big({c_{d+1,d+1}\over \gamma_{\rm c} c_\Phi}\Big)^2\le 1\, ,\quad  \where \quad \gamma_{\rm c}=\sqrt{2\over (d-1)(d+2)}\, .
\label{c}
\ee 
In this case, the value of $\Am$, which appears in the definition of $\tau_0$, is determined up to a sign~$\epsilon$, 
\be
\Am= \left[{2d\over d^2-4} \, (1+\epsilon\, r) \, c_\Phi\right]^{1\over d-1}\, ,
\label{rA}
\ee
where $c_\Phi>0$ is required and $r$ defined as
\be
r={2\over d}\; \sqrt{1-\Big({c_\bot\over \gamma_{\rm c} c_\Phi}\Big)^2-\Big({c_{d+1,d+1}\over \gamma_{\rm c} c_\Phi}\Big)^2}\, .
\label{rr}
\ee

In condition $(ii)$, the inequality that involves $\O_3$ is always satisfied, as follows from the decrease in $|y_{d+1,d+1}|/  \sqrt{G^{dd}}$, (see Eq.~(\ref{o3})). However, the other constraint may be more intriguing. If $C_\K\neq 0$, it would imply  $\pm J_\pm> \pm {K_\pm \over d-1}$,   
which would restrict the choices of integration constants characterizing the QNSRs (see the next paragraph). However, such a reduction of the set of solutions should not occur, since we have already solved all differential equations and the only remaining piece of information captured by Friedmann equation must be the value of~$\Am$. As we will now check, $C_\K$ actually does vanish. Moreover,  all implicit contributions in Eq.~(\ref{cii}) in the dots  should respect the inequality, without imposing further constraints on the existence of QNSRs, as will be checked numerically in~Sects~\ref{simu} and~\ref{global}.


\vskip .2cm
\noindent {\large \em Determination of $K_\pm$ and $J_\pm$}

\noindent Using the value of $\Am$, Eq.~(\ref{dphi})  yields 
\be
d\alpha\dot \Phi\sim  -\left(2+\epsilon \, {r(d^2-4)\over 2(1+\epsilon \, r)}\right)\!{1\over t-t_\pm} \, ,
\ee
where the overall coefficient is to be identified with $-(2+{K_\pm\over d-1})$, as follows from Eq.~(\ref{h2a}). The fact that $\pm K_\pm>0$ fixes $\epsilon =\pm$, and we obtain
\be
\label{drop}
e^{d\alpha\Phi}\sim{e^{d\alpha\Phi_{\pm}}\over [\Ms(t-t_\pm)]^{2+{K_\pm\over d-1}}}\, ,\qquad \where \qquad K_\pm=\pm  {r(d^2-4)\over 2(1\pm r)}
\ee
and $\Phi_\pm$ is an integration constant. 

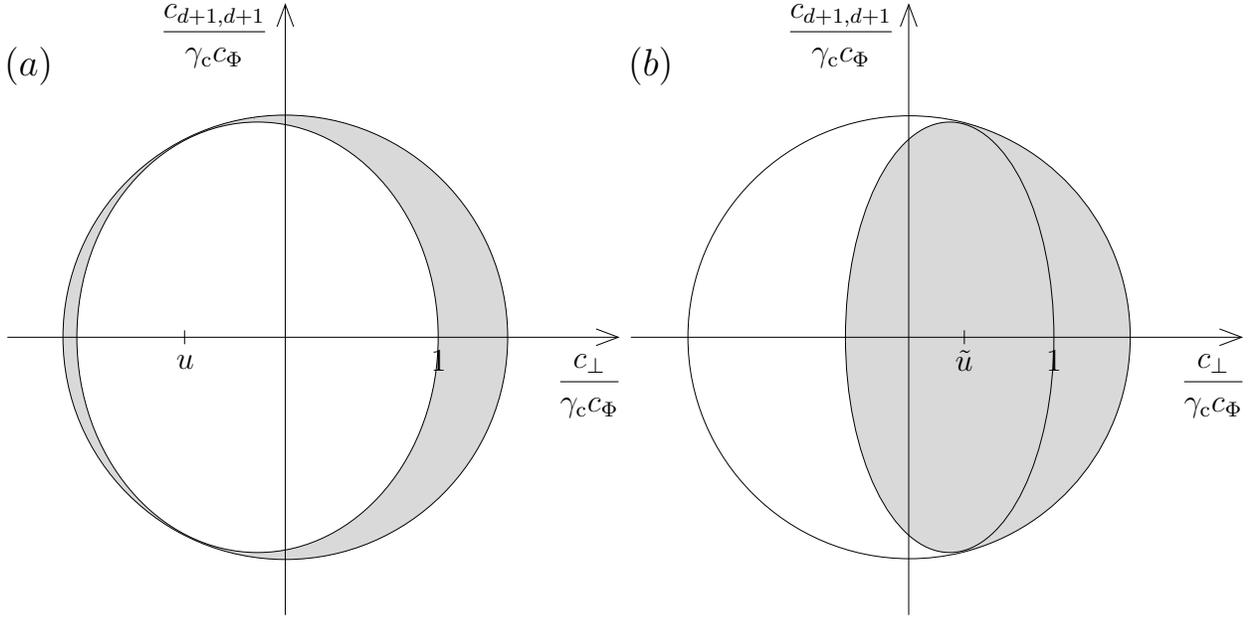
\begin{figure}[!h]
\vspace{.4cm}
\begin{center}
\includegraphics[width=8.15cm]{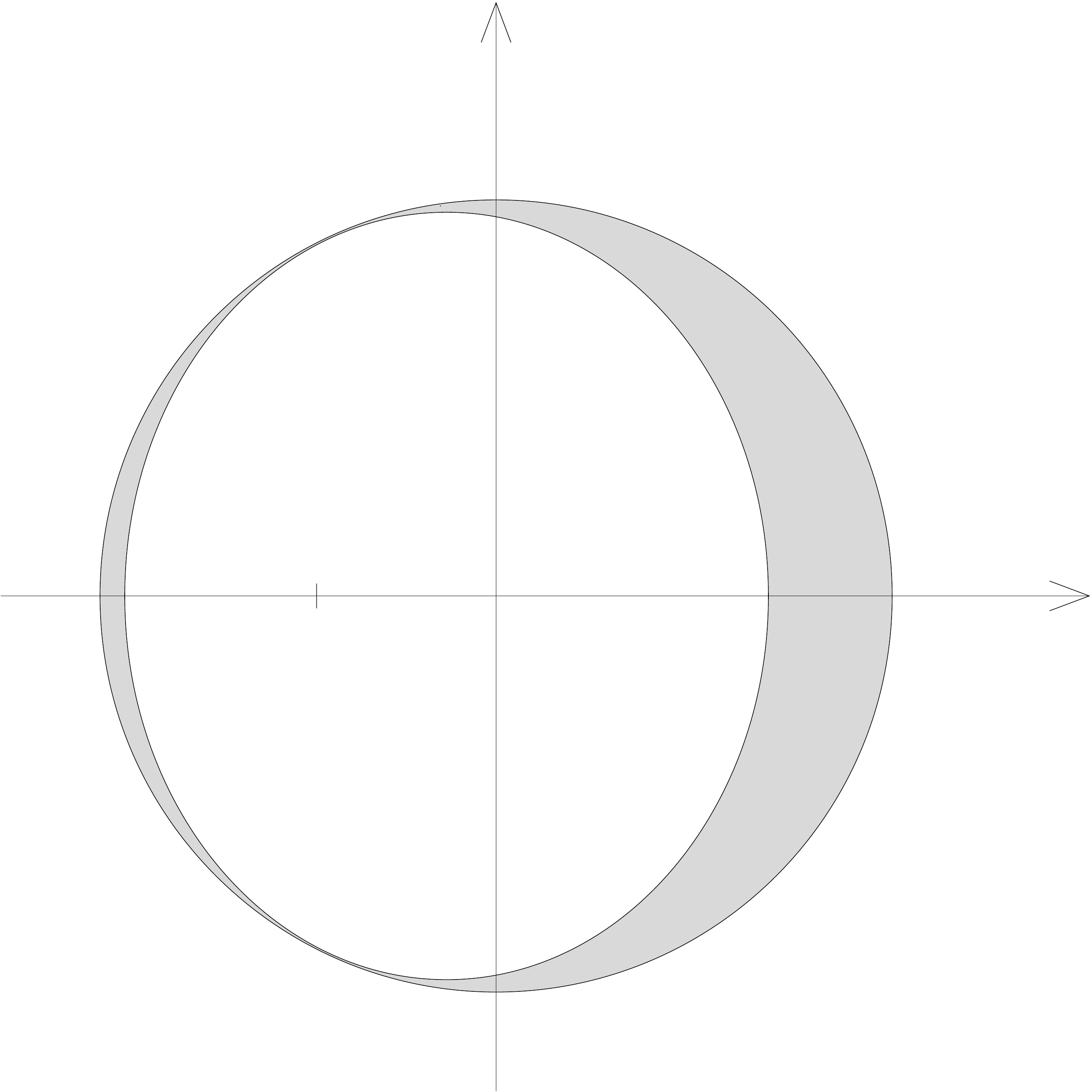}
\includegraphics[width=8.15cm]{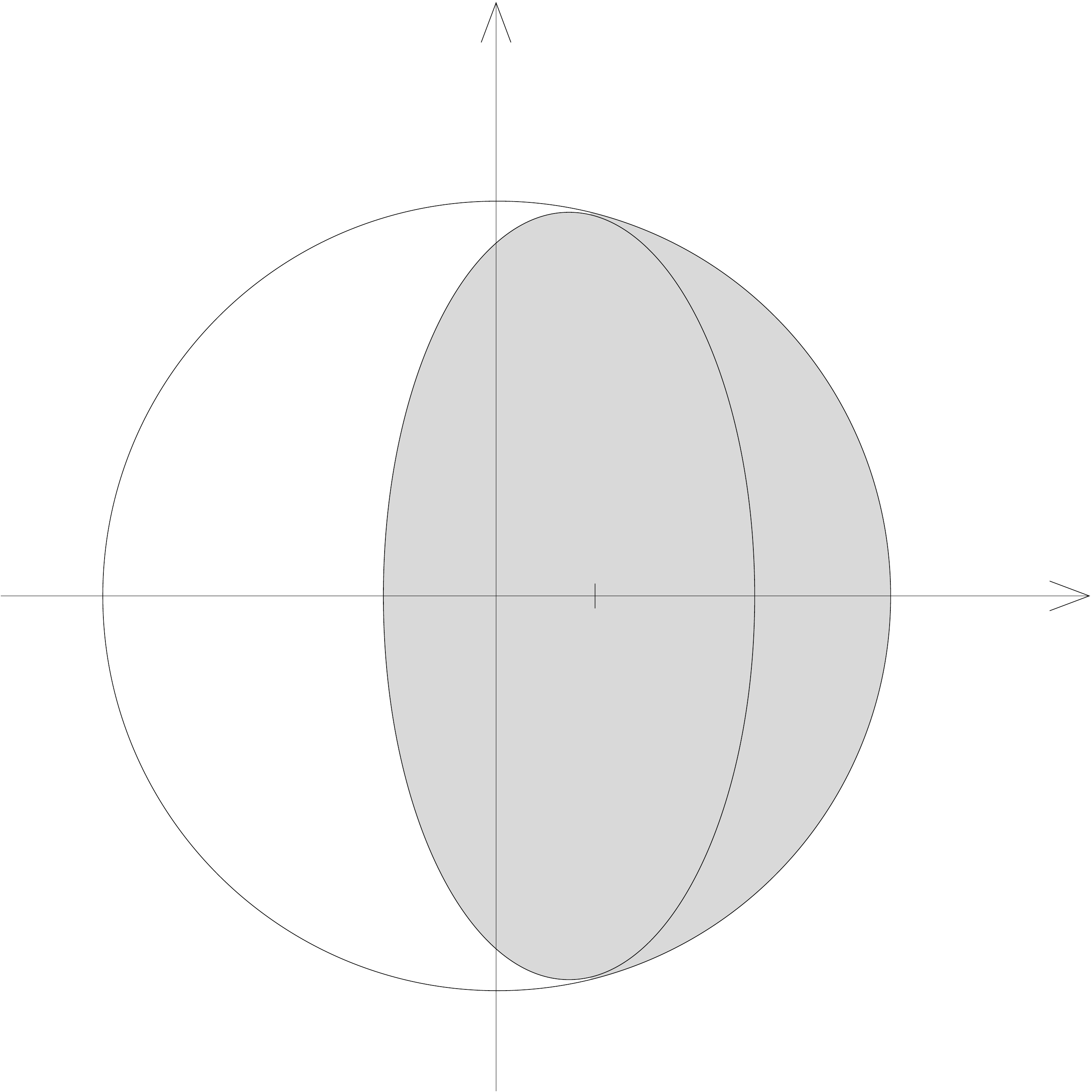}
\end{center}
\begin{picture}(0,0)
\put(209,113){$\dis {c_\bot\over \gamma_{\rm c} c_\Phi}$}
\put(445,113){$\dis {c_\bot\over \gamma_{\rm c} c_\Phi}$}
\put(59,245.5){$\dis {c_{d+1,d+1}\over \gamma_{\rm c} c_\Phi}$}
\put(296,245.5){$\dis {c_{d+1,d+1}\over \gamma_{\rm c} c_\Phi}$}
\put(65,121){$u$}
\put(359.7,120){$\tilde u$}
\put(161.3,120){1}
\put(394,120){1}
\put(00,232){\large($a$)}
\put(236,232){\large($b$)}
\end{picture}
\vspace{-.9cm}
\caption{\footnotesize \em  The points $\big({c_\bot\over \gamma_{\rm c}c_\Phi},{c_{d+1,d+1}\over \gamma_{\rm c}c_\Phi}\big)$ of the disk of  radius 1 that yield a QNSR $t-t_+\to +\infty$ sit in the left crescent of figure $(a)$. Those in the right crescent lead to a QNSR $t-t_-\to0 _+$. The former are always perturbative, while the latter are compatible with weak string coupling when $d\ge 3$  if $\big({c_\bot\over \gamma_{\rm c}c_\Phi},{c_{d+1,d+1}\over \gamma_{\rm c}c_\Phi}\big)$ is also located in the shaded era of figure $(b)$.}
\label{domaines}
\end{figure}

The coefficient $J_\pm$ can be determined in a similar way by using the linear relation between  $(\ln G^{dd})^{\dis \cdot}$, $\dot \Phi$ and $\dot\phi_\bot$. The result is 
\be
\label{j+}
J_\pm = {d\over \alpha^2}\Big(\big( 1-\alpha^2\sqrt{2\over d-1} \, {c_\bot\over c_\Phi}\big){1-{4\over d^2}\over 1\pm r}-1\Big)\, ,
\ee
which leads as anticipated to $C_\K=0$.
  With the expression of $J_\pm$, we are ready to solve the only non-trivial consistency condition for QNSRs to exist. The points $\big({c_\bot\over \gamma_{\rm c}c_\Phi},{c_{d+1,d+1}\over \gamma_{\rm c}c_\Phi}\big)$ of the disk of radius 1, Eq.~(\ref{c}), compatible with the constraint  
  \be
  \pm J_\pm > 0
  \ee 
  sit  outside an ellipse 
\begin{align}
&\Big({c_\bot\over \gamma_{\rm c}c_\Phi}-{u\over 1+v}\Big)^2+{v\over 1+v}\Big({c_{d+1,d+1}\over \gamma_{\rm c}c_\Phi}\Big)^2\ge {v\over (1+v)^2}(1+v-u^2)\, ,\nonumber \\
\where \qquad &u=-{2\over \sqrt{d+2}} \, , \qquad v= {d^2\over d+2} \, . \esp
\end{align}
As shown in Fig.~\ref{domaines}$(a)$, for arbitrary dimension $d>2$, this ellipse  is located in the interior of the disk and is tangential to it at ${c_\bot\over \gamma_{\rm c}c_\Phi}=u$. The points in the left crescent \big(${c_\bot\over \gamma_{\rm c}c_\Phi}\le u$\big) allow a QNSR $t-t_+\to +\infty$, while those in the right crescent \big(${c_\bot\over \gamma_{\rm c}c_\Phi}\ge u$\big) yield a regime $t-t_-\to 0_+$. In reality, the left crescent is more tiny than the one shown on the qualitative Fig.~\ref{domaines}$(a)$. Its width at   ${c_{d+1,d+1}\over \gamma_{\rm c}c_\Phi}=0$ is  3--$11\cdot10^{-3}$ for $3\le d\le 9$, and actually vanishes when $d\to 2_+$. 
Thus, we have $r\simeq 0$ in the left crescent, so that $M^d$ and $H^2$ evolve approximately at the same cosmological speed. 

Finally, we can make some remark about $\Cm_{d+1,d+1}$, which is related to $c_{d+1,d+1}$ in Eq.~(\ref{Cc}) and must be small, as required by our assumption on $y_{d+1,d+1}$ given in  Eq.~(\ref{small}).
If the left and right crescents allow $\Cm_{d+1,d+1}$ to be as small as desired, its maximal value is reached for $\big({c_\bot\over \gamma_{\rm c}c_\Phi},{c_{d+1,d+1}\over \gamma_{\rm c}c_\Phi}\big)=(0,1)$, which yields 
\be
|\Cm^{\rm max}_{d+1,d+1}|=\sqrt{2(d+2)\over d-1} \, {d-2\over d}\, . 
\ee 
This expression being of order~1, it is consistent as a limiting case. 


\vskip .2cm
\noindent {\large \em Perturbative condition}

\noindent At this stage, we have found time-dependent fields that extremize the 1-loop effective action in the limit $t-t_+\to \infty$ or $t-t_-\to 0_+$. To make sense, however, this analysis requires string perturbation theory to be valid in these regimes. Expressing $\dot \phi$ as a linear combination of  $\dot \Phi$ and $\dot\phi_\bot$, one obtains
\be
e^{2d\alpha^2\phi}\sim{e^{d\alpha\Phi_{\pm}}\, e^{d\sqrt{d-1}\, \phi_{\bot\pm}}\over [\Ms(t-t_\pm)]^{P_\pm\over d-1}}\, ,
\ee
where $\phi_{\bot\pm}$ is the constant arising by integration of Eq.~(\ref{solbot}) and 
\be
\label{p+-}
P_\pm={K_\pm\over (1-{4\over d^2})\, r}\!\left[r\pm \Big({4\over d^2}-\big(1-{4\over d^2}\big)\sqrt{2(d-1)}\; {c_\bot\over c_\Phi}\Big)\right].
\ee

The QNSR $t-t_+\to +\infty$  happens to be perturbative, since $P_+>0$ is always satisfied when the point $\big({c_\bot\over \gamma_{\rm c}c_\Phi},{c_{d+1,d+1}\over \gamma_{\rm c}c_\Phi}\big)$ sits in the left crescent in Fig.~\ref{domaines}$(a)$. For the regime $t-t_-\to 0_+$, string perturbation theory is valid when  $P_-<0$. If $d\ge d_{\rm c}\simeq 2.90$, this condition is satisfied for ${c_\bot\over \gamma_{\rm c}c_\Phi}\ge \tilde u$ or when $\big({c_\bot\over \gamma_{\rm c}c_\Phi},{c_{d+1,d+1}\over \gamma_{\rm c}c_\Phi}\big)$ sits in the interior of an ellipse 
\begin{align}
&\Big({c_\bot\over \gamma_{\rm c}c_\Phi}-{\tilde u\over 1+\tilde v}\Big)^2+{\tilde v\over 1+\tilde v}\Big({c_{d+1,d+1}\over \gamma_{\rm c}c_\Phi}\Big)^2\le {\tilde v\over (1+\tilde v)^2}(1+\tilde v-\tilde u^2)\, ,\nonumber \\
\where \qquad &\tilde u={2\over (d-2)\sqrt{d+2}} \, , \qquad \tilde v= {d^2\over (d-2)^2(d+2)}\, . \esp
\label{ell2}
\end{align}
As shown in Fig.~\ref{domaines}$(b)$, this second ellipse is inside the disk of radius 1 and tangential to it at ${c_\bot\over \gamma_{\rm c}c_\Phi}=\tilde u$. As a result, the right part of the right crescent in Fig.~\ref{domaines}$(a)$ yields perturbative QNSRs $t-t_-\to 0_+$.  If $2<d<d_{\rm c}$, $\tilde u$ being greater than 1, the  condition $P_-<0$ is true only inside the ellipse~(\ref{ell2}), which now sits entirely in the interior of the disk. However,  the intersection of this perturbative domain with the right crescent of in Fig~\ref{domaines}$(a)$ is always non-empty. It is only in the limit $d\to 2_+$, where the ellipse~(\ref{ell2}) vanishes, that the QNSRs  $t-t_-\to 0_+$ are always formal because non-perturbative (unless we fine tune $\big({c_\bot\over \gamma_{\rm c}c_\Phi},{c_{d+1,d+1}\over \gamma_{\rm c}c_\Phi}\big)$ to be exactly $(0,0)$).  

To summarize, the QNSRs we have found in arbitrary dimension $d>2$ depend on 5~velocity parameters $c_\Phi>0$, $c_\bot$, $c_{d+1,d+1}$,  $c_{d,d+1}$, $c_{d+1,d}$, 5~zero modes $\Phi_\pm$, $\phi_{\bot\pm}$, $t_0$, $y^{(0)}_{d,d+1}$, $y^{(0)}_{d+1,d}$, and the last constant $t_\pm$ arising by integration of the scale factor. Therefore, they are limit behaviors of generic solutions, even if  the left crescent in Fig.~\ref{domaines}$(a)$ is tiny. 


\section{Simulations of QNSRs at small Wilson lines}
\label{simu}

From now on, we study numerically the dynamics of the scale factor~$a$, no-scale modulus~$\Phi$ and scalar~$\phi_\bot$, in the presence of the moduli fields~$y_{d,d+1}$, $y_{d+1,d}$, $y_{d+1,d+1}$. Our goal in the present section is to check the validity of QNSRs described in Sect.~\ref{QNSR}, where the Wilson lines implement small deformations of the initial background. We fix in the analysis the spacetime dimension to be $d=4$ and focus only on the expanding solutions where $t\to +\infty$ (we set $t_+=0$). 
In all simulations, we take $\langle \phi_{\rm dil}\rangle=0$ so that $\kappa^2=1/\Ms^2$, thus identifying the Planck mass with the string scale. This has the advantage of matching the weak string coupling condition with the negativity of $\phi$. Once the range of time compatible with perturbation theory is identified, it is always possible to restore a sensible value of the Planck mass by shifting the dilaton zero-mode.

As can be seen in Eq.~(\ref{drop}), one feature of the QNSR $t\to +\infty$ is that the supersymmetry breaking scale $M$ always drops. If this may be expected when $\Vone$ is positive, it may be counterintuitive  when it is negative, since $M$ climbs the potential in this case. Moreover, the behaviors of $y_{d,d+1}$ and $y_{d+1,d+1}$ are independent of the fact that these moduli are massive or tachyonic at 1-loop. To check these striking properties, we simulate solutions of the differential equations of Sect.~\ref{QNSR}, which are valid for small Wilson lines deformations. This is done for 4 initial backgrounds characterized by different signs for $\nF-\nB$ and $c_{\G_1}$:

$(i)$ $\nF-\nB>0$, $c_{\G_1}>0$: This case can be achieved in Example 2 of Sect.~\ref{V1l}. For $d=4$ and $s=4$, the right-moving gauge group is $U(1)\times \G_1\times U(1)^4\times SO(16)\times SO(16)'$, where $\G_1=SU(2)$, which corresponds in the setup described below Eq.~(\ref{Bi}) to $\eta^R_5$ even. In this model, one obtains $\nF-\nB=8\times 6$ and $c_{\G_1}=8\times 2$. 

$(ii)$ $\nF-\nB>0$, $c_{\G_1}<0$: To flip the sign of $c_{\G_1}$, it is enough to choose $\eta^R_5$ odd in setup~$(i)$. This yields $\G_1=U(1)$, with $\nF-\nB=8\times 10$ and $c_{\G_1}=-8\times 2$.

$(iii)$ $\nF-\nB<0$, $c_{\G_1}>0$: This case can be realized in Example 1 of Sect.~\ref{V1l}. The right-moving gauge group for $d=4$ is $U(1)\times \G_1\times SU(2)^4\times E_8\times E_8'$, where $\G_1=SU(2)$, which corresponds to $\eta^R_5$ even. This leads to  $\nF-\nB=-8\times 514$ and $c_{\G_1}=8\times 2$.

$(iv)$ $\nF-\nB<0$, $c_{\G_1}>0$: To flip the sign of $c_{\G_1}$, one can take $\eta^R_5$ odd in setup~$(iii)$. This yields $\G_1=U(1)$,  $\nF-\nB=-8\times 510$ and $c_{\G_1}=-8\times 2$.

To set initial conditions adapted to our purpose, we proceed as follows:

$\bullet$ We  consider the case analyzed in Ref.~\cite{CFP}, where no $y$-deformation is implemented. All trajectories that reach a QNSR $t\to +\infty$ are characterized by two constants $c_{\bot0}$, $c_{\Phi0}$ (defined as $c_\bot$ and $c_\Phi$ in the present work) such that $\left|{c_{\bot0}\over \gamma_cc_{\Phi0}}\right|<1$. In order to allow the Wilson lines to vary, we take $c_{\bot0}$, $c_{\Phi0}$ for the expression of $J_+$ in Eq.~(\ref{j+}) evaluated at $(c_\bot, c_\Phi,c_{55})=(c_{\bot0}, c_{\Phi0},0)$ to be positive. This imposes $\approx \!\!0.9941 <\left|{c_{\bot0}\over \gamma_cc_{\Phi0}}\right|<1$, thus reducing the allowed range of this ratio by approximately a factor of 170. 

$\bullet$ In the presence of Wilson lines, we define dynamical quantities 
\be
c^{\dyn}_{\bot}=\frac{ a^{d-1}}{\sqrt{2}}\, \dot \phi_\bot\, , ~\quad c^{\dyn}_{\Phi}=a^{d-1}\!\left(\alpha\dot{\Phi}+{\alpha^2\over 2}\, d(d-2)H\right), \quad ~ c^{\dyn}_{55}=\frac{ a^{d-1}}{2}\, \dot y_{55}\, ,
\ee
which are expected to converge to the constants $c_\bot$, $c_\Phi$ and $c_{55}$ introduced in Sect. \ref{QNSR}. 

$\bullet$ For the initial conditions at $t=0$, we set $a(0)$ to be of order 1 and $c^{\dyn}_{\bot}(0)=c_{\bot0}$ to fix $\dot \phi_\bot(0)$. We also take  $c^{\dyn}_{\Phi}(0)=c_{\Phi0}$, which can be translated into $\dot\Phi(0)$ by  the knowledge of $H(0)$. The latter, which we take to be positive,  is determined by Friedmann equation at~$t=0$, for a given choice of $\Phi(0)$ and initial conditions for the Wilson lines.  To ensure that the evolution of the system starts close to the QNSR expected to arise at late times,~$\Phi(0)$ is chosen for the quantity
\be
\label{taud}
\tau^{\dyn}\equiv {(d^2-4)(d-1)\over 2dc_\Phi^\dyn} \, H a^{d-1} \, , 
\ee
which is inspired by Eq.~(\ref{tau}), to be at $t=0$ very close to the asymptotic value it would reach when no Wilson lines are introduced. To be specific, this means
\be
\label{r0}
\tau^{\dyn}(0)\simeq 1+r_0, \quad \where\quad r_0={2\over d}\; \sqrt{1-\Big({c_{\bot0}\over \gamma_{\rm c} c_{\Phi0}}\Big)^2}\, ,
\ee
as follows from Eqs~(\ref{taulim}) and~(\ref{rA}).
The choice of $\phi_\bot(0)$ is of order 1 and such that the range of cosmic time compatible  with weak string coupling and $M_{(\sigma)}(t)<c\Ms$ is large in the simulations. 

$\bullet$ The remaining initial conditions are those of the Wilson lines: $y_{45}(0)$, $\dot y_{45}(0)$, $y_{54}(0)$, $\dot y_{54}(0)$ and $y_{55}(0)$, $\dot y_{55}(0)$. Their absolute values are chosen small enough (compared to 1 and~$\Ms$) for the trajectory of $\left.\left({c^{\dyn}_\bot\over \gamma_{\rm c}c^{\dyn}_\Phi},{c^{\dyn}_{55}\over \gamma_{\rm c}c^{\dyn}_\Phi}\right)\!\right|_t$ to be entirely in the left crescent in Fig.~\ref{domaines}$(a)$. The motion of this point is expected to converge to $\left({c_\bot\over \gamma_{\rm c}c_\Phi},{c_{55}\over \gamma_{\rm c}c_\Phi}\right)$, when $t\to +\infty$. 

\begin{figure}[!h]
\vspace{.3cm}
\begin{center}
\includegraphics[width=8.18cm]{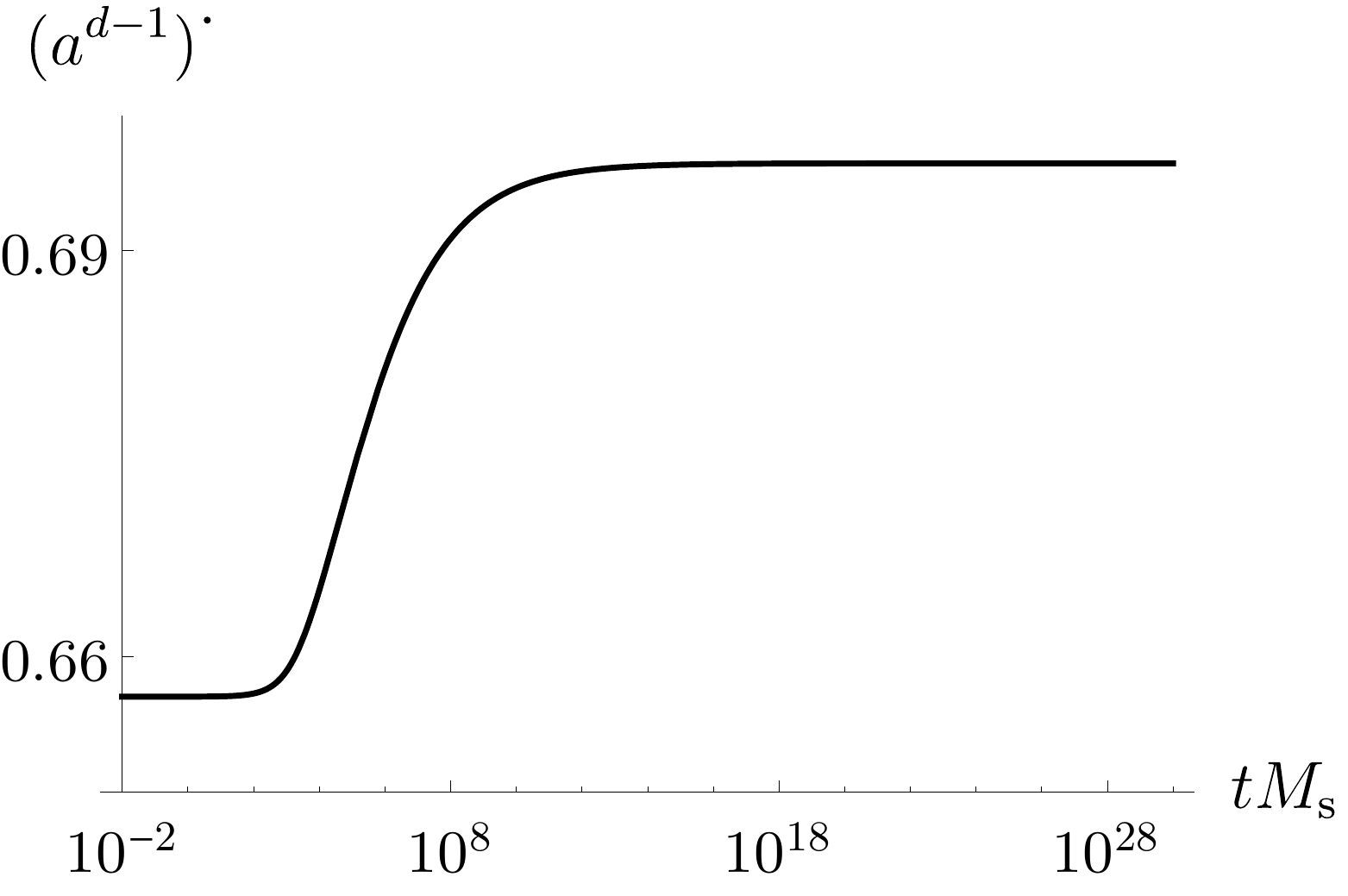}
\includegraphics[width=8.18cm]{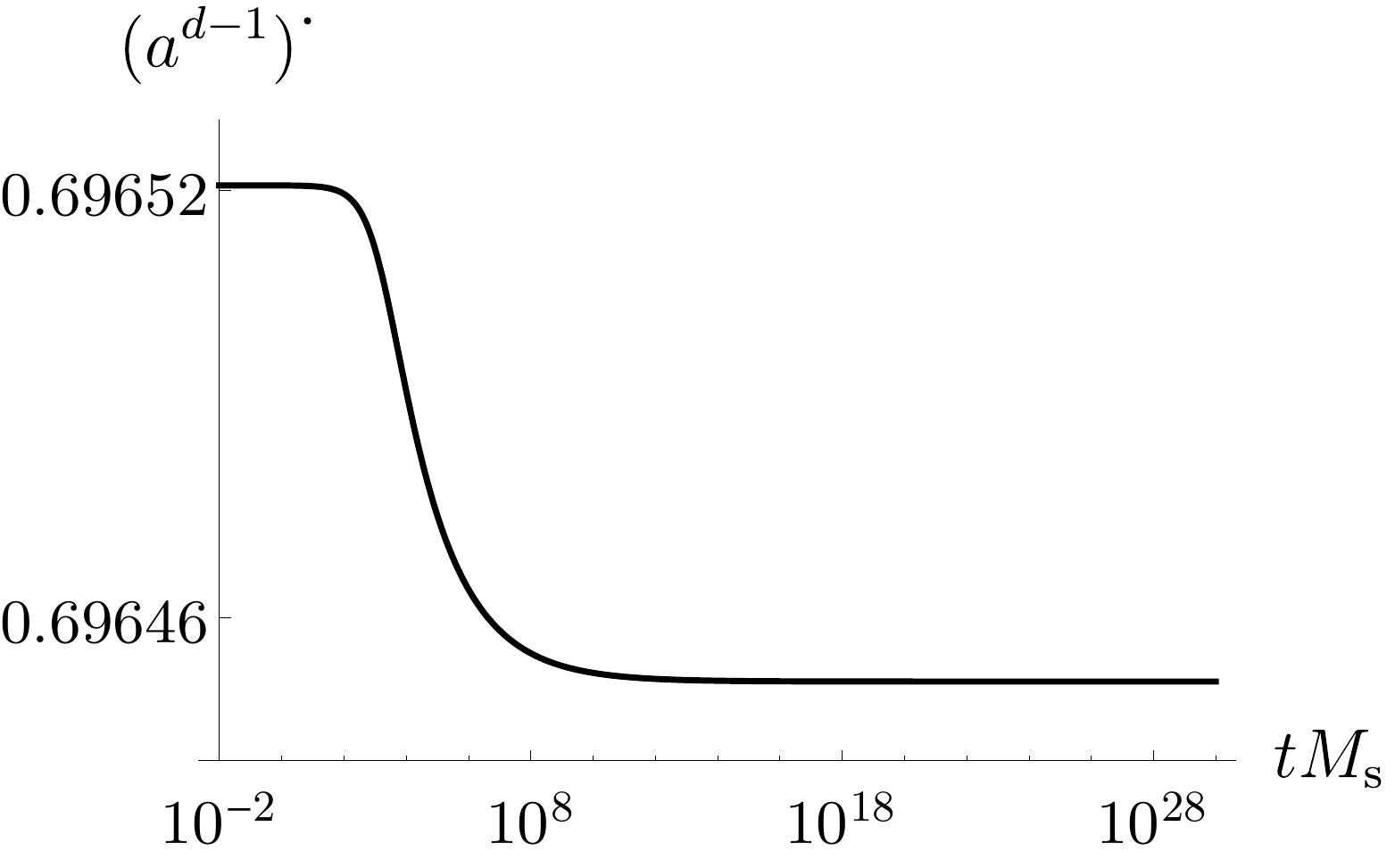}
\end{center}
\begin{picture}(0,0)
\put(105,180){\large($i$)}\put(347,180){\large($iv$)}
\end{picture}
\vspace{-.9cm}
\caption{\footnotesize \em Convergence of  $(a^{d-1})^{\dis\cdot}$ towards its  limit $\Am^{d-1}$, in cases $(i)$ and $(iv)$. The evolution is monotonically increasing or decreasing, depending on the sign of $\nF-\nB$.}
\label{plateau}
\end{figure}
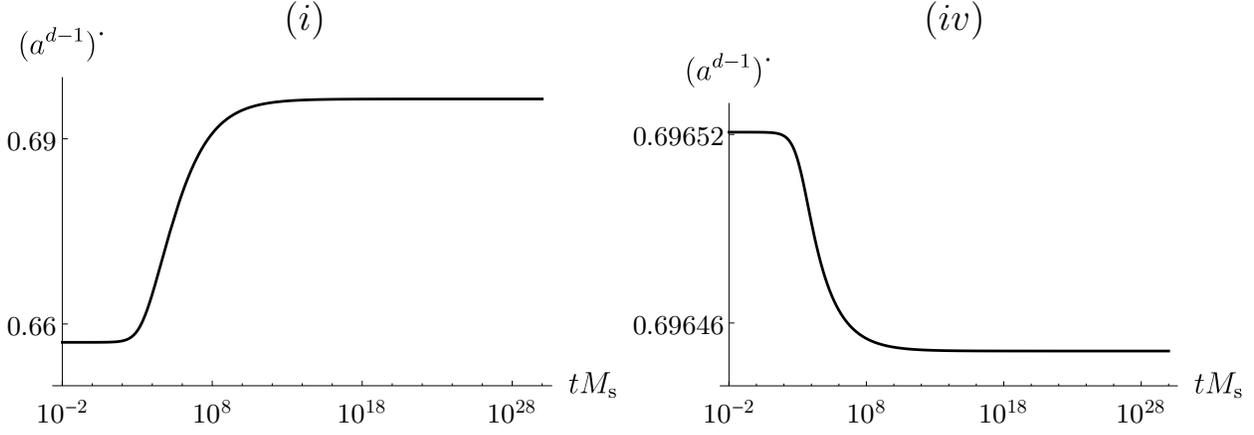

Due to Eq.~(\ref{vseul}), whether $(a^{d-1})^{\dis \cdot}$ increases of decreases with time is determined in full generality by the sign of $\Vone$. In order to discriminate when the universe is in QNSR, the most decisive criterion is the asymptotic behavior of the scale factor, Eq.~(\ref{a0}), which must satisfy
\be
(a^{d-1})^{\dis \cdot}\longrightarrow \Am^{d-1}\, , \qquad \when\qquad t\to +\infty\, . 
\label{pla}
\ee
In all cases $(i)$--$(iv)$, the numerical simulation confirms the above convergence to a constant, either upward or downward depending on the sign of the potential \ie $\nF-\nB$. The plots in Fig.~\ref{plateau} show the evolutions of $(a^{d-1})^{\dis \cdot}$ as a function of $t$ for the backgrounds $(i)$ and $(iv)$. The curves in models $(ii)$ and $(iii)$ are qualitatively similar to those obtained respectively in cases $(i)$ and $(iv)$.

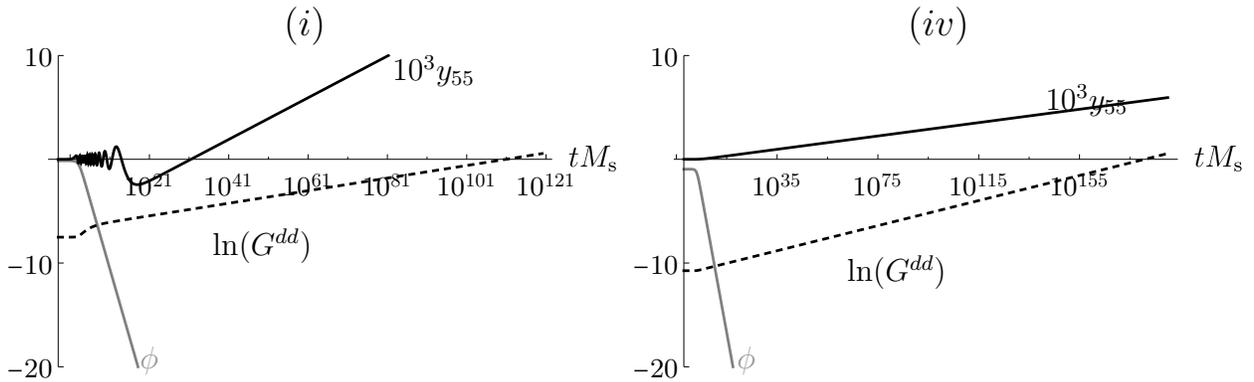
\begin{figure}[!b]
\vspace{0.3cm}
\begin{center}
\includegraphics[width=8.18cm]{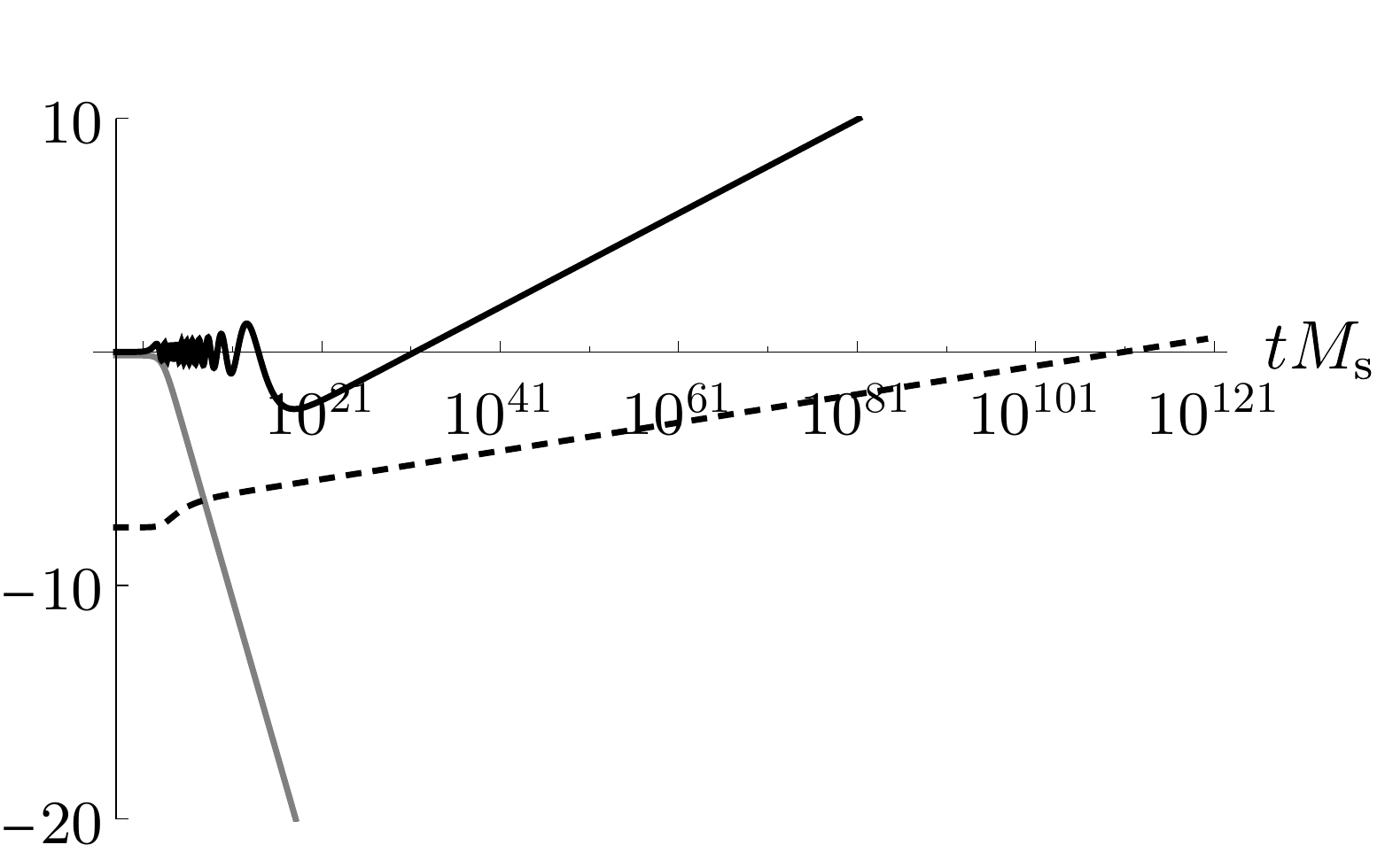} 
\includegraphics[width=8.18cm]{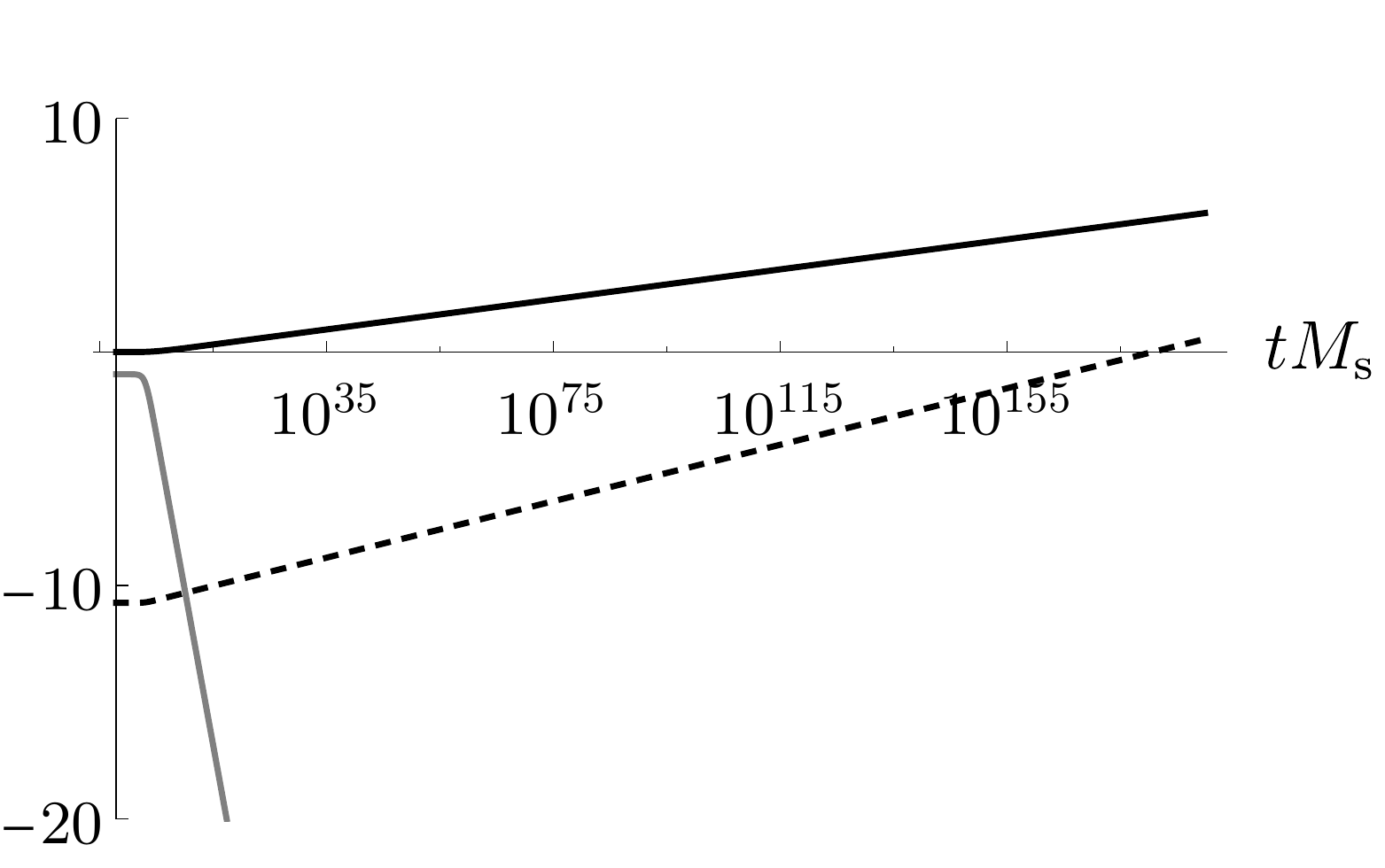}
\end{center}
\begin{picture}(0,0)
\put(50.5,35){\gray$\phi$}\put(276.3,35){\gray$\phi$}
\put(78,78){$\ln(G^{dd})$}\put(318,68){$\ln(G^{dd})$}
\put(146,144){$10^3y_{55}$}\put(393,133){$10^3y_{55}$}
\put(105,161){\large($i$)}\put(341,161){\large($iv$)}
\end{picture}
\vspace{-.9cm}
\caption{\footnotesize \em The behaviors of the dilaton $\phi$ (gray curves), $\ln(G^{dd})$ (dotted curves) and $10^3y_{55}$ (black curves) as functions of cosmic time (in logarithmic scale) are asymptotically linear in cases $(i)$ and $(iv)$. The evolutions can be trusted as long as $\phi(t)<0$ and $\ln\!\big(G^{dd}(t)\big)<0$, for $c=1$.  Oscillations of $y_{55}$ may occur at early times when it is massive, $c_{\G_1}>0$.}
\label{dil_gdd_y55}
\end{figure}

Solving the system of differential equations makes sense as long as the weak coupling condition is fulfilled, $\phi(t)<0$, and the supersymmetry breaking scale measured in $\sigma$-model frame is small, $\sqrt{G^{dd}(t)}<c$. It turns out that the numerical evolutions of $\phi$, $\ln(G^{dd})$ and $y_{55}$ as functions of~$t$ present similar features when $c_{\G_1}>0$ \ie in models~$(i)$ and $(iii)$, and when  $c_{\G_1}<0$ \ie in models~$(ii)$ and $(iv)$. The only qualitative difference may occur at early times, where $y_{55}$ may oscillate when it is massive, $c_{\G_1}>0$. The curves are shown in Fig.~\ref{dil_gdd_y55} in cases~$(i)$ and $(iv)$, where the cosmic times above which the simulations cannot be trusted are respectively $\mbox{$\tf\simeq 10^{110} \Ms^{-1}$}$ and $\tf\simeq 10^{185}\Ms^{-1}$, for $c=1$.  
The fact that $\phi$, $\ln(G^{dd})$ and $y_{55}$ depend asymptotically linearly on $\ln (t\Ms)$ proves that the velocities $\dot \Phi$, $\dot \phi_\bot$ and $\dot y_{55}$ are inversely proportional to cosmic time, \ie that 
\be
(c_\bot^\dyn,c_\Phi^\dyn,c^\dyn_{55})\longrightarrow (c_\bot,c_\Phi,c_{55})\, ,\quad  \when \quad t\to +\infty\, . 
\ee
In particular, we can identify from Eq.~(\ref{G}) the limit reached by the dynamical quantity  
\be
J_\dyn\equiv  t\,(\ln G^{dd})^{\dis \cdot}= t\left(\frac{2}{\alpha}\, \dot \Phi-\frac{2}{\sqrt{d-1}}\, \dot\phi_\bot\right)\!\longrightarrow J_+>0\, , \quad \when \quad t\to +\infty\,.
\ee

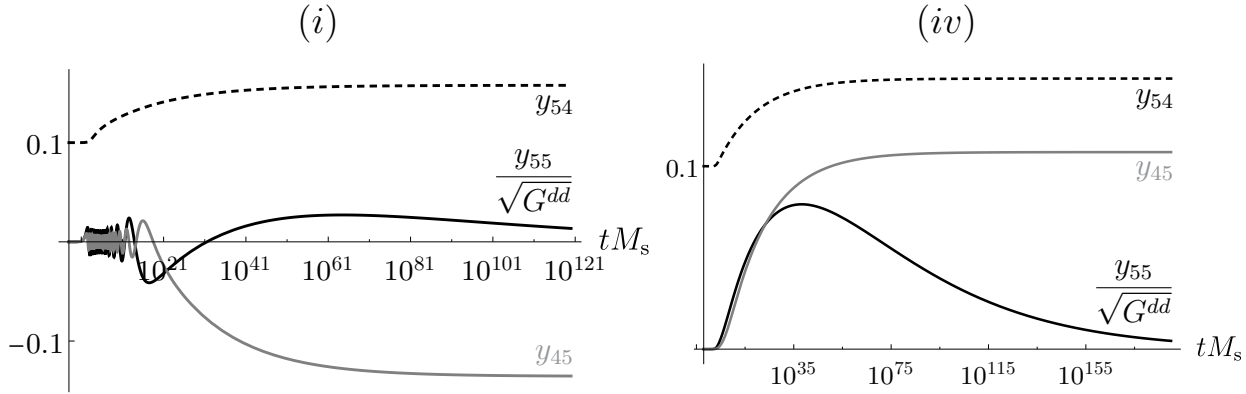
\begin{figure}[!h]
\vspace{.3cm}
\begin{center}
\includegraphics[height=5.05cm]{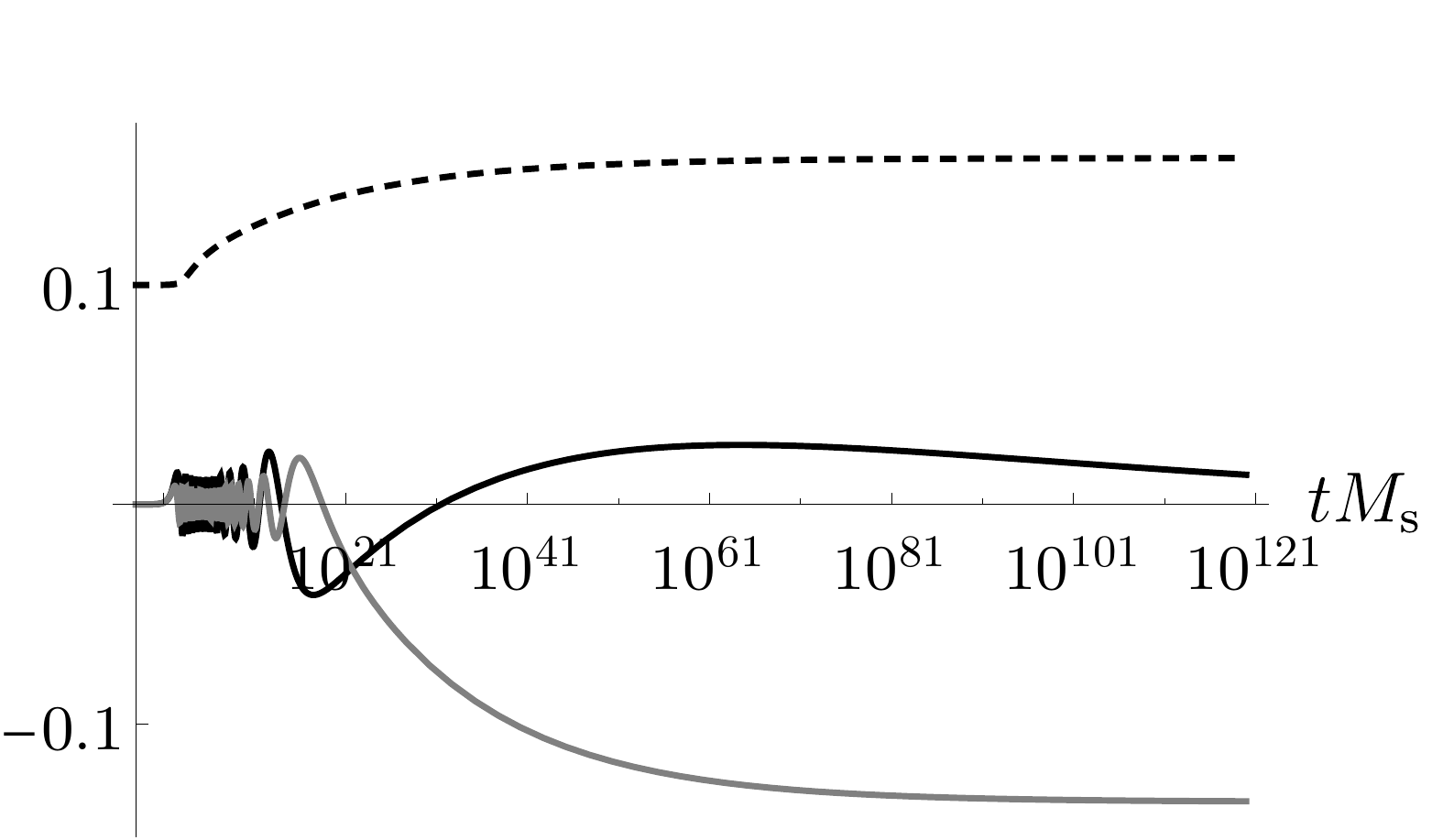}
\includegraphics[height=5.05cm]{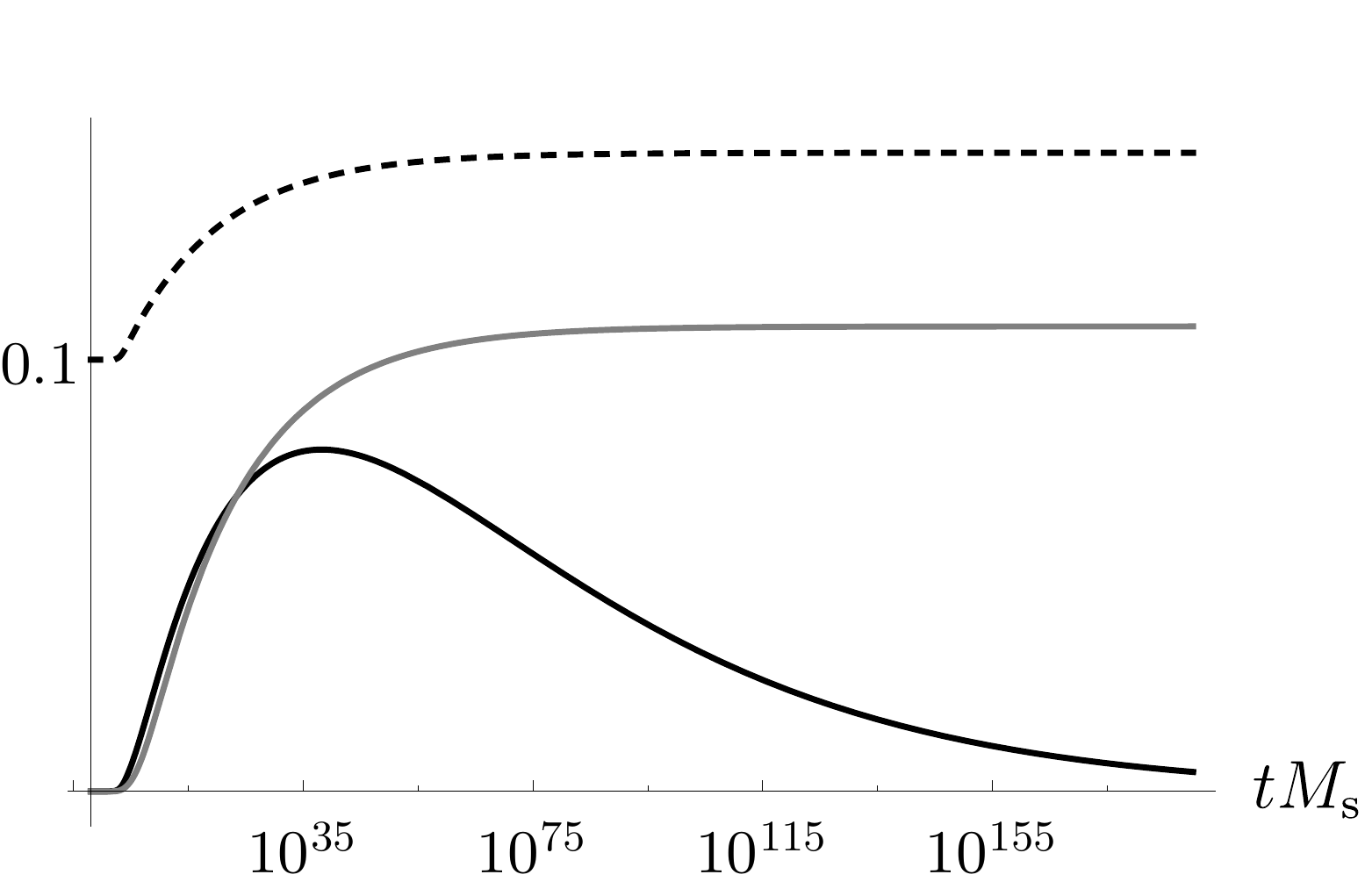}
\end{center}
\begin{picture}(0,0)
\put(184,106.5){$\dis \frac{y_{55}}{\sqrt{G^{dd}}}$}\put(411,64.5){$\dis \frac{y_{55}}{\sqrt{G^{dd}}}$}
\put(200,40.5){\gray$y_{45}$}\put(427,110.0){\gray$y_{45}$}
\put(200,135){$y_{54}$}\put(427,137.5){$y_{54}$}
\put(110,164){\large($i$)}\put(344,164){\large($iv$)}
\end{picture}
\vspace{-.9cm}
\caption{\footnotesize \em Convergences of the Wilson lines $y_{45}$ (gray curves) and  $y_{54}$ (dotted curves) to constants, and of $y_{55}/\sqrt{G^{dd}}$ to $0$, in cases $(i)$ and $(iv)$. Oscillations of $y_{45}$ and $y_{55}$ may occur when they are massive, $c_{\G_1}>0$.}
\label{plot_y}
\end{figure}

What remains to be checked are the behaviors of $y_{45}$ and $y_{54}$, as well as the smallness of all Wilson lines.  Fig.~\ref{plot_y} shows $y_{45}(t)$, $y_{54}(t)$ and $y_{55}(t)/\sqrt{G^{dd}(t)}$ simulated in model~$(i)$ (which is similar to~$(iii)$)  and in model~$(iv)$ (which is similar to~$(ii)$). 
As predicted in Sect.~\ref{QNSR} when $J_+>0$, all curves converge to constants, 
\be
y_{45}\longrightarrow y_{45}^{(0)}\, , \quad y_{54}\longrightarrow y_{54}^{(0)}\, , \quad  {y_{55}\over \sqrt{G^{dd}}}\longrightarrow 0\, , \quad \when\quad t\to +\infty\, ,
\ee
while their upper and lower bounds are small, in the sense of Eq.~(\ref{small}). Let us stress again that even when $c_{\G_1}<0$, contrary to common sense, the tachyonic scalars $y_{45}$ and $y_{55}$ do not induce large destabilizations of the backgrounds, when the universe enters the QNSR regime. These remarks complete our numerical validation of the existence of the QNSR $t\to +\infty$,  demonstrated in the previous section. Note however that even if the  solutions can be trusted all the way until~$\tf$, the universe enters the QNSR only after a certain duration, as can be seen in all Figs~\ref{plateau}--\ref{plot_y}.  During this transient period, the dynamics is affected by the effective potential, as shown on Figs~\ref{dil_gdd_y55} and ~\ref{plot_y}. In fact, when $c_{\G_1}>0$ \ie case $(i)$ and $(iii)$, the Wilson lines $y_{45}$ and $y_{55}$ are massive and oscillate around minima of $\Vone$. It is only when the universe enters the QNSR that the potential is dominated by the canonical kinetic energies of $\Phi$, $\phi_\bot$ and $y_{55}$, so that not only $y_{54}$ but also $y_{45}$ freeze at arbitrary values, while $\Phi$, $\phi_\bot$ and $y_{55}$ behave logarithmically with cosmic time. On the contrary, when $c_{\G_1}<0$ \ie case $(ii)$ and $(iv)$, the tachyonic Wilson lines $y_{45}$ and $y_{55}$ do not oscillate during the early transient regime.  

Before proceeding, we would like to provide comments on the first constraint appearing in Eq.~(\ref{cii}), for a QNSR to be reached. As was argued below Eq.~(\ref{rr}), it is expected to be trivial, a fact that implies $C_\K$ to vanish, which we have verified analytically in Sect.~\ref{QNSR}. In fact, when a QNSR yields $\pm J_\pm>\pm {K_\pm\over d-1}>0$, the kinetic terms of $y_{d,d+1}$ and $y_{d+1,d}$, as well as all terms subdominant compared to 1 in the parentheses appearing in Eqs~(\ref{solbot}) and~(\ref{dphi}) are all individually dominated by $\Vone$. However, when $\pm {K_\pm\over d-1}>\pm J_\pm>0$, it is remarkable that the leading contributions of these terms cancel one another, so that $C_\K=0$. To check that this cancellation is actually exact, we have varied the initial conditions of our simulations of the QNSR $t\to +\infty$, for the characteristic point $\big({c_\bot\over \gamma_{\rm c}c_\Phi},{c_{55}\over \gamma_{\rm c}c_\Phi}\big)$ to explore all of the left crescent in Fig.~\ref{domaines}$(a)$. This means that the condition for the existence of the QNSR $t\to +\infty$ is $J_+>0$, and nothing more. 


\section{Global attractor mechanisms}
\label{global}

The numerical validation of the QNSR $t\to +\infty$  in presence of small Wilson line deformations being established, we would like to consider possible global attractor mechanisms. Our aim is to see whether it is possible to relax, at least in come cases, the  constraint of imposing the trajectories to be entirely in the tiny phase space described in the previous sections. 
It turns out that the kinetic terms in Eq.~(\ref{kinexact}) become quadratic in Wilson lines, when $y_{d+1,d+1}$ is identically frozen at the origin. As a result,  the action~(\ref{S1}) is exact in $y_{d,d+1}$ and $y_{d+1,d}$, provided we set  $y_{d+1,d+1}\equiv 0$ and use the full 1-loop effective potential,
\begin{align}
\Vone= &\, \big(\nF-\nB+(-1)^{\eta^R_{d+1}}\, 8\times 2\big)\,  v_{d}\,  M^d\nonumber \\
&\, -(-1)^{\eta^R_{d+1}}\, 8\times 2\; {2M^d\over (2\pi)^{3d+1\over 2}} \,  \sum_{\tilde m_d}\, {\cos \!\big(2\pi( 2\tilde m_d+1) \sqrt{2}\, y_{d,d+1}\big)\over |2\tilde m_d+1|^{d+1}} \, F(0)+\cdots \, ,
\label{VEexact}
\end{align}
where the ellipses stand for the exponentially suppressed contributions we  neglect as before.  In the following, we use this fact to simulate numerically the 1-loop dynamics of the scale factor~$a$, no-scale modulus~$\Phi$ and scalar~$\phi_\bot$, in the presence of arbitrary Wilson lines deformations~$y_{d,d+1}$ and~$y_{d+1,d}$ of the initial background.

We will find that the sign of $\Vone$ plays a critical role. Depending on the  integer $\nF-\nB$, the latter can be fixed, 
\begin{align}
(-1)^{\eta^R_{d+1}}\, (\nF-\nB)\ge  0 \quad &\Longrightarrow\quad (-1)^{\eta^R_{d+1}}\Vone\ge 0\quad  \mbox{for all } y_{d,d+1}\, , \nonumber\\
(-1)^{\eta^R_{d+1}}\, (\nF-\nB)\le  -32 \quad &\Longrightarrow\quad (-1)^{\eta^R_{d+1}}\Vone\le 0\quad  \mbox{for all } y_{d,d+1}\, , 
\end{align}
or varying,
\be
-31\leq (-1)^{\eta^R_{d+1}}\, (\nF-\nB)\le -1 \quad \Longrightarrow\quad\mbox{the sign of $\Vone$ varies with $y_{d,d+1}$}\, .
\ee
Note that since $y_{d,d+1}$ is allowed to explore a large range of values  during its evolution, there is no need to consider separately the cases $\eta^R_{d+1}$ even or odd. For instance, in spacetime dimension $d=4$ we consider from now on, a half-period shift $\sqrt{2}\, y_{45}\to \sqrt{2}\, y_{45}+{1\over 2}$ maps into each other backgrounds~$(i)$ and~$(ii)$ which have $\Vone>0$ for all $y_{45}$, or~$(iii)$ and~$(iv)$ which have $\Vone<0$ for all $y_{45}$. 

In the simulations, we take as initial conditions $y_{45}(0)$, $y_{54}(0)$, $a(0)$ and $c_{\bot0}\equiv c_\bot^\dyn(0)$, $c_{\Phi0}\equiv c_\Phi^\dyn(0)$ to be of order~1. This fixes $\dot\phi_\bot(0)$ and $\dot \Phi(0)$, provided $H(0)$ (which we take to be positive) or equivalently $\tau^\dyn(0)$ (see Eq.~(\ref{taud}))  is known. The latter is related to $\Phi(0)$ \via Friedmann equation at $t=0$, which we write in the following form
\be
\label{frias0}
-{d^2c_{\Phi0}^2\over 2(d-1)(d+2)}\, {\P_0\!\left(\tau^\dyn(0)\right)\over a(0)^{2(d-1)}}= {G^{dd}(0)\over 4}\, \dot y_{45}(0)^2+{G^{dd}(0)\over 4}\, \dot y_{54}(0)^2+\Vone\big(\Phi(0),y_{45}(0)\big)\, ,
\ee
where $\P_0$ is the degree two polynomial
\be
\P_0(\tau)=\tau^2-2\tau+\!\left(1-{4\over d^2}\right)\!\!\Big(1+2\alpha^2\, {c_{\bot0}^2\over c_{\Phi0}^2}\Big)\, . 
\ee
We impose the Wilson lines' kinetic terms at $t=0$ to be of the  order of $\left|\Vone\big(\Phi(0),y_{45}(0)\big)\right|$. This fixes $\dot y_{45}(0)$ and  $\dot y_{54}(0)$, once we make our choices for $\Phi(0)$ and $\phi_\bot(0)$. The latter is determined {\em a posteriori} for the numerical simulation to satisfy the conditions of weak string coupling and low supersymmetry breaking scale $M_{(\sigma)}$,  for a long period of cosmic time.  The last initial  data $\Phi(0)$ is equivalent to choosing $\tau^\dyn(0)$: 

$\bullet$ When $\Vone\big(\Phi(0),y_{45}(0)\big)>0$, Eq.~(\ref{frias0}) imposes 
\be
\left|{c_{\bot0}\over \gamma_cc_{\Phi0}}\right| <1 \qquad \and\qquad 1-r_0<\tau^\dyn(0)<1+r_0\, , 
\label{con}
\ee
where $r_0$ is defined in Eq.~(\ref{r0}). We have already studied in Sect.~\ref{simu} the case where $\mbox{$\tau^\dyn(0)\simeq 1+r_0$}$, which corresponds to a cosmological evolution starting almost in QNSR $t\to +\infty$. Thus, we will consider the two remaining  qualitatively different types of initial conditions~$(a)$ and~$(b)$, defined as follows:

$(a)$ For $\tau^\dyn(0)\simeq 1$, the cosmological evolution is generic, in the sense that the initial kinetic energies of $\Phi$, $\phi_\bot$, $y_{45}$ and $y_{54}$ as well as the potential are all of the order of~$H(0)^2$.

$(b)$ For $\tau^\dyn(0)\gtrapprox 1-r_0$, the potential and Wilson lines' kinetic energies are small compared to $H(0)^2$. This is clear by looking at Eq.~(\ref{frias0}), whose l.h.s. vanishes in the limit  $\tau^\dyn(0)\underset{>}{\to} 1-r_0$. As a result, the motion of the Wilson lines and the effective potential become irrelevant and the cosmological evolution is expected to approach that of the classical theory, with frozen Wilson lines, \ie $(a^{d-1})^{\dis \cdot}\equiv {2dc_{\Phi0}\over d^2-4}(1-r_0)$ \cite{CFP}. 

$\bullet$ When $\Vone\big(\Phi(0),y_{45}(0)\big)<0$, the r.h.s. of Eq.~(\ref{frias0}) can be negative or positive. In the former case, $\tau^\dyn(0)$ is arbitrary if $\left|{c_{\bot0}\over \gamma_cc_{\Phi0}}\right| \ge 1$, while it must satisfy $\tau^\dyn(0)>1+r_0$ or $\tau^\dyn(0)<1-r_0$ if $\left|{c_{\bot0}\over \gamma_cc_{\Phi0}}\right| < 1$.  When the r.h.s. of Eq.~(\ref{frias0}) is positive, condition~(\ref{con}) applies. 

In the models where $\Vone$ is negative for some/all $y_{45}$, which are illustrated by the backgrounds~$(iii)$ or ~$(iv)$, we find that the numerical simulations yield the following scenario: The universe expands, reaches a maximal size and then collapses into a Big Crunch, unless the initial conditions are tuned so that the whole trajectory sits  inside the tiny phase space that yields the ever-expanding QNSR $t\to +\infty$, as  described in Sect.~\ref{simu}. Notice that in Ref.~\cite{CFP}, where the dynamics of the Wilson lines is not taken into account, the initially expanding cosmological solutions arising when $\Vone<0$ are also either attracted to the QNSR $t\to +\infty$, or lead in the end to a Big Crunch. However, we emphasize again that in this case, the attraction to the QNSR follows from initial conditions chosen in a much larger space, namely $\left|{c_{\bot0}\over \gamma_cc_{\Phi0}}\right|<1$, $\tau^\dyn(0)>1+r_0$. In other words, the dynamics of internal moduli fields provides a severe source of instability for a flat, expanding universe, when the quantum potential can reach negative values. 

\begin{figure}[!h]
\vspace{.3cm}
\begin{center}
\includegraphics[width=8.18cm]{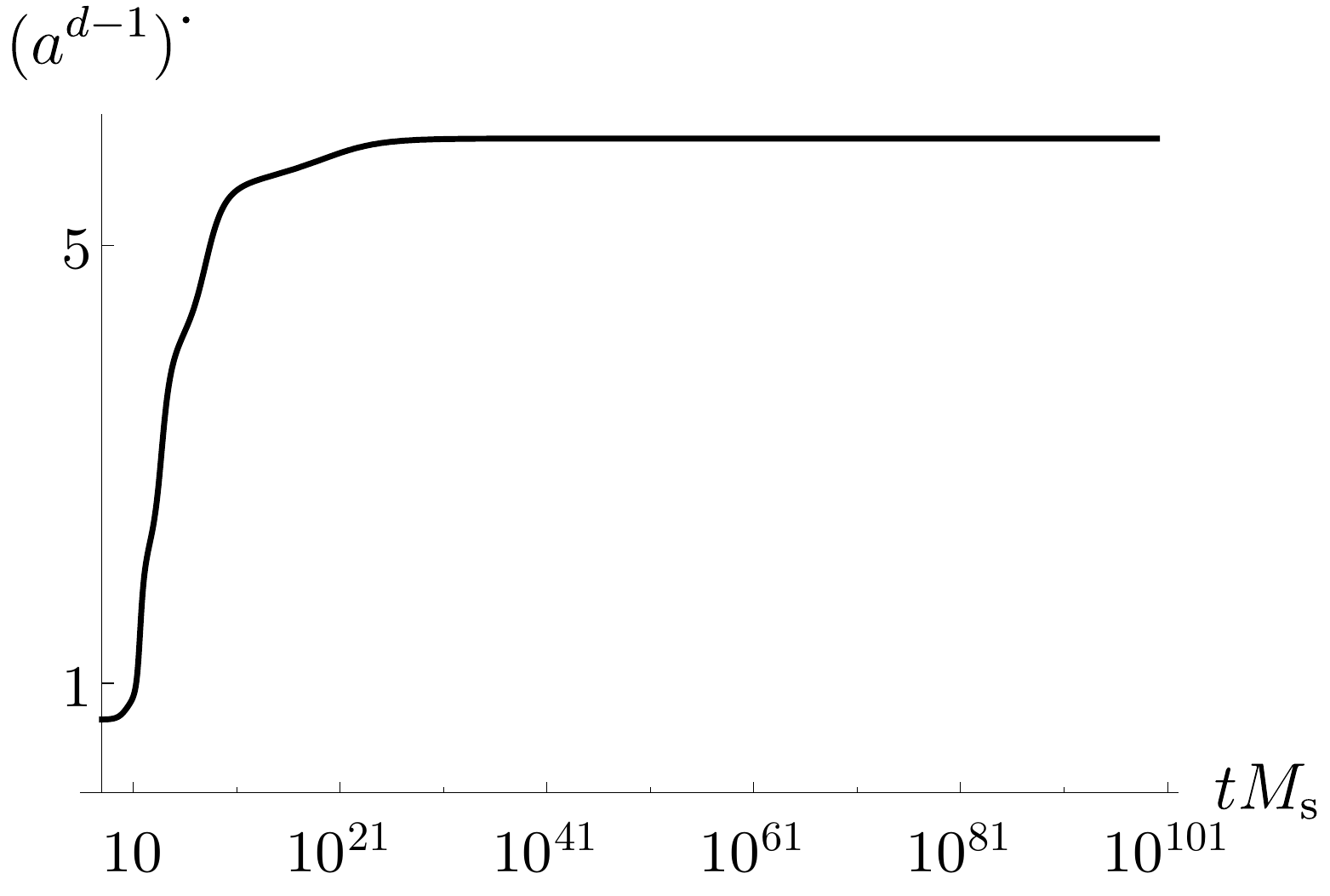}
\includegraphics[width=8.18cm]{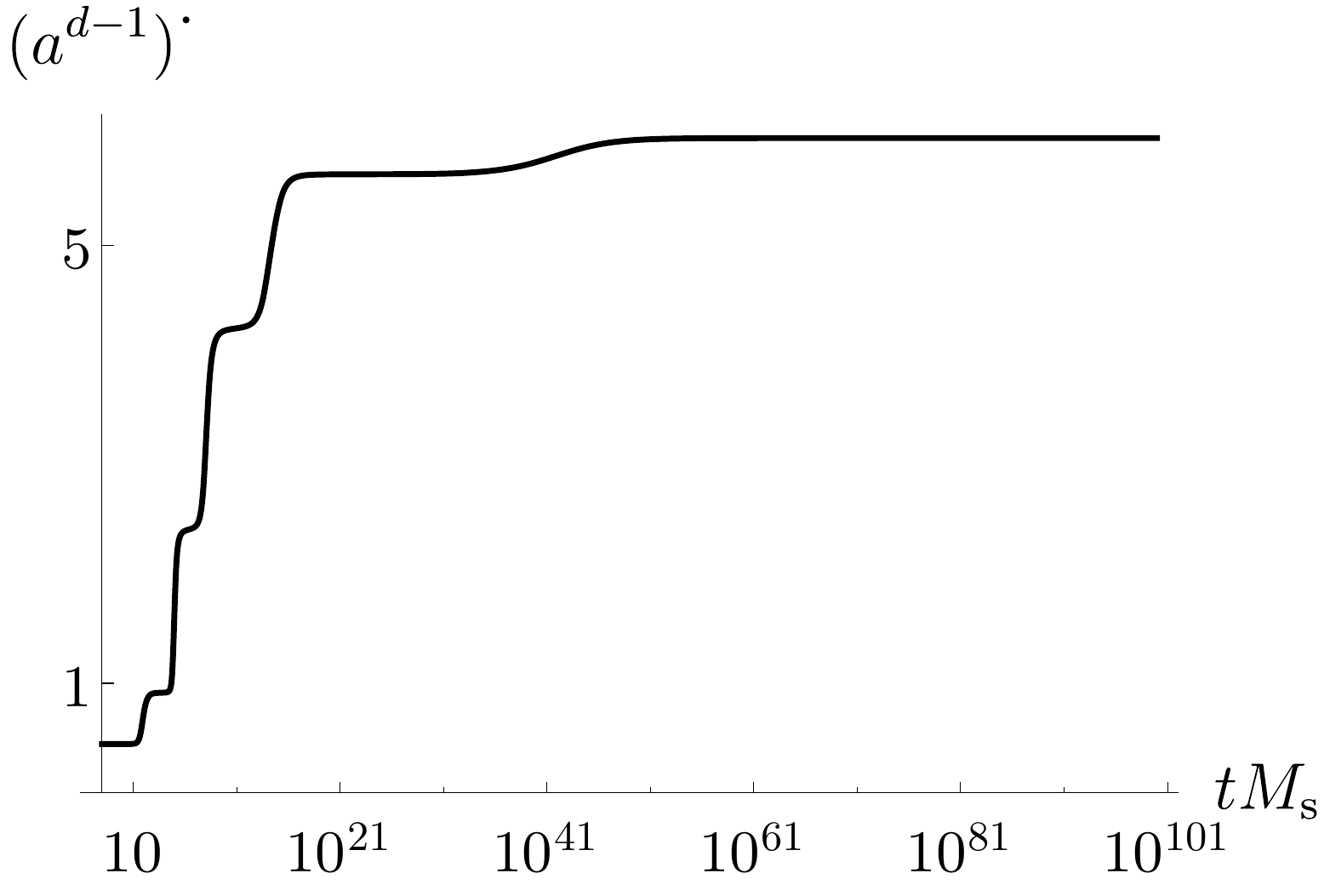}
\end{center}
\begin{picture}(0,0)
\put(105,180){\large($a$)}\put(347,180){\large($b$)}
\end{picture}
\vspace{-.9cm}
\caption{\footnotesize \em Convergence of $(a^{d-1})^{\dis\cdot}$ towards its limit $\Am^{d-1}$ in model~$(ii)$, for initial conditions $(a)$ or $(b)$. Moving from case~$(a)$ to~$(b)$, a structure of plateaux appears.}
\label{multi_plateau}
\end{figure}
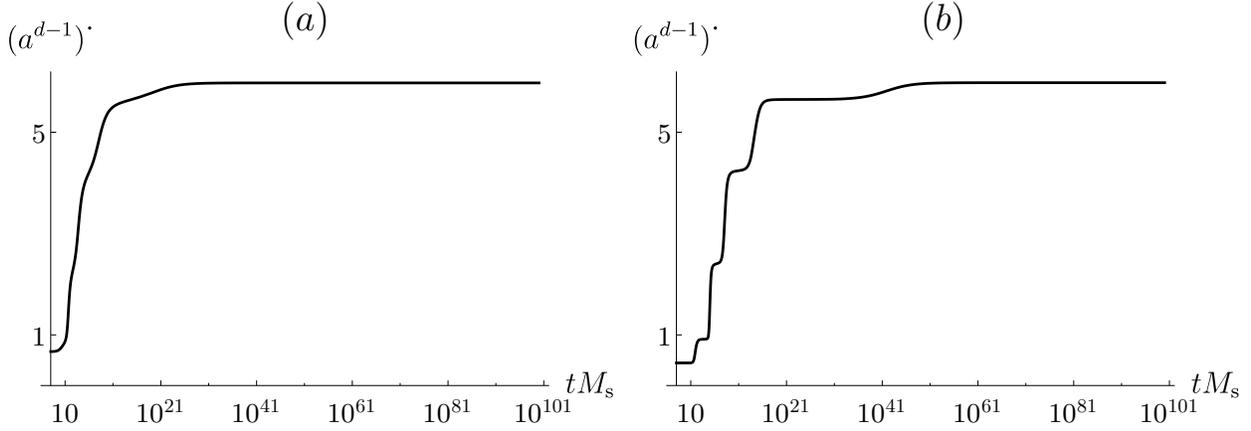

To describe a flat, expanding universe, the numerical simulations show that the models where $\Vone\ge 0$ for all $y_{45}$  are much more appealing, due to a global attraction mechanism to the QNSR $t\to+\infty$. Fig.~\ref{multi_plateau}$(a)$ presents the temporal evolution of $(a^{d-1})^{\dis\cdot}$ obtained in Example~$(ii)$ which has $\eta^R_{5}$ odd, for initial conditions of type~$(a)$. The potential being positive, the curve is monotonically increasing, and turns out to converge to a constant,  as in Eq.~(\ref{pla}). Note the existence of several inflection points, which are not numerical artefacts. Actually, by choosing initial values of type~$(b)$, a structure of plateaux appears, as shown in~Fig.~\ref{multi_plateau}$(b)$.  The latter are longer and longer and, after a finite number of steps, the last one is endless. Comments on this peculiar dynamics will be given at the end of the section. In any case, this phenomenon is the way the trajectory evolves, in order to converge to the straight line encountered in the extreme initial condition $\tau^\dyn(0)\underset{>}{\to} 1-r_0$.

\begin{figure}[!h]
\vspace{0.3cm}
\begin{center}
\includegraphics[width=8.18cm]{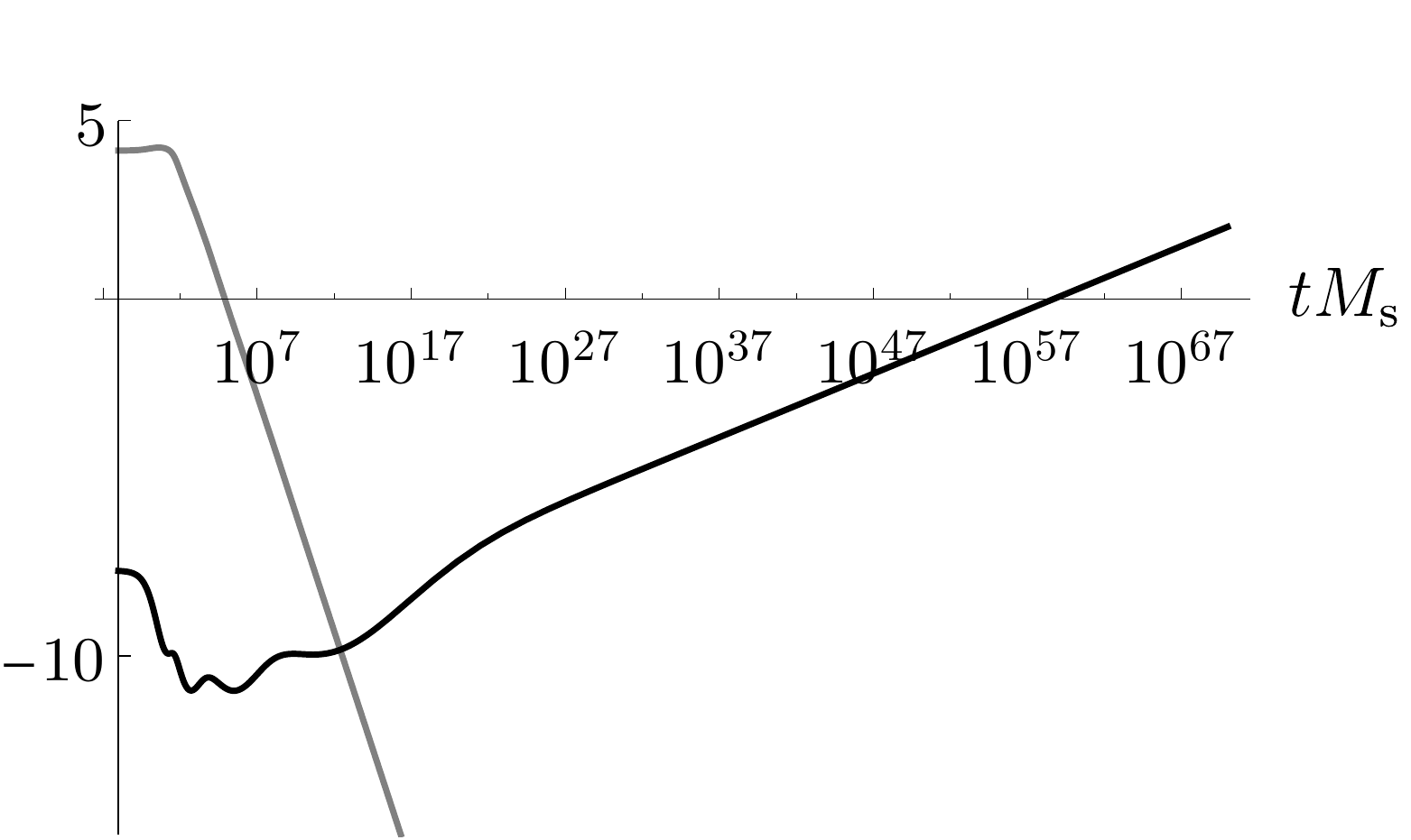} 
\includegraphics[width=8.18cm]{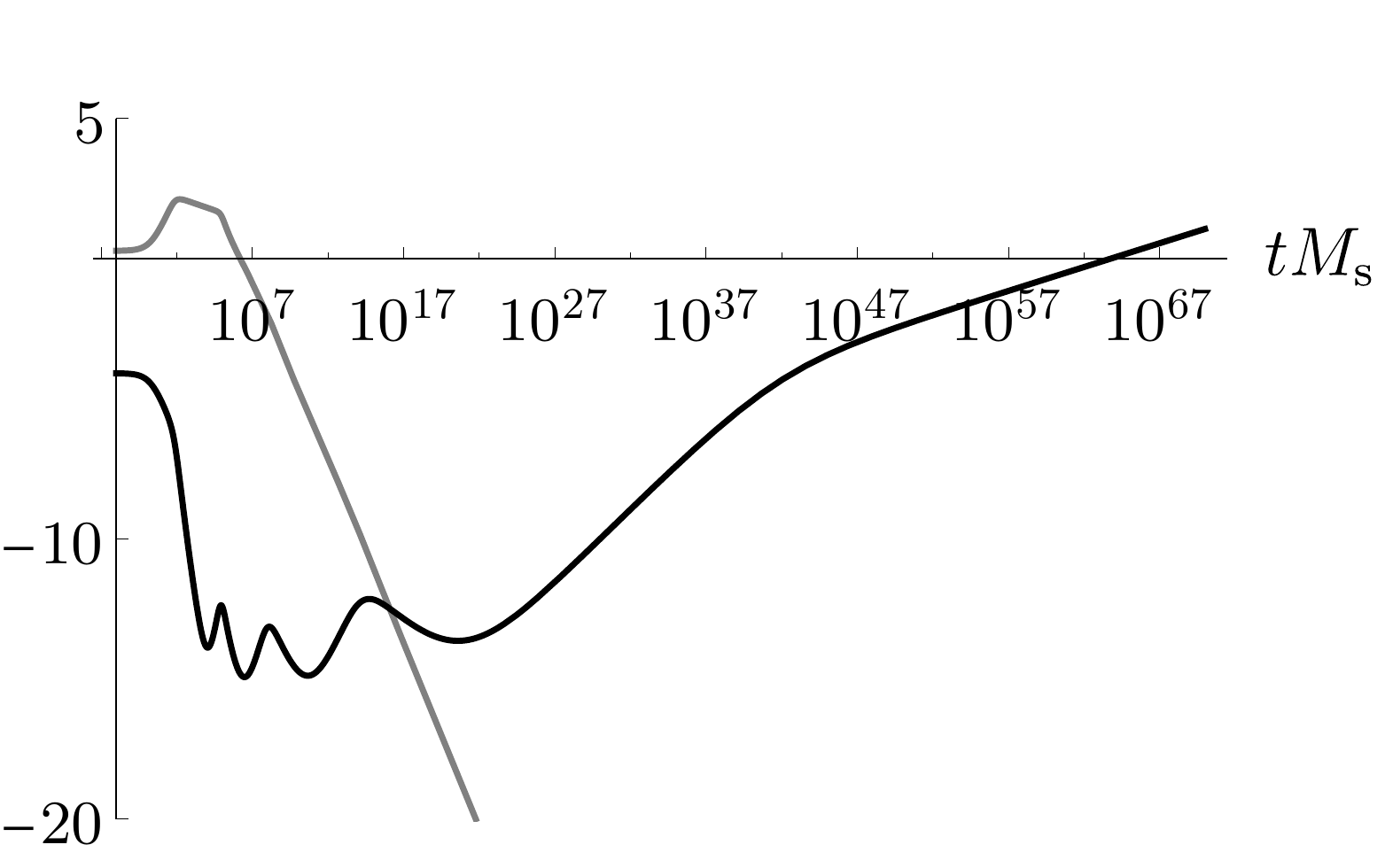}
\end{center}
\begin{picture}(0,0)
\put(66.5,35){\gray$\phi$}\put(316.5,35){\gray$\phi$}
\put(104,77){$\ln(G^{dd})$}\put(347.5,74){$\ln(G^{dd})$}
\put(105,160){\large($a$)}\put(346,160){\large($b$)}
\end{picture}
\vspace{-.9cm}
\caption{\footnotesize \em The behaviors of the dilaton $\phi$ (gray curves), $\ln(G^{dd})$ (black curves) as functions of cosmic time (in logarithmic scale) are asymptotically linear in model~$(ii)$, for initial conditions of type~$(a)$ or~$(b)$. The evolutions can be trusted as long as~$\phi(t)<0$ and~$\ln\!\big(G^{dd}(t)\big)<0$, for $c=1$.}
\label{dil_gdd}
\end{figure}
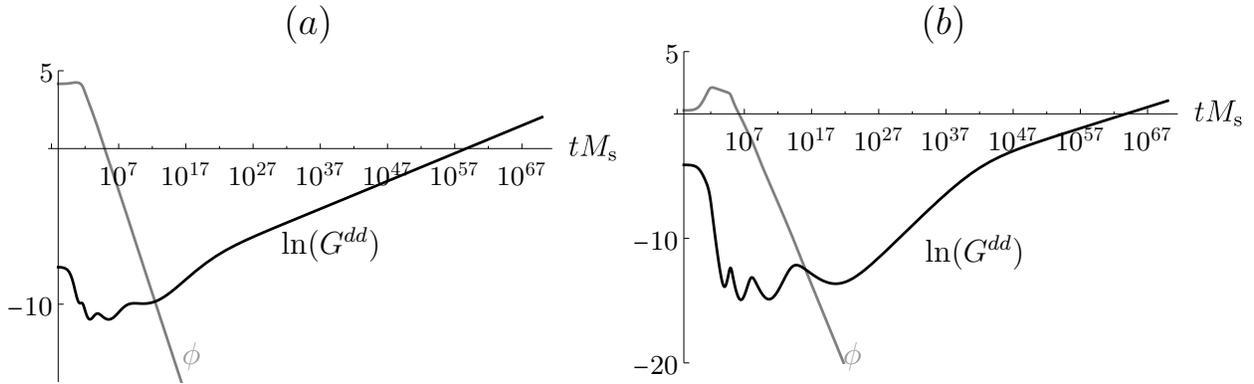

\begin{figure}[!h]
\vspace{0.3cm}
\begin{center}
\includegraphics[width=8.18cm]{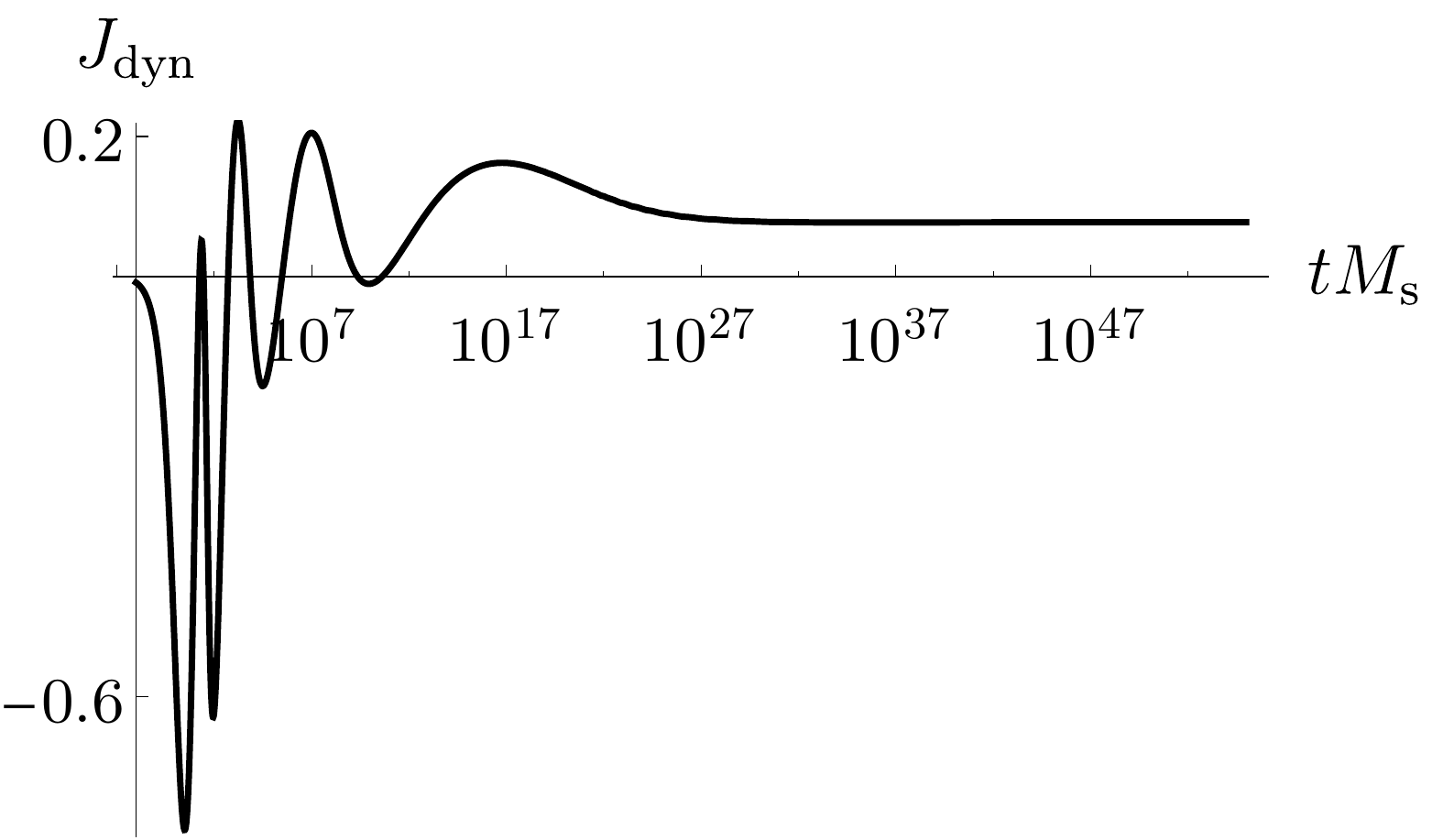}
\includegraphics[width=8.18cm]{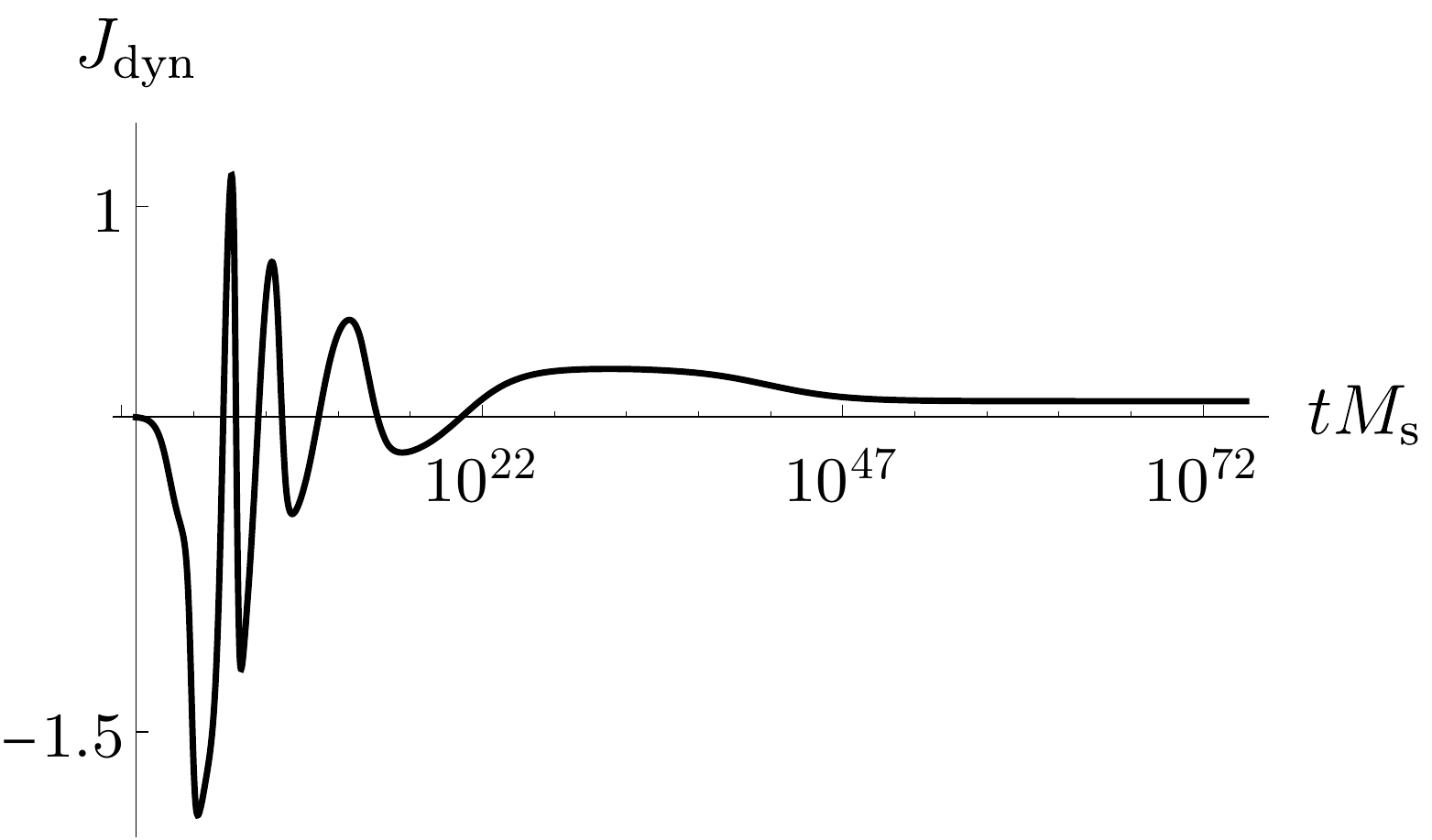}
\end{center}
\begin{picture}(0,0)
\put(107,158){\large($a$)}\put(346,158){\large($b$)}
\end{picture}
\vspace{-.9cm}
\caption{\footnotesize \em Convergence of $J_\dyn$ towards a positive value $J_+$ in model ~$(ii)$, for initial conditions of type~$(a)$ or~$(b)$.}
\label{JDyn}
\end{figure}
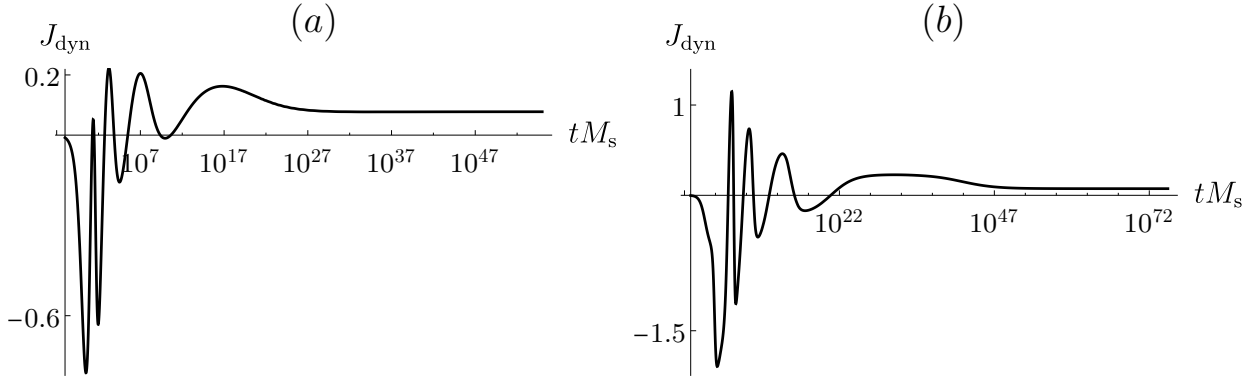

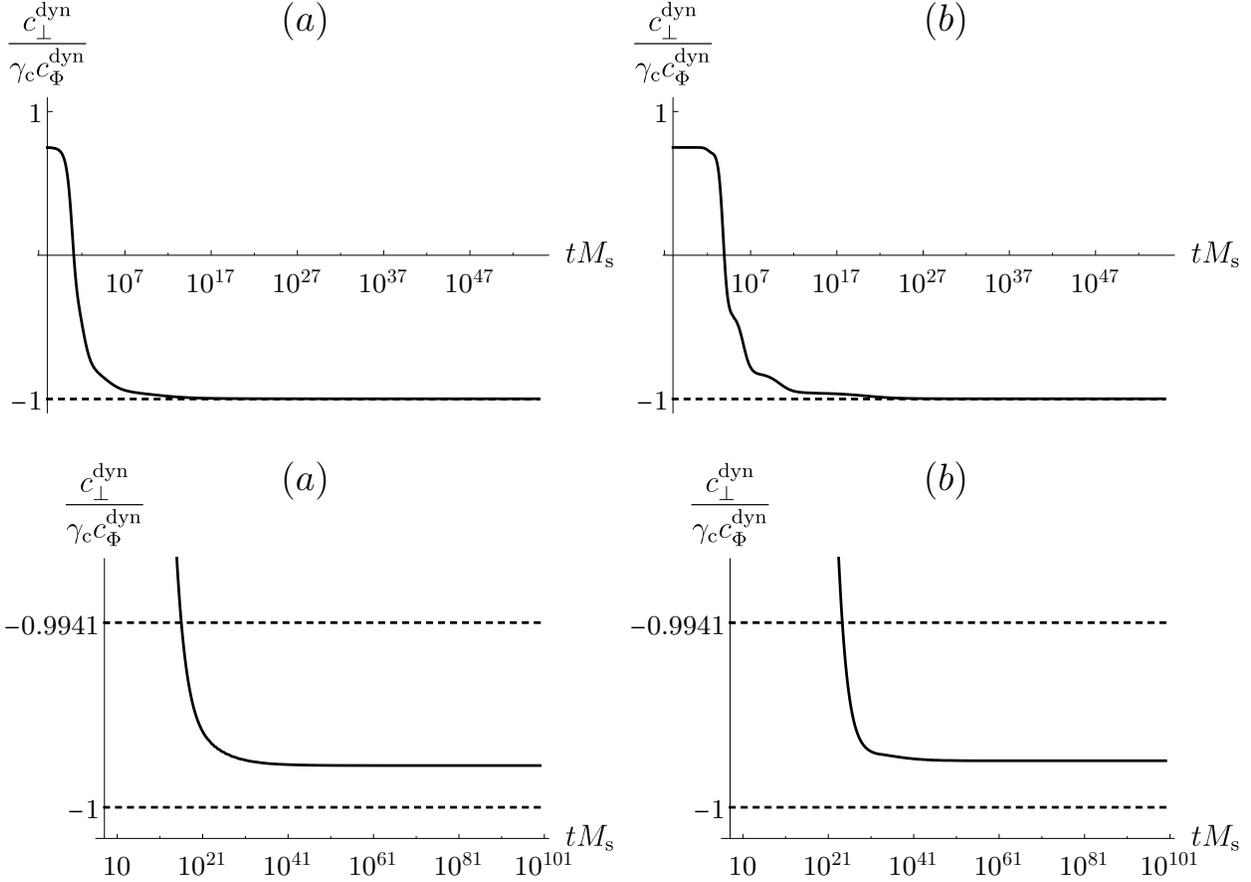
\begin{figure}[!h]
\vspace{0.3cm}
\begin{center}
\includegraphics[width=8.18cm]{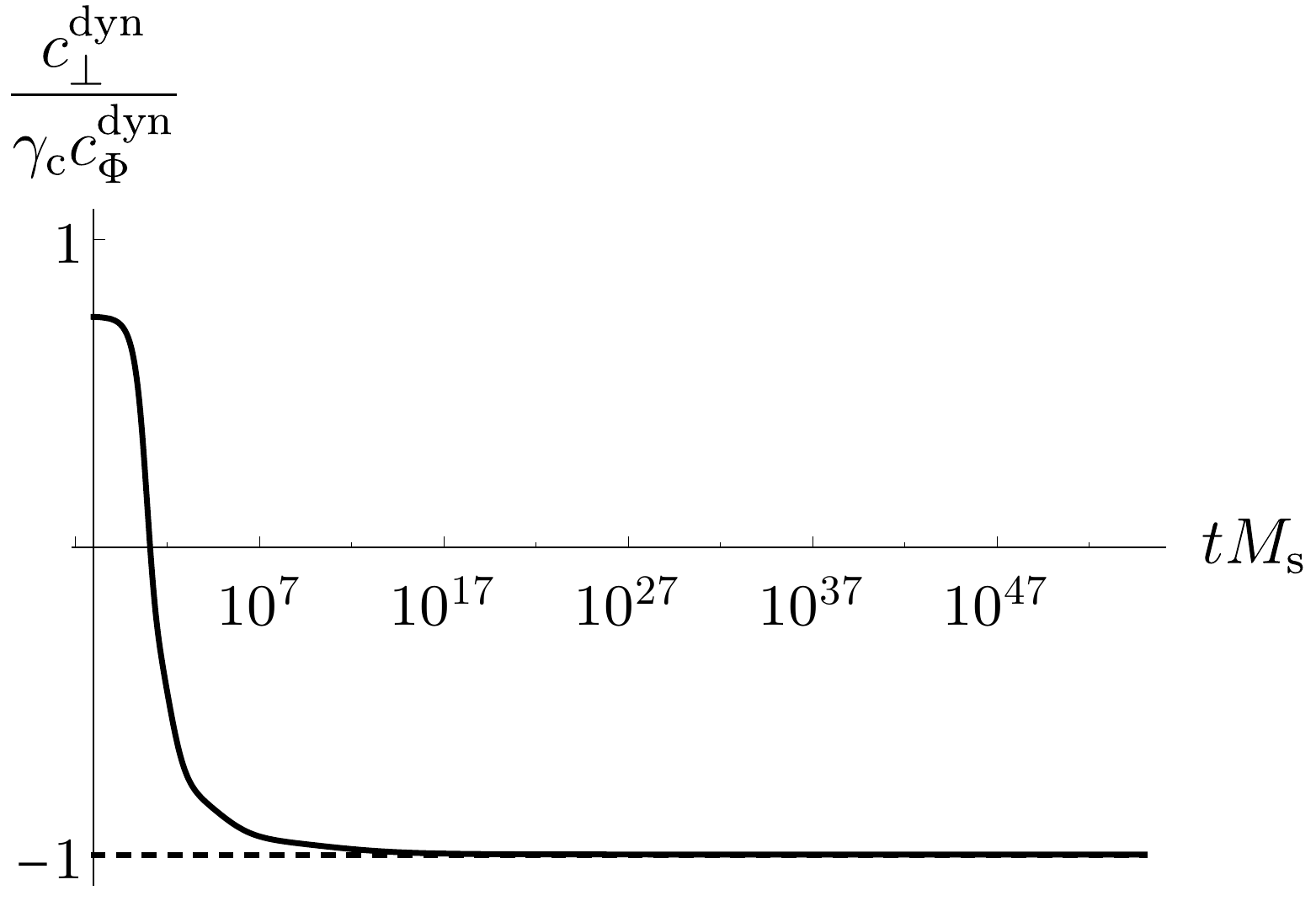}
\includegraphics[width=8.18cm]{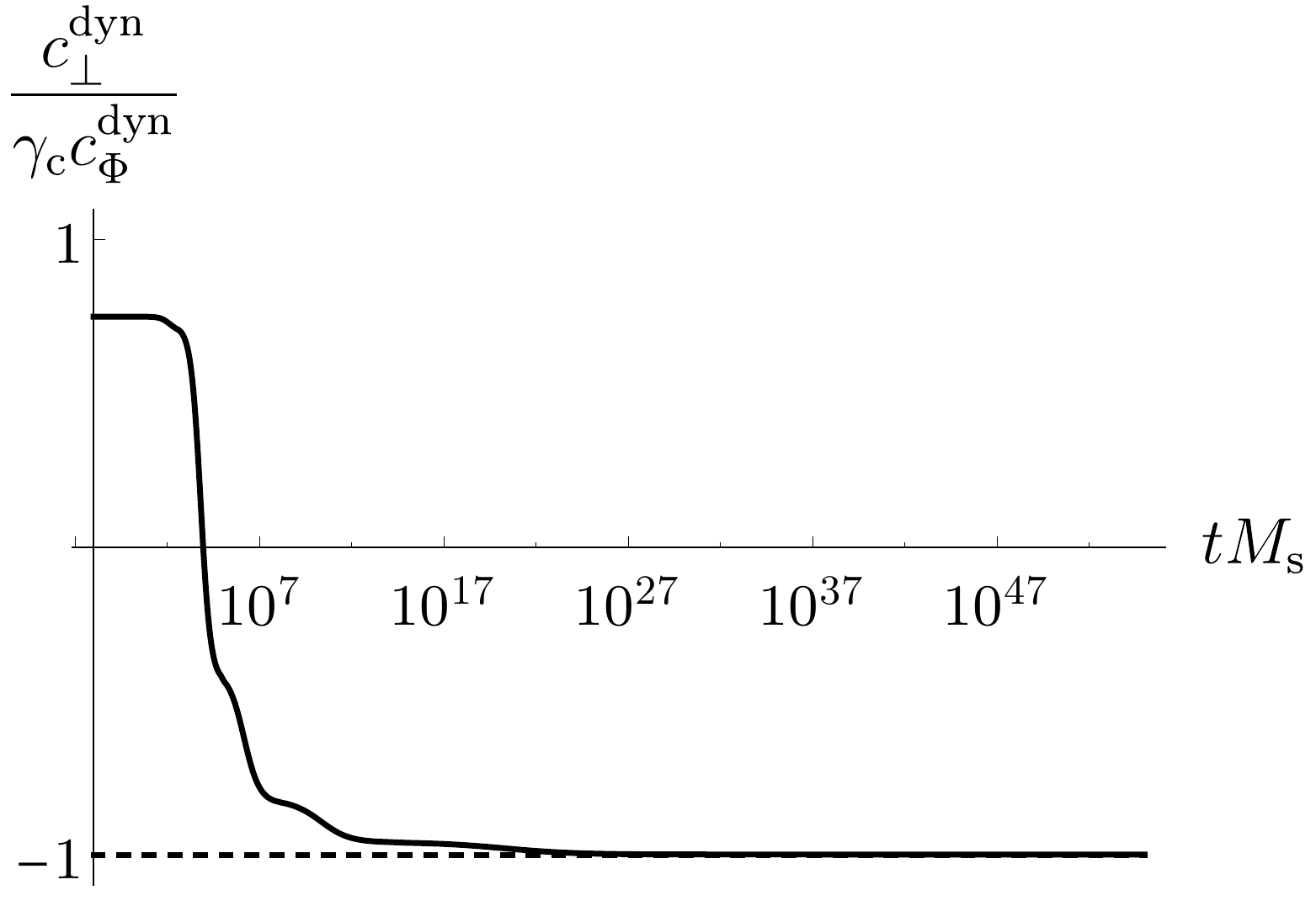}\\ \vspace{0.5cm}
\includegraphics[width=8.18cm]{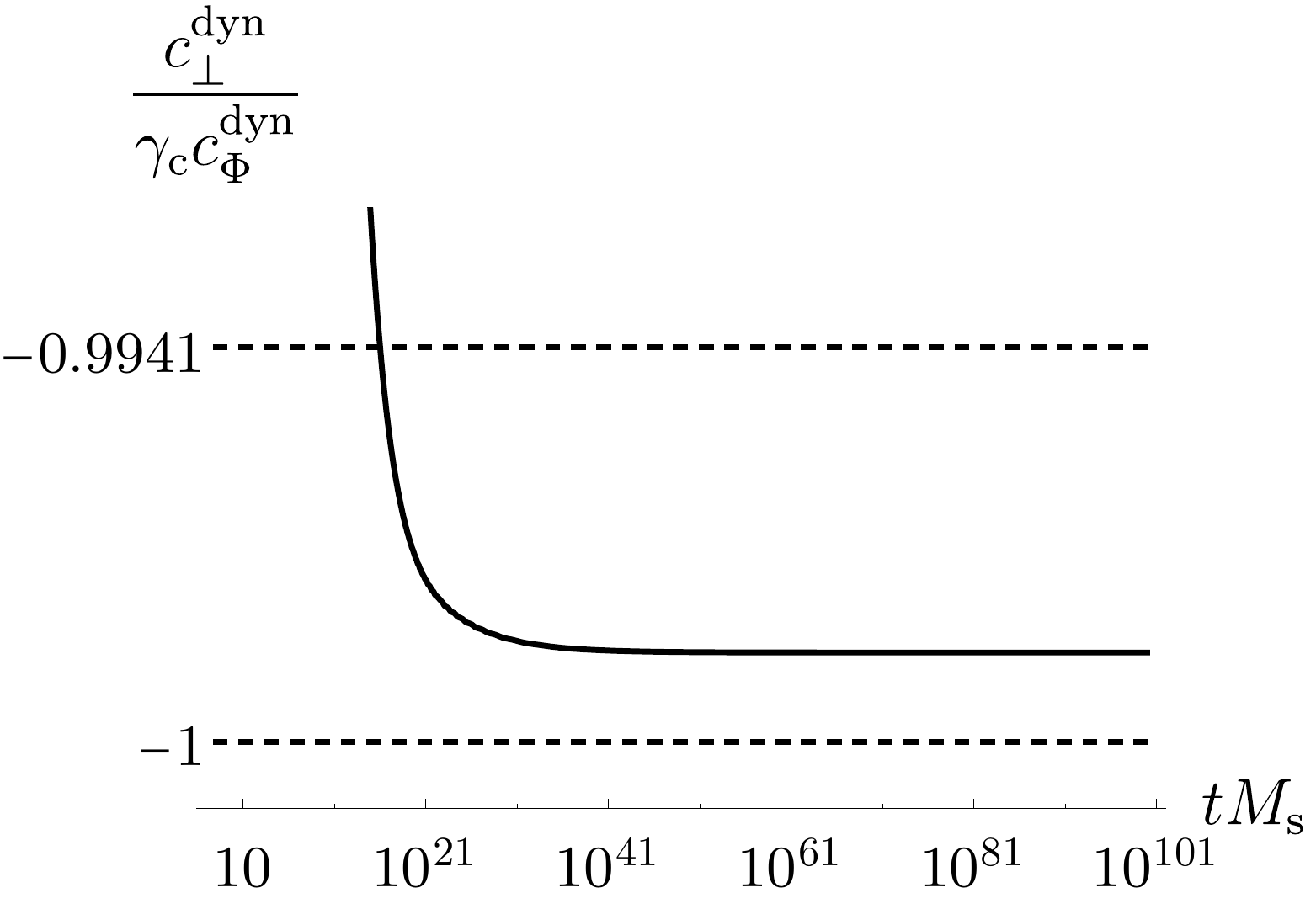}
\includegraphics[width=8.18cm]{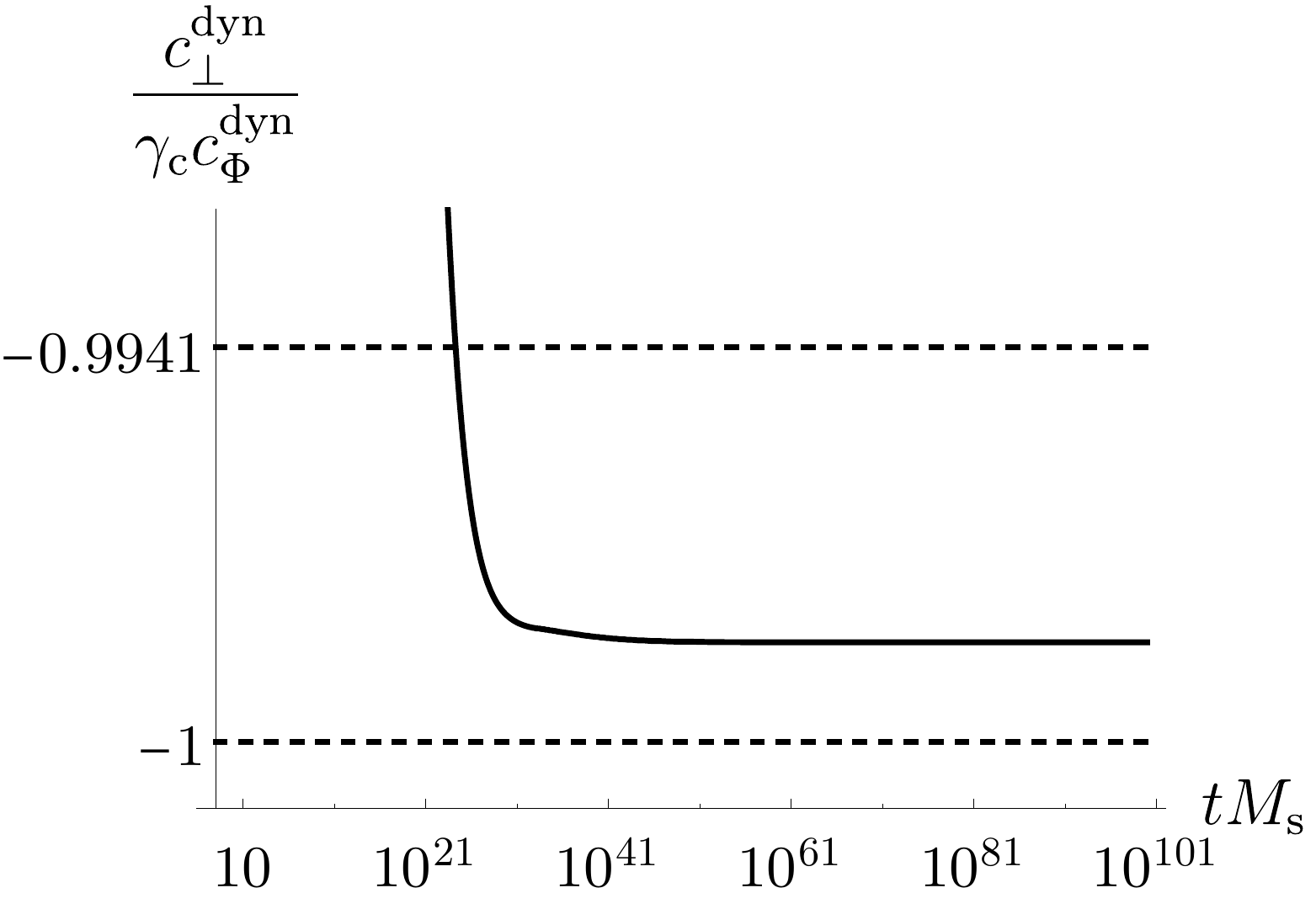}
\end{center}
\begin{picture}(0,0)
\put(105,351){\large($a$)}\put(348,351){\large($b$)}
\put(105,177){\large($a$)}\put(348,177){\large($b$)}
\end{picture}
\vspace{-.9cm}
\caption{\footnotesize \em Attraction of ${c^\dyn\over \gamma_{\rm c}c^\dyn}$ towards the tiny range $[-1,\approx \!\!-0.9941]$ in model~$(ii)$, for initial conditions of type~$(a)$ or~$(b)$ (upper plots). The ratio stabilizes once it enters the interval (lower plots).}
\label{Omega}
\end{figure}

To figure out when string perturbation theory is valid and $M_{(\sigma)}<c\Ms$, we plot in Fig.~\ref{dil_gdd} the dilaton and $\ln(G^{dd})$ as functions of time. The constraints $\phi(t)<0$ and $\ln(G^{dd}(t))<\ln c$ determine the ranges of time  $[\ti,\tf]$ where the simulations can be trusted. In model~$(ii)$, with $c=1$, an example of initial conditions~$(a)$ yields $[\ti,\tf]=[10^5\Ms^{-1},10^{59}\Ms^{-1}]$, while for initial values of type~$(b)$ we obtain  $[\ti,\tf]=[10^6\Ms^{-1},10^{65}\Ms^{-1}]$.  In both simulations, the final asymptotes are reached before~$\tf$.  
At late times (in logarithmic scale), the linearity of the plots and the positivity of the slope of $\ln(G^{dd})$ show  the convergences 
\be 
(c_\bot^\dyn,c_\Phi^\dyn)\longrightarrow(c_\bot,c_\Phi)\quad \and\quad J_\dyn\longrightarrow J_+>0\, ,\quad  \when \quad t\to +\infty\, . 
\ee
Fig.~\ref{JDyn} details the evolution of $J_\dyn(t)$, which describes a transient regime of damped oscillations between positive and negative values, followed by a stabilization at a positive constant~$J_+$.  
In view of our analysis in Sect.~\ref{QNSR}, the sign of $J_+$ suggests that the trajectory of the point $\big({c^{\dyn}_\bot\over \gamma_{\rm c}c^{\dyn}_\Phi},0\big)$ enters the left crescent in Fig.\ref{domaines}$(a)$. This is  confirmed by the upper plots in Fig.~\ref{Omega}.
Even if the initial value~${c_{\bot0}\over \gamma_{\rm c}c_{\Phi0}}$ is far above the tiny range $[-1,\approx \!\!-0.9941]$, the ratio~${c^{\dyn}_\bot\over \gamma_{\rm c}c^{\dyn}_\Phi}$ is inexorably attracted to this interval, where it stabilizes. The lower plots in Fig.~\ref{Omega} zoom  the entrance and freezing of ${c^{\dyn}_\bot\over \gamma_{\rm c}c^{\dyn}_\Phi}$ in the range. 

\begin{figure}[!h]
\vspace{0.3cm}
\begin{center}
\includegraphics[width=8.18cm]{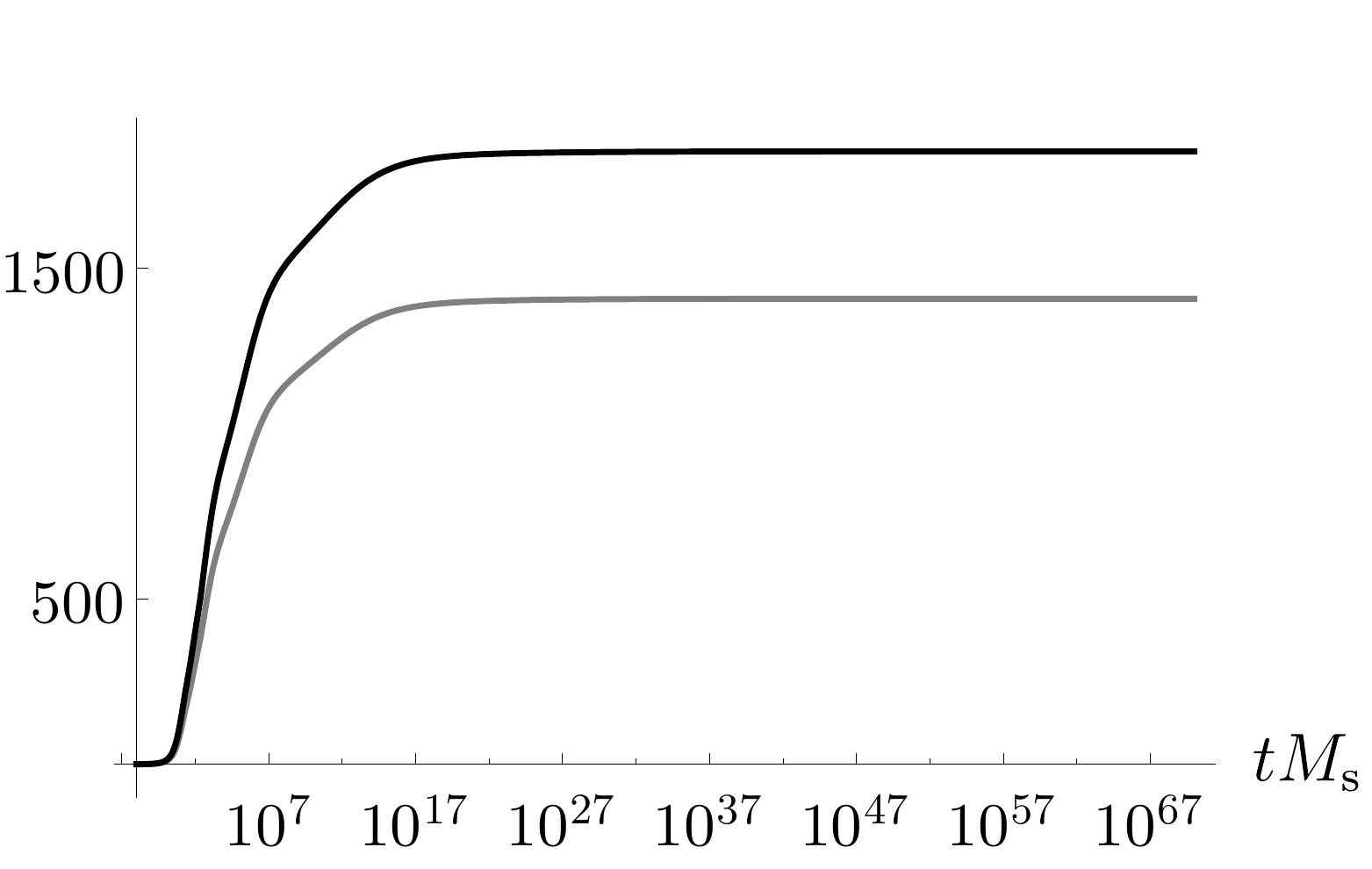}  
\includegraphics[width=8.18cm]{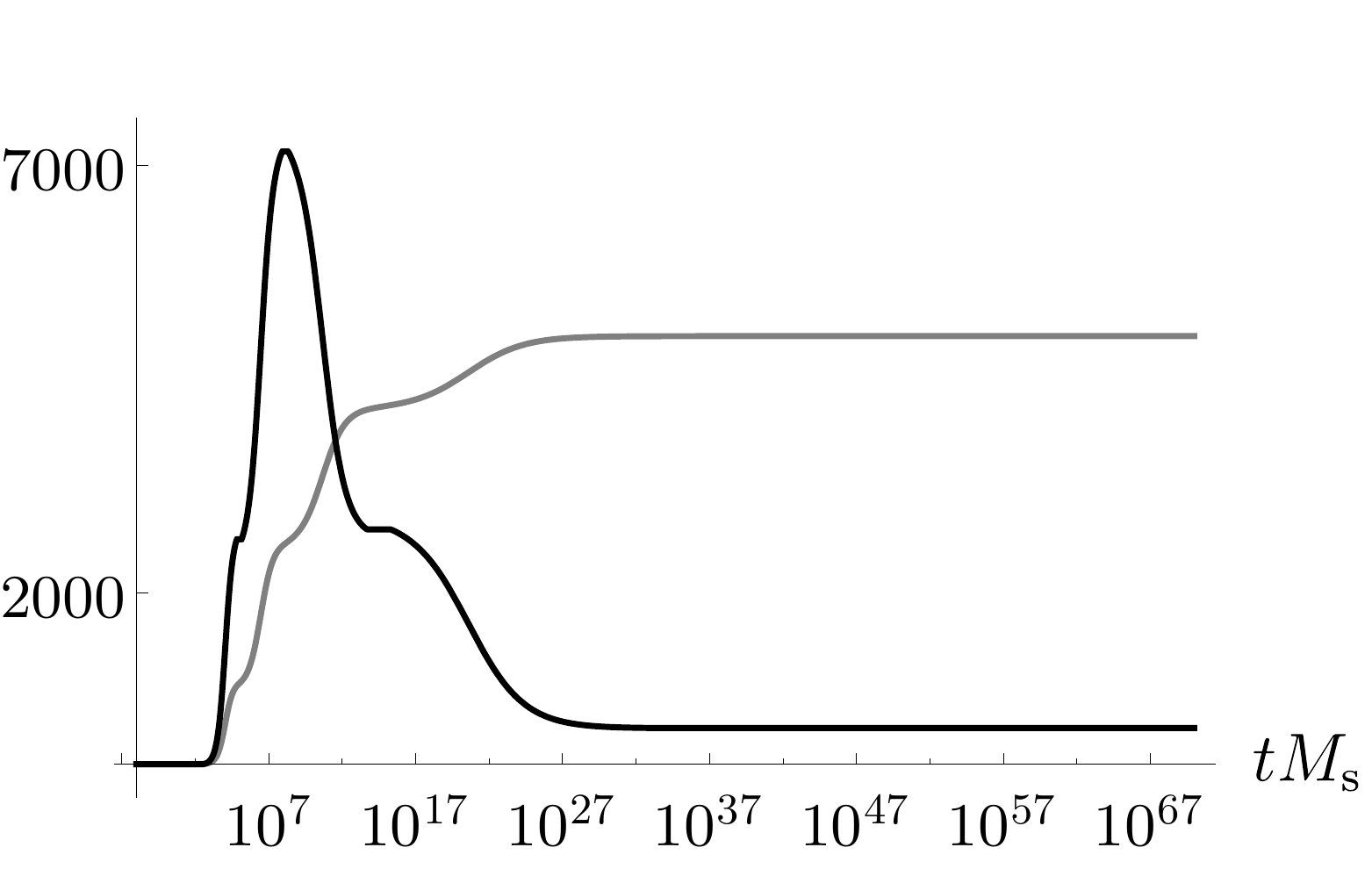}\\ \vspace{0.5cm}
\includegraphics[width=8.18cm]{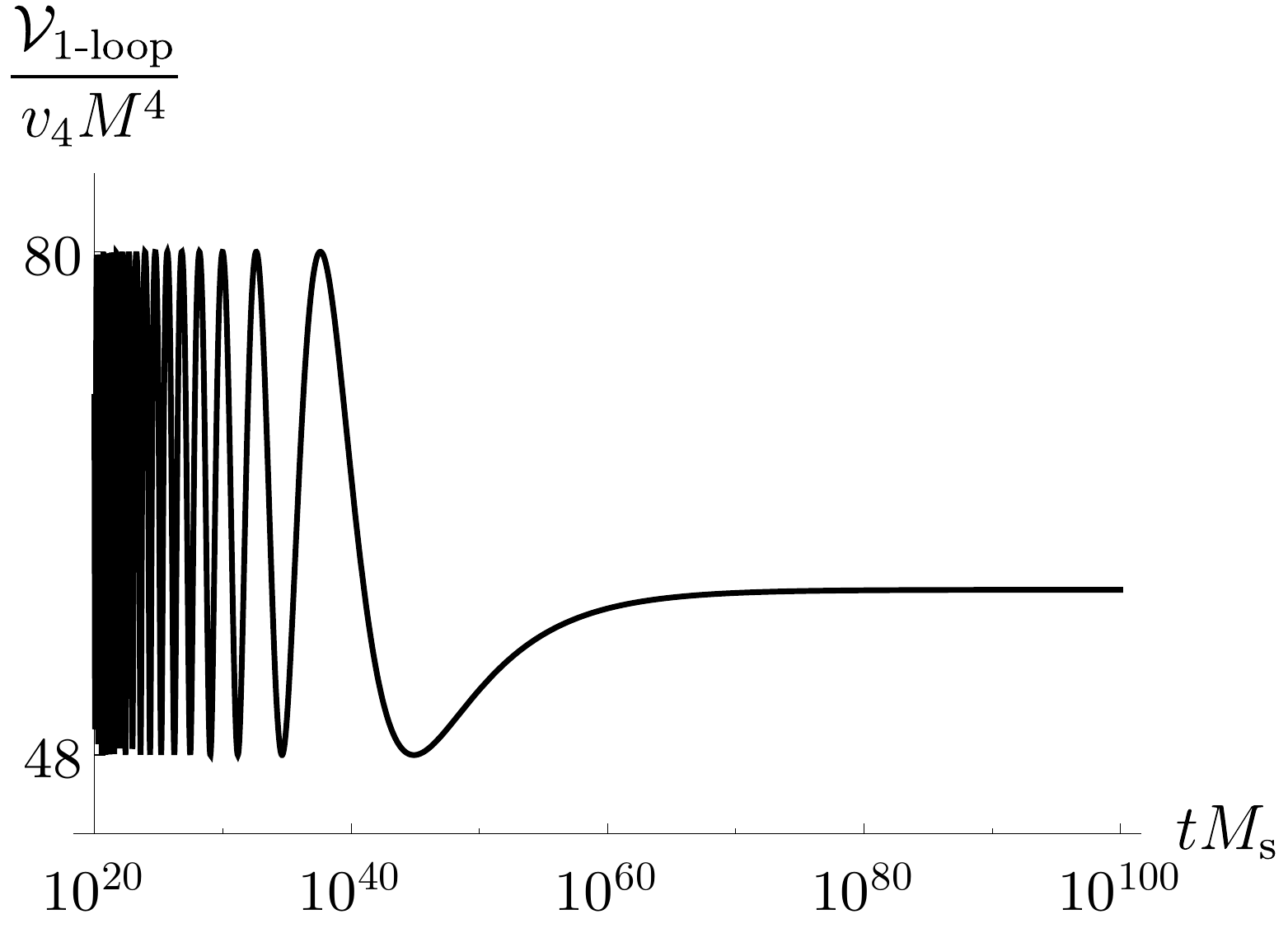} 
\includegraphics[width=8.18cm]{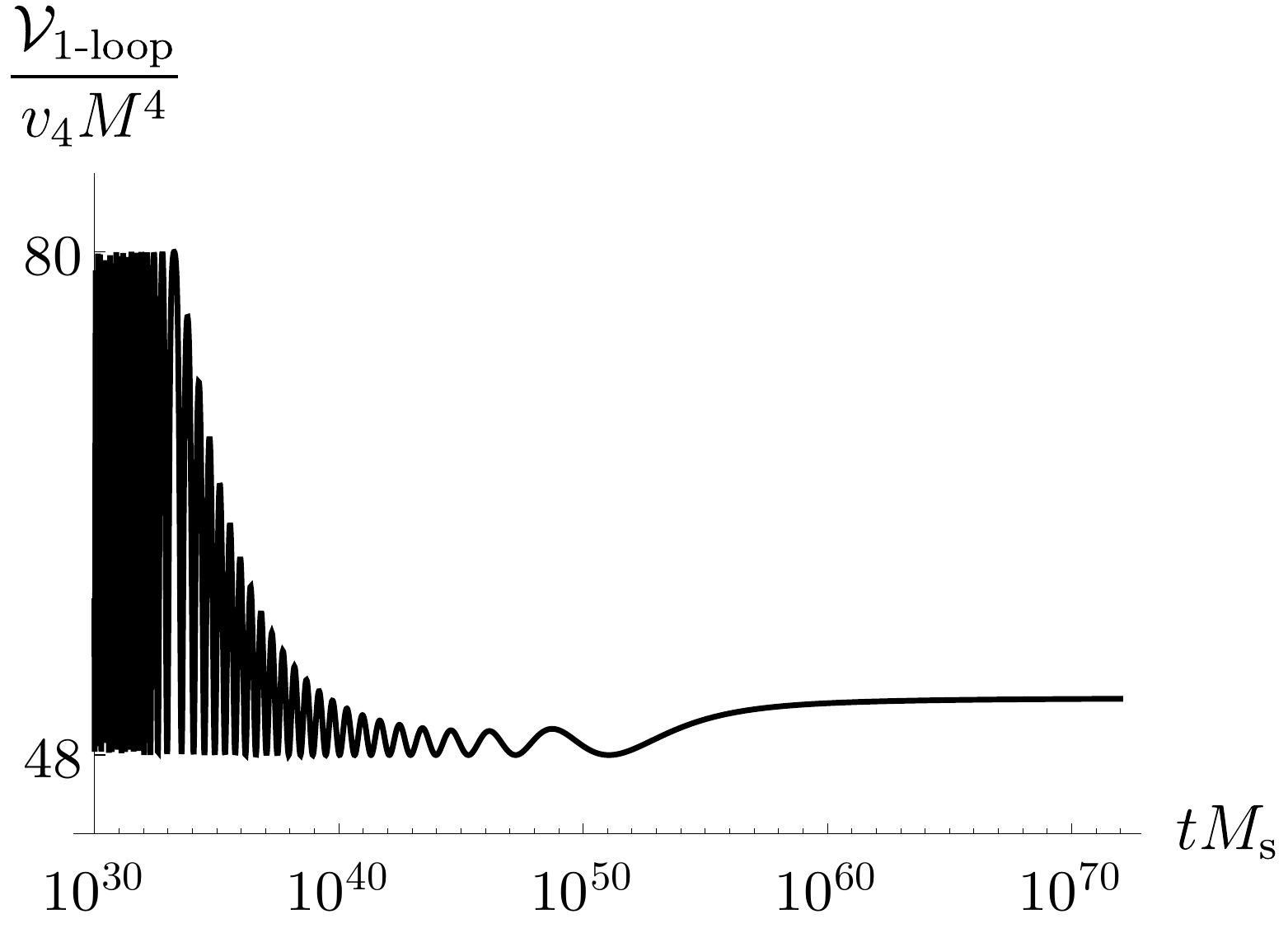}
\end{center}
\begin{picture}(0,0)
\put(189.5,300){\gray$y_{54}$}\put(426,294){\gray$y_{54}$}
\put(189.5,325){$y_{45}$}\put(426,242){$y_{45}$}
\put(105,351){\large($a$)}\put(348,351){\large($b$)}
\put(105,185){\large($a$)}\put(348,185){\large($b$)}
\end{picture}
\vspace{-.9cm}
\caption{\footnotesize \em Convergence of $y_{45}$ (black curves) and $y_{54}$ (gray curves) to their limits $y_{45}^{(0)}$ and $y_{54} ^{(0)}$ in model~$(ii)$, for initial conditions of type~$(a)$ or~$(b)$ (upper plots). If the curve $y_{54}(t)$ is always monotonic, that of $y_{45}(t)$ may not be so in case~$(b)$. The final value $y_{45}^{(0)}$ is random in case~$(a)$, while it is close to a minimum of ${\cal V}_{\mbox{\scriptsize \rm 1-loop}}$ in case~$(b)$, due to the existence of damped oscillations (lower plots).}
\label{yy}
\end{figure}

The remaining numerical behaviors to be described are those of the Wilson lines. The upper plots in Fig.~\ref{yy} show the evolutions of~$y_{45}(t)$ and~$y_{54}(t)$, which converge to constants~$y_{45}^{(0)}$,~$y_{54}^{(0)}$. Due to Eq.~(\ref{r1}), which is exact when $y_{55}\equiv 0$, the curve $y_{54}(t)$ is monotonic. This however may not be  the case for~$y_{45}(t)$, which is not a free field. Actually, accentuating the plateaux structure \ie in case~$(b)$, the magnitudes of both velocities $\dot y_{45}$, $\dot y_{54}$ drop during the transient eras of quasi static $(a^{d-1})^{\dis \cdot}$, and $y_{45}$ may even stop and go backward. Notice that these effects are consistent with the fact that in the limit $\tau^\dyn(0)\underset{>}{\to} 1-r_0$ of the initial conditions, the Wilson lines are expected to become static, $\dot y_{45}\equiv \dot y_{54}\equiv 0$. 

Since $\sqrt{2}\, y_{45}$ can be very large, its convergence to $\sqrt{2} \, y^{(0)}_{45}$ is more accurately accounted for by the effective potential, which is 1-periodic. As shown in the lower plots in Fig.~\ref{yy}, $\Vone\over v_dM^d$ oscillates over time between~$80$ and~$48$, which are the values of $\nF-\nB$ in backgrounds~$(ii)$ and~$(i)$, when they are not deformed. For initial conditions of type~$(b)$, we see that before $\sqrt{2}\, y_{45}$ starts freezing, its kinetic energy is larger than the potential, since $\sqrt{2}\, y_{45}$ evolves quickly, passing easily the maxima and minima. This does not last, however, and $\sqrt{2}\, y_{45}$ ends up oscillating around a minimum, until it stabilizes at a random value close to it. The last fluctuations are those described in Sect.~\ref{simu}, in the massive case in Fig.~\ref{plot_y}$(i)$. As a result,  $\sqrt{2} \, y^{(0)}_{45}\simeq 2k+1$, where $k\in\Z$, \ie the dynamics has driven spontaneously the system from the initial state~$(ii)$ to a slightly deformed background~$(i)$. Remarkably, soon after $y_{45}$ is stabilized, we have checked that its remnant kinetic energy starts again to dominate over the potential, and their ratio even tends to infinity (!).  In the notations used at the end of the previous section, the case at hand leads to $\mbox{${K_+\over d-1}>J_+>0$}$, which  is nevertheless compatible with the stability of the cosmological solution. 
Concerning the modulus $y_{54}$, its monotonicity and the fact that it is a flat direction of the potential when $y_{55}\equiv 0$ imply   its stabilization not to be preceded by oscillations,  and its limit value~$y_{54}^{(0)}$ to be fully arbitrary. As described for $y_{45}$, the remnant kinetic energy of $y_{54}$  is greater than $\Vone$ right before and soon after its stabilization process.  On the contrary, as seen in the lower plot in Fig.~\ref{yy}$(a)$, the attraction of~$y_{45}$ to the neighborhood of a minimum of~$\Vone$ turns out to be inefficient for generic initial conditions of type~$(a)$. This is due to the fact that as soon as the Wilson line cannot pass the next maximum of the potential, it freezes. 

To summarize, the simulated cosmological evolutions in model~$(ii)$ are attracted to the QNSR $t\to +\infty$, for both initial conditions~$(a)$ and~$(b)$. The asymptotic behaviors are reached in a finite number of steps, more visible in case~$(b)$, where the dynamics can be described as follows:

$\bullet$ The plateaux present in Fig.~\ref{multi_plateau}$(b)$ are characterized by almost constant $(a^{d-1})^{\dis\cdot}$. 

$\bullet$ At the beginning of each plateau, the kinetic energy of $y_{45}$ and $y_{54}$ is lower than the effective potential. The motion of these moduli is slowing down, especially for $y_{45}$ whose time-derivative may change sign. As a result, the universe enters an approximate QNSR (see Ref.~\cite{CFP} for the limit case of frozen Wilson lines, which yields a global attraction to the QNSR $t\to +\infty$). 

$\bullet$ However, except at the last step, $J_\dyn $ reaches  negative minimum values at the beginning of each plateau, as seen in Fig.~\ref{JDyn}$(b)$. As a result, the domination of the potential over the Wilson lines' kinetic energies does not last. When the latter  become greater than the canonical kinetic energies of $\Phi$ and $\phi_\bot$, the approximate QNSR is destabilized and $(a^{d-1})^{\dis\cdot}$ leaves its current plateau.

$\bullet$ We have checked analytically that there is no power-like asymptotic solution that describes an ever-expanding flat universe, dominated by the Wilson lines' kinetic energies.\footnote{\label{bcy}Power-like limit behaviors describing a Big Crunch (or Big Bang) dominated by the Wilson lines' kinetic energies however exist.} Thus,~$y_{45}$ and~$y_{54}$ have to release their kinetic energies to the rest of the system, so that the universe is again  attracted to an approximate QNSR. In other words, $(a^{d-1})^{\dis\cdot}$ has moved from one plateau to the next.

$\bullet$ This process of climbing steps ends when the system enters a plateau where $J_\dyn$ is positive. In this case, the Wilson lines' kinetic energies may soon dominate over the effective potential (when ${K_+\over d-1}>J_+$), but never over the canonical kinetic energies of $\Phi$ and $\phi_\bot$ (because $J_+>0$).  As a result, the universe remains in QNSR for good.

\noindent For generic initial conditions of type~$(a)$, even if the plateaux of $(a^{d-1})^{\dis\cdot}$ and the slowdowns of the Wilson lines are less pronounced, the correspondence between the negative minima of~$J_\dyn$ and the transient dominations of the potential over the kinetic energies of $y_{45}$ and $y_{54}$ remains valid. 


\section{Conclusion}
\label{cl}

In this work, we have shown that the notion of QNSR introduced in Ref.~\cite{CFP} for toy models involving only the scale factor~$a$, the supersymmetry breaking scale~$M\equiv e^{\alpha \Phi}$ and the dilaton~$\phi$ can be extended to full string theories. This has been done at the 1-loop level in toroidally compactified heterotic string at weak coupling, where a Scherk-Schwarz mechanism involving a single internal direction $X^d$ breaks spontaneously all supersymmetries. The key point is the presence of a bunch of marginal deformations: $\sqrt{G^{dd}}$ and $\phi$, which are equivalent to the canonical no-scale modulus $\Phi$ and scalar $\phi_\bot$, and the Wilson lines $y_{d\Upsilon}$,  $y_{id}$, $y_{i\Upsilon}$, for $\Upsilon\in\{d+1,\dots, 25\}$, $i\in\{d+1,\dots, 9\}$. If we have analyzed in great details the dynamics involving the moduli where $\Upsilon=i=d+1$, our results should be more general. In the QNSRs describing an ever-expanding universe or a Big Bang, the kinetic energies of $\Phi$, $\phi_\bot$, $y_{i\Upsilon}$ are expected to dominate over those of $y_{d\Upsilon}$,  $y_{id}$ and the effective potential $\Vone$. As a result, the classical no-scale structure is restored at the quantum level during the cosmological evolution of the flat universe. 

The existence of the QNSRs is independent of the characteristics of $\Vone$. Denoting the values at $t=0$ of the scale factor and moduli fields as $a(0)$ and  $(\Phi(0),\phi(0),y(0))$, the initial time derivatives $(\dot \Phi(0),\dot \phi(0),\dot y(0))$ can always be set in a phase space region for the universe to be attracted to a QNSR. This turns out to be the case whether $\Vone(\Phi(0),\phi(0),y(0))$ is positive, negative or null, as well as maximal, minimal or at a saddle point. 

Global effects, however,  depend drastically on the sign of $\Vone$. We have ``shown numerically'' in dimension~4 that when~$y_{45}$ and~$y_{54}$ vary arbitrarily, while keeping~$y_{55}$ frozen at a point of extended light spectrum, the initially growing universes always end in the QNSR $a\to +\infty$, provided $\Vone\ge 0$ for all~$y_{45}$. Allowing all $y$-deformations to be dynamical,  we expect this attraction to be true when the trajectory does not explore regions in moduli space where $\Vone< 0$.\footnote{In any case, the evolutions along which $\Vone\ge 0$ are ever-expanding, due to the monotonicity of $(a^{d-1})^{\dis \cdot}>0$ (see Eq.~(\ref{vseul})), which  forbids $H$ to vanish.} As noticed before, this sufficient condition is not necessary, when the initial conditions are tuned in tiny intervals. On the contrary, when an initially growing cosmological evolution does not converge to the QNSR $a\to +\infty$, which requires $\Vone$ to reach negative values, the simulated expansion of the scale factor comes to a halt and the universe eventually collapses. In Ref.~\cite{CFP}, such a Big Crunch can be realized in two ways: With the QNSR $a\to 0$ (by applying time reversal on the Big Bang solution), or as an evolution dominated by the no-scale modulus kinetic and potential energies. As noticed in Footnote~\ref{bcy}, taking into account the Wilson lines' dynamics, the kinetic energies of the scalars $y_{d\Upsilon}$, $y_{id}$ may also dominate. It would be interesting to extend the analysis of the system to derive an overview of all possible limit behaviors of the solutions and associated attractor mechanisms. 

An important consequence of  the above remarks is that a flat expanding universe is more naturally described by a model with positive potential, while Big Crunch solutions arise in most cases when the potential is negative. This result is in the spirit of Refs~\cite{critical, attra, attraM/T,attT}, where hot universes are considered, \ie when finite temperature $T$ is switched on in addition to the implementation of the spontaneous breaking of supersymmetry.  In this case, when the zero-temperature effective potential is positive, the trajectory of the flat universe at finite~$T$ is attracted to an expanding solution satisfying proportionality properties~\cite{attra,attraM/T},
\be
{1\over a(t)}\sim \#\,  {M(t)\over \Ms} \sim \#\, {T(t)\over \Ms}\sim  \#\, e^{2\alpha^2\phi(t)}\sim {\#\over (t\Ms)^{2\over d}}\, .
\ee
The above asymptotic evolution is said to be ``radiation-like'', due to the state equation $\mbox{$\rho_{\rm tot}\sim (d-1)P_{\rm tot}$}$ satisfied by the total energy density $\rho_{\rm tot}$ and pressure $P_{\rm tot}$ present in the universe. The latter take into account the thermal contributions derived from the 1-loop free energy, as well as the kinetic energy of the no-scale modulus~$\Phi$. On the contrary, when the zero-temperature 1-loop potential is negative, the universe at finite $T$ collapses into a Big Crunch, where temperature effects tend to be screened, $T/M\to 0$. 
 


\section*{Acknowledgement}
 
We are grateful to  Steve Abel, Carlo Angelantonj, Keith Dienes, Emilian Dudas, Sergio Ferrara, Claude Fleming, Lucien Heurtier, Alexandros Kahagias and Costas Kounnas  for fruitful discussions. 
The work of H.P. is partially supported by the Royal Society International Cost Share Award. H.P. thanks the C.E.R.N. Theoretical Physics Department, the Simons Center for Geometry and Physics, and the IPPP in Durham University for hospitality.


\section*{Appendix: Moduli dependence of the effective potential}
\label{wl}
\renewcommand{\theequation}{A.\arabic{equation}}
\renewcommand{\thesection}{A}
\setcounter{equation}{0}

For the present work to be self-content, let us  review how discrete deformations responsible for the total spontaneous breaking of supersymmetry, as well as continuous Wilson lines, can be introduced in maximally supersymmetric heterotic string. Our final goal is to derive an expression of the effective potential valid when the supersymmetry breaking scale is low, compared to the string scale $\Ms$.


\subsection{The deformations}
\label{a1}

In the 1-loop partition function, the relevant deformed conformal block to be considered turns out to be
\begin{align}
\cZ[\vec a, \vec b,G,B,\vec Y]=&\,{\sqrt{\det G}\over \tau_2^{10-d\over 2}}\sum_{\substack{\tilde m_d,\dots,\tilde m_9 \\ n_d,\dots, n_9}} e^{-{\pi\over \tau_2}(\tilde m_I+n_I\bar \tau)(G+B)_{IJ}(\tilde m_I+n_I\tau)}\nonumber \\
&\, \times e^{i\pi n_I\vec Y_I\cdot (\vec b-\tilde m_J\vec Y_J)}\prod_{\A=1}^{4}\theta\!\left[{}^{a^L_\A-2n_IY^L_{I\A}}_{b^L_\A-2\tilde m_IY^L_{I\A}}\right]\!(\tau)\prod_{\J=10}^{25}\bar \theta\!\left[{}^{a^R_\J-2n_IY^R_{I\J}}_{b^R_\J-2\tilde m_IY^R_{I\J}}\right]\!(\bar\tau)\, ,
\label{block}
\end{align}
where sums over repeated indices $I,J\in\{d,\dots, 9\}$ are understood. Our notations are as follows:

$\bullet$ The first line is the contribution of the zero modes of the $10-d$ bosonic coordinates compactified on a torus, whose metric and antisymmetric tensor are $G_{IJ}$, $B_{IJ}$, $I,J\in\{d,\dots,9\}$.  Written in Lagrangian form, this expression involves a discrete sum over the integers $\tilde m_I$ and winding numbers $n_I$. 

$\bullet$ In the second line, the holomorphic Jacobi $\theta$ functions arise from the partition functions $\theta/\eta$ of the 4 complex left-moving fermions of the superstring, where $\eta$ is the Dedekind function\footnote{Our conventions for $\theta$ and $\eta$ functions can be found in Ref.~\cite{KiritsisBook}.}. In their brackets, $a^L_\A$ and $b^L_\A$, $\A\in\{1,\dots,4\}$,  define their boundary conditions before deformation, along the cycles $z\to z+1$ and $z\to z+\tau$ of the genus-1 worldsheet parameterized by $z$ and of Teichmüller parameter $\tau\equiv \tau_1+i\tau_2$. Similarly, the antiholomorphic~$\bar \theta$ functions arise from the contributions $\bar \theta/\bar \eta$ associated with the 16 complex right-moving fermions of the bosonic string, with  boundary conditions before deformation determined by~$a^R_\J$ and~$b^R_\J$, $\J\in\{10,\dots,25\}$. In the derivation to come, $\vec a\equiv(\vec a^L,\vec a^R)$ and $\vec b\equiv (\vec b^L,\vec b^R)$ can have arbitrary real entries. However, modular invariance of the entire model imposes constraints on the set of values they can take. For instance, in our maximally supersymmetric case of interest, we have $a^L_1=\dots=a^L_4$ and $b^L_1=\dots=b^L_4$. However, we will keep the 4-components of~$\vec a^L$ and~$\vec b^L$ independent, since this can be useful when dealing with non-maximally supersymmetric models. (See Ref.~\cite{SNSM2} for an example in 4 dimensions realizing the $\N=2\to 0$ spontaneous breaking.)

$\bullet$ Beside the torus moduli $(G+B)_{IJ}$, we introduce deformations of the left- and right-moving (super)-conformal theories,
\be
Y^L_{I\A}\, ,\:  Y^R_{I\J}\, , \quad I\in\{d,\dots, 9\}\,,  \; \A\in\{1,\dots, 4\}\,,  \; \J\in\{10,\dots, 25\}\, .
\ee
For the holomorphic supercurrent to be preserved, the left-moving ones are quantized~\cite{SSstring}, $Y^L_{I\A}\in\Z$. Thus, different choices of $Y^L_{I\A}$'s yield different models. On the contrary, the right-moving $Y^R_{I\J}$'s are arbitrary marginal deformations. In each given model, they are moduli fields that can be interpreted as Wilson lines along $T^{10-d}$ of a rank 16 gauge group $\G_{16}$. 

$\bullet$ In the second line of Eq.~(\ref{block}), the overall phase uses the following  definition of scalar product:   For two vectors  $\vec v\equiv(\vec v^L,\vec v^R)$ and $\vec w\equiv(\vec w^L,\vec w^R)$ in $\R^{4,16}$, we write 
\be
\vec v\cdot \vec w = \vec v^L\cdot \vec w^L-\vec v^R\cdot \vec w^R=\sum_{\A=1}^4 v^L_\A w^L_\A-\sum_{\J=10}^{25} v^R_\J w^R_\J\, .
\ee
The phase is introduced for the following modular transformations of the entire conformal block to be independent of $\vec Y_I\equiv (\vec Y_I^L,\vec Y_I^R)$:
\begin{align}
\label{motrans}
\tau\to -{1\over \tau} &\Longleftrightarrow \left\{ \begin{array}{l}
(n_I,\tilde m_I)\to (n_I,\tilde m_I) \S \espD\\ 
(a^L_\A,b^L_\A)\to (a^L_\A,b^L_\A)\S\, , \; (a^R_\J,b^R_\J)\to (a^R_\J,b^R_\J)\S\, 
\end{array}
\right.\nonumber \\ 
 \tau\to \tau+1 &\Longleftrightarrow \left\{ \begin{array}{l}
(n_I,\tilde m_I)\to (n_I,\tilde m_I) \T\espD\\
(a^L_\A,b^L_\A)\to (a^L_\A,b^L_\A+a^L_\A-1)\, , \; (a^R_\J,b^R_\J)\to(a^R_\J,b^R_\J+a^R_\J-1)\, 
\end{array}
\right.\nonumber \\
&\esp \mbox{where} \qquad \S=\left(\!\!\begin{array}{cc} 0 &-1\\ 1 & 0\end{array}\!\!\right)\, , \qquad \T=\left(\!\!\begin{array}{cc} 1 &1\\ 0 & 1\end{array}\!\!\right) .
\end{align}
Thus, any 1-loop partition function, which is modular invariant for $\vec Y_I=\vec{0}$, $I\in\{d,\dots,9\}$, remains consistent when arbitrary $Y$-deformations are switched on.

Another way to write Eq.~({\ref{block}) clarifies the spectrum interpretation of the conformal block, at the cost of obscuring the modular transformation $\tau\to -1/\tau$. It is obtained by inserting in $\cZ[\vec a, \vec b,G,B,\vec Y]$ the definition of the $\theta$ functions in terms of a sum over $N\in \Z$, 
\be
\theta[^a_b](\tau)=\sum_Nq^{{1\over 2}(N-{a\over 2})^2}e^{-bi\pi (N-{a\over2})} ,\quad \where \quad q\equiv  e^{2i\pi\tau}\, ,
\ee
and applying a Poisson summation over the integers $\tilde m_d,\dots,\tilde m_9$. The result is the Hamiltonian form \cite{421},
\be
\cZ[\vec a, \vec b,G,B,\vec Y]=\sum_{\substack{m_d,\dots,m_9 \\ n_d,\dots, n_9}} \sum_{\vec N}e^{-i\pi\vec b\cdot \vec Q}\, q^{{1\over 4}\left[ P^L_IG^{IJ}P^L_J+2(\vec Q^L+n_I\vec Y^L_I)^2\right]}\, \bar q^{{1\over 4}\left[ P^R_IG^{IJ}P^R_J+2(\vec Q^R+n_I\vec Y^R_I)^2\right]},
\label{Ham}
\ee
where  we have defined $\vec N\equiv(\vec N^L,\vec N^R)$, $G^{IJ}\equiv (G^{-1})_{IJ}$ and 
\begin{align}
P^L_I&=m_I-\vec Y\cdot \vec Q-{1\over 2}\, \vec Y_I\cdot n_J\vec Y_J+(B+G)_{IJ}n_J\, , \quad I\in\{d,\dots, 9\}\, , \nonumber \\
P^R_I&=m_I-\vec Y\cdot \vec Q-{1\over 2}\, \vec Y_I\cdot n_J\vec Y_J+(B-G)_{IJ}n_J\, , \nonumber \\
\vec Q&\equiv(\vec Q^L,\vec Q^R)=\vec N-{\vec a\over 2}\, .
\end{align}

The genus-1 partition function of a model takes the following form
\be
\label{defi}
Z[G,B,\vec Y]={1\over \tau_2^{d-2\over 2}}\, {1\over \eta^{12}\, \bar \eta^{24}}\, {1\over |\Xi|}\sum_{\vec a, \vec b \in\Xi}\C\big[{}^{\vec a}_{\vec b}\big] \cZ[\vec a, \vec b,G,B,\vec Y]\, , 
\ee
where $\Xi$ is the set of spin structures $\vec a$ and $\vec b$ take, $|\Xi|$ is the cardinal of $\Xi$, and $\C\big[{}^{\vec a}_{\vec b}\big]$ are complex numbers of modulus 1,  so that $Z[G,B,\vec{\boldsymbol{0}}]$ is modular invariant. Expanding
\be
{1\over \eta^{12}\, \bar \eta^{24}}={1\over q^{1\over 2}\, \bar q}\sum_{\ell_L,\ell_R\ge 0}c^L_{\ell_L}c^R_{\ell_R}\, q^{\ell_L}\, \bar q^{\ell_R}, 
\ee
we obtain
\be
Z[G,B,\vec Y]={1\over \tau_2^{d-2\over 2}}{1\over |\Xi|}\sum_{\vec a, \vec b \in\Xi}\sum_{\vec N}\C\big[{}^{\vec a}_{\vec b}\big]e^{-i\pi\vec b\cdot \vec Q}  \sum_{\ell_L,\ell_R\ge 0}c^L_{\ell_L}c^R_{\ell_R} \sum_{\substack{m_d,\dots,m_9 \\ n_d,\dots, n_9}}  q^{{1\over 4}{M_L^2/\Ms^2}}\, \bar q^{{1\over 4}{M_R^2/\Ms^2}},
\label{Ha}
\ee
in terms of left- and right-moving squared masses 
\begin{align}
M_L^2&=\Ms^2\left[ P_IG^{IJ}P_J+2(\vec Q^L+n_I\vec Y^L_I)^2+4\ell_L-2\right] ,\nonumber \\
 M_R^2&=\Ms^2\left[ \bar P_IG^{IJ}\bar P_J+2(\vec Q^R+n_I\vec Y^R_I)^2+4\ell_R -4\right] .
\end{align}
In Eq.~(\ref{Ha}), the sum over $\vec b\in\Xi$ divided by $|\Xi|$ implements the generalized GSO projection. Since 
\be
q^{{1\over 4}{M_L^2/\Ms^2}}\, \bar q^{{1\over 4}{M_R^2/\Ms^2}}=e^{2i\pi\tau_1{M_L^2-M_R^2\over 4\Ms^2}}\, e^{-\pi\tau_2{M_L^2+M_R^2\over 2\Ms^2}}\, , 
\ee
invariance under $\tau_1\to\tau_1+1$ implies 
\be
\label{lmc}
{M_L^2-M_R^2\over 4\Ms^2}\equiv m_In_I+{1\over 2} \big(\vec Q^2+1\big)\!+\ell_L-\ell_R\in \Z\, ,
\ee 
for all states that survive the GSO projection. Among them, the physical ones are those which contribute to the integral over $\tau_1\in[-{1\over 2},{1\over 2}]$  \ie whose squared masses satisfy $M^2=M_L^2=M_R^2$.


\subsection{The \bm  $SO(32)$ and $E_8\times E_8'$ heterotic string}
\label{a2}

As a warm up, let us recover the massless spectrum of the $SO(32)$ and $E_8\times E_8$ heterotic string compactified on $T^{10-d}$. In the former case, the partition function  is obtained with 
\begin{align}
&\Xi=\big\{ \mbox{$(4,16)$-tuples } (a,\dots, a; \gamma,\dots ,\gamma) \, ,\;  \where\; \;  a,\gamma\in \Z_2\big\}\; \Longrightarrow\; |\Xi|=2^2\, , \nonumber \\
&\C[^{a;\gamma}_{b\, ;\delta}]=(-1)^{a+b+ab}\, ,\nonumber\\
&\vec Y^L_I= \vec{0} \, , \; \vec Y^R_I=\vec{0}\, , \quad  I\in\{d,\dots, 9\}\, .
\end{align}
The lightest physical states have, on the left-moving side, $\ell_L=0$ and $P^L_I=0$, $I\in\{d,\dots,9\}$. The sector $a=0$ yields  spacetime bosons, whose charges $\vec Q^L$ are the weights of the {\bm $8_{\rm v}$} vectorial representation of the $SO(8)$ affine Lie algebra, while $a=1$ leads to fermions in the  {\bm $8_{\rm s}$} spinorial representation. All are massless. On the right-moving side, this implies $\gamma=0$. Moreover, at oscillator level $\ell_R=1$, we have  $P^R_I=0$, $I\in\{d,\dots,9\}$, and  $\vec Q^R=\vec 0$, corresponding to $c^R_1=24$ modes. For $\ell_R=0$, the charges $\vec Q^R$ are either the roots of $\G_{16}=SO(32)$ with $P^R_I=0$, $I\in\{d,\dots,9\}$, or $\vec Q^R=\vec 0$ with ${1\over 2}P^R_IG^{IJ}P^R_J=2$. In the latter case, the modes have charges equal to the roots of a gauge group $\G_{10-d}$ of rank $10-d$. Writing $24=(d-2)+(10-d)+16$, the massless states are organized as follows, 
\be
(\boldsymbol{8_{\rm v}}\oplus \boldsymbol{8_{\rm s}})\otimes \big([d-2]\oplus {\rm Adj}_{\G_{10-d}}\oplus {\rm Adj}_{\G_{16}}\big) ,
\label{spect}
\ee
corresponding to a supergravity multiplet in $d$ dimensions, coupled to a vector multiplet in the adjoint representation of $\G_{10-d}\times \G_{16}$.

The 1-loop partition function of the $E_8\times E_8'$ heterotic strings is realized with
\begin{align}
&\Xi=\big\{ \mbox{$(4,8+8)$-tuples } (a,\dots, a; \gamma,\dots ,\gamma,\gamma',\dots,\gamma') \, ,\;  \where\; \;  a,\gamma,\gamma'\in \Z_2\big\}\; \Longrightarrow\; |\Xi|=2^3\, , \nonumber \\
&\C[^{a;\gamma,\gamma'}_{b\, ;\delta,\, \delta'}]=(-1)^{a+b+ab}\, ,\nonumber\\
&\vec Y^L_I= \vec{0} \, , \; \vec Y^R_I=\vec{0}\, , \quad  I\in\{d,\dots, 9\}\, .
\end{align}
The left-moving side of the lightest physical states is identical to that encountered in the $SO(32)$ case. It is massless, which implies $(\gamma,\gamma')\neq (1,1)$ on the right-moving side. At oscillator level $\ell_R=1$, there are again $c^R_1=24$ modes with $P^R_I=0$, $I\in\{d,\dots,9\}$, and  $\vec Q^R=\vec 0$ (implying $(\gamma,\gamma')=(0,0)$). For $\ell_R=0$, the charges $\vec Q^R$ in the sector $(\gamma,\gamma)=(0,0)$ are either the roots of $SO(16)\times SO(16)'$ with $P^R_I=0$, $I\in\{d,\dots,9\}$, or $\vec Q^R=\vec 0$ with ${1\over 2}P^R_IG^{IJ}P^R_J=2$, corresponding to states whose charges are the roots of a gauge group $\G_{10-d}$. For $(\gamma,\gamma')=(1,0)$ or $(0,1)$, $\vec Q^R$ is a weight of the spinorial representation of $SO(16)$ or  
$SO(16)'$ with   $P^R_I=0$, $I\in\{d,\dots,9\}$. Noticing that the adjoint of $E_8$ can be decomposed into the adjoint plus spinorial representation of $SO(16)$, $[248]_{E_8}=[120]_{SO(16)}\oplus[128]_{SO(16)}$, the massless spectrum is given in Eq.~(\ref{spect}), with $\G_{16}=E_8\times E_8'$. 


\subsection{Spontaneous supersymmetry breaking and Wilson lines}
\label{a3}

The spontaneous breaking of all supersymmetries can be realized by a suitable choice of discrete left-moving deformations. For instance, for a stringy Scherk-Schwarz~\cite{SSstring,Kounnas-Rostand} mechanism implemented along a single internal direction $X^d$, we can take
\be
Y^L_{I\A}=\delta_{Id}\, \delta_{\A 1}\, ,\; Y^R_{I\J} \mbox{ arbitrary}\, , \quad  I\in\{d,\dots, 9\}\,,  \; \A\in\{1,\dots, 4\}\,,  \; \J\in\{10,\dots, 25\}\, ,
\label{nosusy}
\ee
in terms of which the conformal block in Eq.~(\ref{block}) becomes
\begin{align}
{\sqrt{\det G}\over \tau_2^{10-d\over 2}}\sum_{\substack{\tilde m_d,\dots,\tilde m_9 \\ n_d,\dots, n_9}} & e^{-{\pi\over \tau_2}(\tilde m_I+n_I\bar \tau)(G+B)_{IJ}(\tilde m+n_I \tau)}\, e^{-i\pi(\tilde m_d a_1^L-n_db^L_1+\tilde m_d n_d)}\nonumber \\
&\, \times e^{-i\pi n_I\vec Y^R_I\cdot (\vec b^R-\tilde m_J\vec Y^R_J)}\prod_{\A=1}^{4}\theta\!\left[{}^{a^L_\A}_{b^L_\A}\right]\!(\tau)\prod_{\J=10}^{25}\bar \theta\!\left[{}^{a^R_\J-2n_IY^R_{I\J}}_{b^R_\J-2\tilde m_IY^R_{I\J}}\right]\!(\bar\tau)\, .
\label{block2}
\end{align}
Compared to $\cZ[\vec a, \vec b,G,B,(\vec{\boldsymbol{0}}{}^L,\vec Y^R)]$, a pure phase appears in the first line. Consistently, the latter is invariant under the modular transformations (\ref{motrans}). The key point is that it depends on $a^L_1$, which determines the Neveu-Schwarz ($a^L_1=0$) or Ramond ($a^L_1=1$) boundary condition of the light cone worldsheet fermions $\psi^2, \psi^3$, \ie the bosonic or fermionic nature of the states. By Poisson summation, the momentum along the direction $X^d$ being shifted by $a^L_1/2$, the boson/fermion degeneracy is lifted. 

To see how this works explicitly, we consider the Hamiltonian form given in Eq.~(\ref{Ham}). It is convenient to write the latter using redefined internal metric, antisymmetric tensor and Wilson lines,  
\begin{align}
&(G'+B')_{IJ}=\left(\!\!\begin{array}{lr}4(G+B)_{dd} & 2(G+B)_{dj}\\2(G+B)_{id}&(G+B)_{ij}\end{array}\!\!\right) , \;\; \vec Y^{\prime R}_d = 2\vec Y^R_d  \, , \; \vec Y^{\prime R}_i = \vec Y^R_i  \, , \;\;  i,j\in\{d+1,\dots,9\}\, ,
\label{orbi}
\end{align}
as well as new momenta, winding numbers and indices $N'$'s,
\begin{align}
&m_d'=2(m_d-N^L_1)+a_1^L-n_d\, , \; m_i'=m_i\, , \quad i\in\{d+1,\dots, 9\}\nonumber \\
&n_d'={n_d\over 2}\, , \; n_i'=n_i\, , \nonumber \\
&N^{\prime L}_1=N^L_1+n_d\, , \; N^{\prime L}_\A=N^L_\A\, ,  \; N^{\prime R}_\J=N^R_\J\, , \quad \A\in\{2,3, 4\}\, , \; \J\in\{10,\dots, 25\}\, ,
\end{align}
where $n'_d\in \Z\cup (\Z+{1\over 2})$. Given the above notations, we set
\begin{align}
P^{\prime L}_I&=m'_I+\vec Y_I^{\prime R}\cdot \vec Q^{\prime R}+{1\over 2}\, \vec Y_I^{\prime R}\cdot n'_J\vec Y^{\prime R}_J+(B'+G')_{IJ}n'_J\, , \quad I\in\{d,\dots, 9\}\, ,\nonumber \\
P^{\prime R}_I&=m'_I+\vec Y_I^{\prime R}\cdot \vec Q^{\prime R}+{1\over 2}\, \vec Y^{\prime R}_I\cdot n'_J\vec Y^{\prime R}_J+(B'-G')_{IJ}n'_J\, , \nonumber \\
\vec Q'&\equiv(\vec Q^{\prime L},\vec Q^{\prime R})=\vec N'-{\vec a\over 2}\, ,
\end{align}
in terms of which the conformal block (\ref{Ham}) becomes
\be
\sum_{\substack{m_d\\ n_d}}e^{2i\pi b^L_1n'_d}\sum_{\substack{m'_{d+1},\dots,m'_9 \\ n'_{d+1},\dots, n'_9}} \sum_{\vec N'}e^{-i\pi\vec b\cdot \vec Q'} 
 \, q^{{1\over 4}\left[ P^{\prime L}_IG^{\prime IJ}P^{\prime L}_J+2(\vec Q^{\prime L})^2\right]}\,  \bar q^{{1\over 4}\left[ P^{\prime R}_IG^{\prime IJ}P^{\prime R}_J+2(\vec Q^{\prime R}+n'_I\vec Y^{\prime R}_I)^2\right]}.
\ee
As a result, the partition function takes the suggestive form
\begin{align}
Z'[G',B',\vec Y^{\prime R}]
=&{1\over \tau_2^{d-2\over 2}}{1\over |\Xi|}\sum_{\vec a, \vec b \in\Xi}\sum_{\vec N'}\C\big[{}^{\vec a}_{\vec b}\big]e^{-i\pi\vec b\cdot \vec Q'}  \sum_{\ell_L,\ell_R\ge 0}c^L_{\ell_L}c^R_{\ell_R} \nonumber \\
&\, \times \sum_{n_d'\in\Z\, \cup \, (\Z+{1\over 2})} \!\!\!e^{2i\pi b^L_1n'_d}\; \delta_{a^L_1,m'_d-2n'_d\, \rm mod\,  2}\sum_{\substack{m'_d,\dots,m'_9 \\ n'_{d+1},\dots, n'_9}}  q^{{1\over 4}{{M'}_{\!\!L}^2 /\Ms^2}}\, \bar q^{{1\over 4}{{M'}_{\!\!R}^2 /\Ms^2}},
\label{za}
\end{align}
where the left- and right-masses satisfy
\begin{align}
\label{Ha'}
{M'}_{\!\!L}^2 &=\Ms^2\left[ P^{\prime L}_IG^{\prime IJ}P^{\prime L}_J+2(\vec Q^{\prime L})^2+4\ell_L-2\right] ,\nonumber \\
{M'}_{\!\!R}^2 &=\Ms^2\left[ P^{\prime R}_IG^{\prime IJ}P^{\prime R}_J+2(\vec Q^{\prime R}+n'_I\vec Y^{\prime R}_I)^2+4\ell_R-4\right] .
\end{align}
Comparing $Z'[G',B',\vec Y^{\prime R}]$ with the supersymmetric partition function $Z[G,B,(\vec{\boldsymbol{0}}{}^L,\vec Y^R)]$, we observe three modifications: 

$\bullet$ The exchange of the moduli fields, as shown in Eq~(\ref{orbi}).

$\bullet$ The winding number $n_d'$ along the Scherk-Schwarz breaking direction can be integer or half-integer. In the latter case, the GSO-projection is reversed. 

$\bullet$ The fermionic number $a^L_1$ is restricted to be equal to the parity of $m_d'-2n_d'$, which shows that there is no more boson/fermion degeneracy. 

\noindent Note however that when the compact direction $X^d$ is large compared to the string scale, and the right-moving Wilson lines are small compared to $\sqrt{G^{\prime dd}}$, the states lighter than the KK mass $M'_{(\sigma)}=\Ms\sqrt{G^{\prime dd}}$ have vanishing momentum and winding numbers, $m_d'=n_d'=0$. Therefore, they are bosons, $a_1^L=0$, while their superpartners,  $m_d'=1$, $n_d'=0$, acquire a KK mass $M'_{(\sigma)}$ identified with the supersymmetry breaking scale. For light fermions to exist, we consider in the following more general patterns of supersymmetry breaking.  


\subsection{\!\!\!Supersymmetry breaking, discrete and continuous Wilson lines}
\label{a4}

In the setup described in the previous section, when the supersymmetry breaking scale $M'_{(\sigma)}=\Ms\sqrt{G^{\prime dd}}$ is low compared to the string scale $\Ms$, one way to have fermions lighter than $M'_{(\sigma)}$ is to introduce large $Y$-deformations. For this purpose, we consider  left- and right-moving discrete Wilson lines along the direction $X^d$,
\begin{align}
Y^L_{I\A}=\delta_{Id}\, \delta_{\A 1}\, ,\; &\, Y^R_{I\J} = \delta_{Id}\, \eta^R_\J+y^R_{I\J}\, , \quad  I\in\{d,\dots, 9\}\,,  \; \A\in\{1,\dots, 4\}\,,  \; \J\in\{10,\dots, 25\}\, ,\nonumber\espD\\
 \where \quad &\, \vec \eta^R\cdot a^R\in\Z\, , \quad y^R_\J \mbox  { arbitrary}\, .
 \label{ba}
\end{align}
In the above notations, $\vec \eta^R$ can be interpreted as a constant background, while $\vec y^R$ plays the role of  continuous Wilson lines. Using properties of the $\theta$ functions\footnote{\label{foot}We use the fact that $\vec \eta^R$ has integer entries. More specifically, in a consistent model, the components of $\vec a^R$ are  of the form $a^R_\J=k/\chi_\J$, $\J\in\{10,\dots,25\}$, where $\chi_\J\in\natural^*$ and $k$ spans the set $\{0,\dots,\chi_\J-1\}$. Thus, $\eta^R_\J\in \chi_\J \Z$.}, the conformal block~(\ref{block}) becomes
\begin{align}
{\sqrt{\det G}\over \tau_2^{10-d\over 2}}\sum_{\substack{\tilde m_d,\dots,\tilde m_9 \\ n_d,\dots, n_9}} & e^{-{\pi\over \tau_2}(\tilde m_I+n_I\bar \tau)(G+B)_{IJ}(\tilde m+n_I \tau)}\,  e^{i\pi \vec \eta^R\cdot\vec y_i^R(n_d\tilde m_i-n_i\tilde m_d) }\nonumber \\
&\, \times e^{-i\pi\left[\tilde m_d (a_1^L-\eta^R\cdot \vec a^R)-n_d(b^L_1-\eta^R\cdot \vec b^R)+\tilde m_d n_d(1-(\vec \eta^R)^2)\right]}\nonumber\\
&\, \times e^{-i\pi n_I\vec y^R_I\cdot (\vec b^R-\tilde m_J\vec y^R_J)}\prod_{\A=1}^{4}\theta\!\left[{}^{a^L_\A}_{b^L_\A}\right]\!(\tau)\prod_{\J=10}^{25}\bar \theta\!\left[{}^{a^R_\J-2n_Iy^R_{I\J}}_{b^R_\J-2\tilde m_Iy^R_{I\J}}\right]\!(\bar\tau)\esp\, ,
\label{blockB2}
\end{align}
where sums over repeated indices $i\in\{d+1,\dots, 9\}$ are understood. The phase in the second line generalizes that found in Eq.~(\ref{block2}), and will be shown to yield a new pattern of spontaneous breaking of supersymmetry in the light spectrum. In the first line, another phase, which is $\vec \eta^R$-dependent, has appeared. Consistently, it is invariant under the transformations~(\ref{motrans}), so that the modular properties of the whole conformal block are not spoiled. Actually, since 
\be
e^{i\pi \vec \eta^R\cdot\vec y_i^R(n_d\tilde m_i-n_i\tilde m_d) }=e^{-{\pi\over \tau_2}(\tilde m_I+n_I\bar \tau)\Delta B_{IJ}(\tilde m_J+n_J \tau)}, 
\ee
where $\Delta B_{IJ}$ is antisymmetric and defined as
\be
\Delta B_{dj}={1\over 2}\, \eta^R\cdot \vec y_j^R\, , \quad \Delta B_{ij}=0\, , \quad i,j\in\{d+1,\dots,9\}\, ,
\label{Db}
\ee
it is natural to write Eq.~(\ref{blockB2}) in the following form
\begin{align}
{\sqrt{\det G}\over \tau_2^{10-d\over 2}}\sum_{\substack{\tilde m_d,\dots,\tilde m_9 \\ n_d,\dots, n_9}} & e^{-{\pi\over \tau_2}(\tilde m_I+n_I\bar \tau)(G+B+\Delta B)_{IJ}(\tilde m+n_I\tau)}\nonumber \\
&\, \times e^{-i\pi\left[\tilde m_d (a_1^L-\eta^R\cdot \vec a^R)-n_d(b^L_1-\eta^R\cdot \vec b^R)+\tilde m_d n_d(1-(\vec \eta^R)^2)\right]}\nonumber\\
&\, \times e^{-i\pi n_I\vec y^R_I\cdot (\vec b^R-\tilde m_J\vec y^R_J)}\prod_{\A=1}^{4}\theta\!\left[{}^{a^L_\A}_{b^L_\A}\right]\!(\tau)\prod_{\J=10}^{25}\bar \theta\!\left[{}^{a^R_\J-2n_Iy^R_{I\J}}_{b^R_\J-2\tilde m_Iy^R_{I\J}}\right]\!(\bar\tau)\esp\, .
\label{blockB3}
\end{align}

The Hamiltonian form of the above result can be expressed in terms of 
\begin{align}
&\, (G'+B')_{IJ}=\left(\!\!\begin{array}{lr}4(G+B)_{dd} & 2(G+B)_{dj}\\2(G+B)_{id}&(G+B)_{ij}\end{array}\!\!\right) , \;\; \vec y^{\, \prime R}_d = 2\vec y^R_d  \, , \; \vec y^{\, \prime R}_i = \vec y^R_i  \, , \;\;  i,j\in\{d+1,\dots,9\}\, ,\nonumber \\
&\,  \Delta B'_{IJ} \mbox{ antisymmetric}\, , \quad \Delta B'_{dj}=\eta^R\cdot \vec y_j^{\, \prime R}\, , \quad \Delta B'_{ij}=0\, ,\esp
\label{orbi'}
\end{align}
and by redefining the momenta, winding numbers and indices $N'$'s as 
\begin{align}
&m_d'=2(m_d-N^L_1+\vec \eta^R\cdot \vec N^R)+a_1^L-\vec \eta^R\cdot \vec a^R-(1-(\vec \eta^R)^2) \, n_d\, , \; m_i'=m_i\, , \;\; i\in\{d+1,\dots, 9\}\, , \nonumber \\
&n_d'={n_d\over 2}\, , \; n_i'=n_i\, , \nonumber \\
&N^{\prime L}_1=N^L_1+n_d\, , \; N^{\prime L}_\A\in N^L_\A\, ,  \; N^{\prime R}_\J=N^R_\J+n_d \, \eta^R_\J\, , \quad  \A\in\{2,3, 4\}\, , \; \J\in\{10,\dots, 25\}\, ,
\end{align}
where $n'_d\in \Z\cup (\Z+{1\over 2})$. With these conventions, we set
\begin{align}
P^{\prime L}_I&=m'_I+\vec y_I^{\, \prime R}\cdot \vec Q^{\prime R}+{1\over 2}\, \vec y^{\, \prime R}_I\cdot n'_J\vec y^{\, \prime R}_J+(B'+\Delta B'+G')_{IJ}n'_J\, , \quad I\in\{d,\dots, 9\}\, ,\nonumber \\
P^{\prime R}_I&=m'_I+\vec y_I^{\, \prime R}\cdot \vec Q^{\prime R}+{1\over 2}\, \vec y^{\, \prime R}_I\cdot n'_J\vec y^{\, \prime R}_J+(B'+\Delta B'-G')_{IJ}n'_J\, ,\nonumber \\
\vec Q'&\equiv(\vec Q^{\prime L},\vec Q^{\prime R})=\vec N'-{\vec a\over 2}\, ,
\label{momen}
\end{align}
and write the conformal block (\ref{Ham}) as 
\be
\sum_{\substack{m_d\\ n_d}}e^{2i\pi (b^L_1-\vec \eta^R\cdot \vec b^R)n'_d}\sum_{\substack{m'_{d+1},\dots,m'_9 \\ n'_{d+1},\dots, n'_9}} \sum_{\vec N'}e^{-i\pi\vec b\cdot \vec Q'} 
 \, q^{{1\over 4}\left[ P^{\prime L}_IG^{\prime IJ}P^{\prime L}_J+2(\vec Q^{\prime L})^2\right]}\,  \bar q^{{1\over 4}\left[ P^{\prime R}_IG^{\prime IJ}P^{\prime R}_J+2(\vec Q^{\prime R}+n'_I\vec y^{\, \prime R}_I)^2\right]}\, .
\ee
As a result, when the condition $ \vec \eta^R\cdot \vec a^R\in\Z$ is satisfied for all sectors $\vec a\in \Xi$, the partition function takes the final form
\begin{align}
Z'[G',B',\vec y^{\,\prime R}]_{\vec \eta^R}
=&{1\over \tau_2^{d-2\over 2}}{1\over |\Xi|}\sum_{\vec a, \vec b \in\Xi}\sum_{\vec N'}\C\big[{}^{\vec a}_{\vec b}\big]e^{-i\pi\vec b\cdot \vec Q'}  \sum_{\ell_L,\ell_R\ge 0}c^L_{\ell_L}c^R_{\ell_R} \nonumber \\
&\, \times  \sum_{\substack{m'_d\\ n_d'\in\Z\, \cup \, (\Z+{1\over 2})}} \!\!\!e^{2i\pi (b^L_1-\vec \eta^R\cdot b^R)n'_d}\; \delta_{a^L_1,m'_d+\vec \eta^R\cdot a^R-(1-(\vec \eta^R)^2)2n'_d\, \rm mod\,  2}\nonumber \\
&\, \times  \sum_{\substack{m'_{d+1},\dots,m'_9 \\ n'_{d+1},\dots, n'_9}}  q^{{1\over 4}{{M'}_{\!\!L}^2 /\Ms^2}}\, \bar q^{{1\over 4}{{M'}_{\!\!R}^2 /\Ms^2}},
\label{parti}
\end{align}
where the left- and right-masses satisfy
\begin{align}
\label{Ha''}
{M'}_{\!\!L}^2 &=\Ms^2\left[ P^{\prime L}_IG^{\prime IJ}P^{\prime L}_J+2(\vec Q^{\prime L})^2+4\ell_L-2\right] ,\nonumber \\
{M'}_{\!\!R}^2 &=\Ms^2\left[ P^{\prime R}_IG^{\prime IJ}P^{\prime R}_J+2(\vec Q^{\prime R}+n'_I\vec y^{\, \prime R}_I)^2+4\ell_R-4\right] .
\end{align}
Comparing with the case $\vec \eta^R=\vec 0^R$ given in Eq~(\ref{za}), there are two differences:

$\bullet$ There is a  shift $\Delta B'_{IJ}$ of the antisymmetric tensor. In the core of the present paper, we however denote $B+\Delta B$ as $B$ (or $B'+\Delta B'$ as $B'$), using a field redefinition. 

$\bullet$ The states $m_d'=n_d'=0$ have fermionic number $a^L_1$ equal to the parity of $\vec \eta^R\cdot a^R$, which may be even or odd. When the compact direction $X^d$ is large compared to the string scale, and the components of the Wilson line vector  $\vec y^{\, \prime R}$ are small compared to $\sqrt{G^{\prime dd}}$, the lightest states may therefore be bosons or fermions, while their superpartners acquire a KK mass $M'_{(\sigma)}=\Ms\sqrt{G^{\prime dd}}$ (see next section).  

Additional discrete Wilson lines can also be switched on as follows. Without loss of generality, we can split the internal metric plus antisymmetric tensor component $(d,J)$ into a quantized background and  continuous Wilson line deformations: 
\be
(G'+B'+\Delta B')_{dJ}=\eta^R_J+\sqrt{2}\,  y^{\prime R}_{dJ}\, ,\quad J\in\{d,\dots,9\}\, , \quad \where \quad \eta^R_J\in\Z\, , \quad y^{\prime R}_{dJ}\mbox{ arbitrary}\, .
\label{ydis}
\ee
In terms of the redefined momentum quantum number $\hat m'_d=m_d'+\eta^R_Jn_J'$, we obtain new expressions for 
\begin{align}
P^{\prime L}_d&=\hat m'_d+\vec y_d^{\, \prime R}\cdot \vec Q^{\prime R}+{1\over 2}\, \vec y^{\, \prime R}_d\cdot n'_J\vec y^{\, \prime R}_J+\sqrt{2}\, y^{\prime R}_{dJ}n'_J\, , \nonumber \\
P^{\prime R}_d&=\hat m'_d+\vec y_d^{\, \prime R}\cdot \vec Q^{\prime R}+{1\over 2}\, \vec y^{\, \prime R}_d\cdot n'_J\vec y^{\, \prime R}_J+(\sqrt{2}\, y^{\prime R}-2G')_{dJ}n'_J\, ,
\label{momen2}
\end{align}
and the partition function takes the alternative form
\begin{align}
Z'[G',B',\vec y^{\,\prime R}]_{\eta^R_d,\dots,\eta^R_9,\vec \eta^R}
=&{1\over \tau_2^{d-2\over 2}}{1\over |\Xi|}\sum_{\vec a, \vec b \in\Xi}\sum_{\vec N'}\C\big[{}^{\vec a}_{\vec b}\big]e^{-i\pi\vec b\cdot \vec Q'}  \sum_{\ell_L,\ell_R\ge 0}c^L_{\ell_L}c^R_{\ell_R} \nonumber \\
&\, \times  \sum_{\substack{\hat m'_d\\ n_d'\in\Z\, \cup \, (\Z+{1\over 2})}} \!\!\!e^{2i\pi (b^L_1-\vec \eta^R\cdot b^R)n'_d}\; \delta_{a^L_1,\hat m'_d-\eta^R_Jn'_J+\vec \eta^R\cdot a^R-(1-(\vec \eta^R)^2)2n'_d\, \rm mod\,  2}\nonumber \\
&\, \times  \sum_{\substack{m'_{d+1},\dots,m'_9 \\ n'_{d+1},\dots, n'_9}}  q^{{1\over 4}{{M'}_{\!\!L}^2 /\Ms^2}}\, \bar q^{{1\over 4}{{M'}_{\!\!R}^2 /\Ms^2}},
\label{parti2}
\end{align}
where one argument of the Kronecker symbol is shifted. To conclude, we stress that the above partition function is equivalent to the initial one in Eq.~(\ref{defi}), with left-moving discrete Wilson lines given in Eq.~(\ref{nosusy}). It is therefore independent of $\eta^R_d,\dots,\eta^R_9,\vec \eta^R$. However, in order to describe the  light spectrum encountered in a given region of moduli space, it may be helpful to choose $\eta^R_d,\dots,\eta^R_9,\vec \eta^R$ suitably. 
 

\subsection{Low supersymmetry breaking scale}
\label{a5}

From now on in the Appendix, we consider the case where $\sqrt{G'_{dd}}\gg 1$, while all other components of the internal metric and antisymmetric tensor are of order 1. We find convenient to set $\eta^R_d=0$ and use only the notation $G'_{dd}$ (rather than $y^{\prime R}_{dd}$), in terms of which the supersymmetry breaking scale $M'_{(\sigma)}=\Ms\sqrt{G^{\prime dd}}=\O(\Ms/\sqrt{G'_{dd}})$ is low. In the following, our goal is to derive in these conditions expressions of the effective potential, 
\be
\label{vid}
\Vone^{(\sigma)}\equiv -{\Ms^d\over (2\pi)^d}\int_\F {d^2\tau\over 2\tau_2^2}\, Z'[G',B',\vec y^{\,\prime R}]_{0,\eta^R_{d+1},\dots,\eta^R_9,\vec \eta^R}\, . 
\ee

In the Hamiltonian form of the partition function~(\ref{parti2}), the left- and right-moving masses satisfy 
\be
q^{{1\over 4}{{M'}_{\!\!L}^2 /\Ms^2}}\, \bar q^{{1\over 4}{{M'}_{\!\!R}^2 /\Ms^2}}=\O\big(e^{-\pi \tau_2 \left(G'_{dd}n^{\prime 2}_d+\O(1)\right)}\big)\, , \quad \where \quad \tau_2>{\sqrt{3}\over 2}\, .  
\ee
Therefore, all strings stretched non-trivially along the very large direction $X^d$ are supermassive (even much more than oscillator states) and yield  contributions to $\Vone^{(\sigma)}$ that are exponentially suppressed in $G'_{dd}$. Thus, we proceed by focusing  on the modes having trivial winding number, $n'_d=0$. 

Next, we note that in the Langrangian formulation of the conformal block~(\ref{blockB2}), the phase responsible for the spontaneous breaking of supersymmetry is trivial when  $n_d=2n_d'$ and $\tilde m_d$ are even. Hence, in the sector $n_d=2n_d'=0$, the only non-vanishing contributions arise for $\tilde m_d$ odd. In this case, denoting  $\tilde m_d=2\tilde k_d+1$, each term of the integrand in Eq.~(\ref{vid}) contains a $\tau_2$-dependent factor
\be
{1\over\tau_2^{2+{10-d\over 2}}}\, e^{-{\pi\over \tau_2}[(2\tilde k_d+1)^2G_{dd}+\O(1)]}\, e^{-\pi\tau_2\O(1)}\, .
\ee
The latter allows an extension of the integration over the fundamental domain $\F$ to the ``upper half strip'', 
\be
\int_\F d\tau_1  d\tau_2\longrightarrow \int_{-{1\over 2}}^{1\over 2}d\tau_1\int_0^{+\infty}d\tau_2 \, , 
\ee
at the price of introducing an error exponentially suppressed in $G'_{dd}$. Note that no ultraviolet divergence occurs as $\tau_2\to 0$.

Switching back to the Hamiltonian picture, the integration over $\tau_1$ projects out the non-level matched states. Therefore, we obtain
\be
\Vone^{(\sigma)}=-{\Ms^d\over (2\pi)^d}\int_0^{+\infty} {d\tau_2\over 2\tau_2^{d+2\over 2}}\sum_s(-1)^{\hat m_d'-\eta^R_jn'_j+\vec \eta^R\cdot \vec a^R}\, e^{-\pi\tau_2{M'}_{\!\!L}^2 /\Ms^2}+\O\big(e^{-\#G'_{dd}}\big)\, ,
\ee
where $\#$ is an order 1 positive coefficient and the discrete sum is over all physical states~$s$ having $n'_d=0$. To be explicit, they belong to some sector labeled by $\vec a^R\in \Xi$ and have arbitrary quantum numbers $\hat m'_d,m'_{d+1},\dots,m'_9$, $n'_{d+1},\dots, n'_9$ and $\ell_L,\ell_R$. Notice that we have used the spin-statistics theorem as well as the Kronecker symbol appearing in the partition function~(\ref{parti2}) to fix the sign of the contribution of $s$. Since the level matching condition 
\be
\label{lmc2}
{1\over 4}({M'}_{\!\!L}^2-{M'}_{\!\!R}^2)\equiv \Ms^2\Big[(\hat m'_d-\eta^{\prime R}_In'_I)n'_d+m'_in'_i+{1\over 2} \big(\vec Q^2+1\big)\!+\ell_L-\ell_R\Big]\!\in \Z
\ee 
is independent of $\hat m_d'$ when $n_d'=0$, the states $s$ are actually organized in KK towers of modes of arbitrary momentum $\hat m_d'$, so that we may write
\be
\Vone^{(\sigma)}=-{\Ms^d\over (2\pi)^d}\sum_{s_0}(-1)^{\vec \eta^R\cdot \vec a^R-\eta^{\prime R}_jn'_j}\int_0^{+\infty} {d\tau_2\over 2\tau_2^{d+2\over 2}}\sum_{\hat m_d'}(-1)^{\hat m_d'}\, e^{-\pi\tau_2{M'}_{\!\!L}^2 /\Ms^2}+\O\big(e^{-\#G'_{dd}}\big)\, ,
\label{for}
\ee
where the first discrete sum is over the physical states $s_0$ having $\hat m'_d=n'_d=0$.  In the above expression, $M_L'$ is the mass of the KK mode of momentum $\hat m_d'$,
 \be
{M'}_{\!\!L}^2 =\Ms^2\!\left[ \hat m_d^{\prime 2}   G^{\prime dd}+2\hat m'_d\zeta^d\right]+M_{L0}^{\prime 2}\, ,
\label{MKKgene}
\ee
where $M_{L0}'$ is that of the zero-momentum state $s_0$, 
\be
{M'}_{\!\!L0}^2 =\Ms^2\left[\xi^2_dG^{\prime dd}+2\xi_dG^{\prime dj}P^{\prime L}_j+P^{\prime L}_iG^{\prime ij}P^{\prime L}_j+2(\vec Q^{\prime L})^2+4\ell_L-2\right] , 
\label{M0}
\ee
and $\xi_d,\zeta^d$ are introduced for notational convenience, 
\be
\label{zgene}
\xi_d=\vec y_d^{\, \prime R}\cdot \vec Q^{\prime R}+{1\over 2}\, \vec y^{\, \prime R}_d\cdot n'_j\vec y^{\, \prime R}_j+\sqrt{2}\, y^{\prime R}_{dj}n'_j\, ,\qquad \zeta^d=G^{\prime dd}\xi_d+G^{\prime dj}P_j^{\prime L}\,.
\ee

By Poisson summation over $\hat m_d'$, one obtains a mixed Lagrangian/Hamiltonian form of the effective potential, where the order of the integral and the discrete sum can be inverted. One finds
\begin{align}
\Vone^{(\sigma)}= -{\Ms^d\over (2\pi)^d}\sum_{s_0}(-1)^{\vec \eta^R\cdot \vec a^R-\eta^R_jn'_j}\sum_{\tilde m_d'} e^{i\pi(2\tilde m_d'+1){\zeta^d\over G^{\prime dd}}} {1\over \sqrt{G^{\prime dd}}}&\, \int_0^{+\infty}{d\tau_2\over 2\tau_2^{d+3\over 2}}\, e^{-\pi {(\tilde m'_d+{1\over 2})^2\over \tau_2G^{\prime dd}}}\, e^{-\pi\tau_2{\M'}^2_{\!\!L0}}\nonumber \\
&\qquad\quad\;\;\;+\O\big(e^{-\#G'_{dd}}\big)\, , 
\end{align}
where $\M'_{L0}$ is a characteristic mass associated with the KK tower labeled by $s_0$,
\be
\label{Mgene}
{\M'}_{\!\!L0}^2\equiv M_{L0}^{\prime 2}-{(\zeta^d\Ms)^2\over G^{\prime dd}}=\Ms^2\!\left[P^{\prime L}_iG^{\prime ij}P^{\prime L}_j-{(G^{\prime dj}P'_j)^2\over G^{\prime dd}}+2(\vec Q^{\prime L})^2+4\ell_L-2\right]\ge 0\, .
\ee
The result of the integration over $\tau_2$ can be formulated in terms of a function
\begin{align}
\label{Ffunc}
F(z)\equiv z^{d+1\over 2}\, K_{d+1\over 2}(z)& =2^{d-1\over 2}\Gamma\Big({d+1\over 2}\Big)\!\!\left[1-{z^2\over 2(d-1)}+\O(z^4)\right] , \quad \when\quad  z\to 0\, , \nonumber \espD\\
& \sim z^{d\over 2}\, e^{-z}\sqrt{\pi\over 2}\, , \quad \when\quad  z\to +\infty\, , 
\end{align}
where $K_{d+1\over 2}$ is a modified Bessel function of the second kind. If $F$ is finite at $z=0$, it happens to be exponentially suppressed for $z\gg 1$, thus implying that only a finite number of towers yields significant contributions. To be specific, let us consider the  $\nB$ (or $\nF$) KK towers having $\M'_{L0}$ lower than a few times $M'_{(\sigma)}$ (say $\varrho M'_{(\sigma)}$ for some  $\varrho=\O(1)$), and such that $\vec \eta^R\cdot \vec a^R-\eta^R_jn'_j$ is even (or  odd). Defining $c\Ms$ to be the lowest mass $\M'_{L0}$ of the infinite number of heavier KK towers, 
we obtain\footnote{Choosing for instance $\varrho=3$, the  most unfavorable configuration, which corresponds to $c\Ms=\varrho M'_{(\sigma)}$, implies the non-explicit terms in the second line of Eq.~(\ref{Lf}) to be about 1\% of the contribution of a KK tower with vanishing characteristic mass. Of course, $\Ms\ge c\Ms>\varrho M'_{(\sigma)}$ yields much lower errors.}  
\begin{align}
\Vone^{(\sigma)}=-{{M'}^d_{\!\!\!(\sigma)}\over (2\pi)^{3d+1\over 2}}\sum_{s_0=1}^{\nB+\nF}(-1)^{\vec \eta^R\cdot \vec a^R-\eta^R_jn'_j}\sum_{\tilde m_d'}&\,{\cos\!\left(\pi(2\tilde m_d'+1){\zeta^d\Ms^2\over {M'}^2_{\!\!\!(\sigma)}}\right) \over |\tilde m'_d+{1\over 2}|^{d+1}}\, F \bigg(\pi |2\tilde m'_d+ 1|{\M'_{L0}\over M'_{(\sigma)}}\bigg)\nonumber\espDD \\
&\, \qquad +\O\big((c\Ms M'_{(\sigma)})^{d\over 2}e^{-\pi{c\Ms/M'_{(\sigma)}}}\big)\, .
\label{Lf}
\end{align}
In fact, the KK towers with characteristic masses $\M'_{L0}>\varrho M'_{(\sigma)}$ are almost supersymmetric and do not contribute significantly to the effective potential. Going back to the Hamiltonian picture given in Eq.~(\ref{for}), we also have 
\begin{align}
\Vone^{(\sigma)}=-{\Ms^d\over (2\pi)^d}\sum_{s_0=1}^{\nB+\nF}(-1)^{\vec \eta^R\cdot \vec a^R-\eta^R_jn'_j}\int_0^{+\infty} &\, {d\tau_2\over 2\tau_2^{d+2\over 2}} \sum_{\hat m_d'}(-1)^{\hat m_d'}\, e^{-\pi\tau_2{M'}_{\!\!L}^2 /\Ms^2}\nonumber \\
&\, \qquad +\O\big((c\Ms M'_{(\sigma)})^{d\over 2}e^{-\pi{c\Ms/M'_{(\sigma)}}}\big)\, .
\label{Hf}
\end{align}
Some remarks are in order:

$\bullet$  The gauge symmetry arising from the $10-d+16$ internal directions and extra dimensions of the right-moving bosonic string yield an $U(1)\times \G_{9-d+16}$ gauge symmetry, where  the rank of $\G_{9-d+16}$ is $9-d+16$. When $\G_{9-d+16}$ is ``maximally enhanced'', \ie contains no $U(1)$ factor, we have in particular $y^{\prime R}_{dj}=0$, $j\in\{d+1,\dots,9\}$, $\vec y^{\, \prime R}_d=\vec 0$, so that $\xi_d=0$.\footnote{An arbitrary component $(G'+B')_{id}$, $i\in\{d+1,\dots,9\}$, is however allowed.} Moreover, except the KK scale $M'_{(\sigma)}$ itself, there is no mass scale between 0 and $\Ms$.
As a result, there are $\nB+\nF$ KK towers with exactly vanishing characteristic masses, $\M'_{L0}=0$, while all other  towers are very heavy, $\M'_{L0}=\O(\Ms)$. The former satisfy $P^{\prime L}_i=0$, $i\in\{d+1,\dots,9\}$, $(\vec Q^{\prime L})^2=1$, $\ell_L=0$, so that $\zeta^d=0$. From the Hamiltonian point of view, the  $\nB+\nF$ zero-modes $s_0$ are massless, $M'_{L0}=0$, their KK counterparts satisfy $M'_{L}\ge M'_{(\sigma)}$, and the string states belonging to other KK towers satisfy $M'_L=\O(\Ms)$. 

$\bullet$ The situation presents mild differences when some moduli fields are switched on and bring the model slightly away from the background where $\G_{9-d+16}$ is maximally enhanced. This happens when deviations of $(G'+B'+\Delta B')_{ij}$, $i,j\in\{d+1,\dots,9\}$, from the initial background are smaller (in absolute value) than $\varrho \sqrt{G^{\prime dd}}$, or when some non-vanishing $|y^{\prime R}_{dj}|, |y^{\prime R}_{d\J}|$, $j\in\{d+1,\dots,9\}$, $\J\in\{10,\dots, 25\}$, are lower than 1. In both cases, new scales lower than $\varrho \sqrt{G^{\prime dd}}\Ms$ are introduced, whose effects are to induce small Higgs masses to some of the $\nB+\nF$ initially massless modes $s_0$. However, the KK towers they belong to remain light, in the sense that their characteristic masses still satisfy $\M'_{L0} < \varrho  M'_{(\sigma)}$. 

$\bullet$ When a deviation $\varsigma$ of some $(G'+B'+\Delta B')_{ij}$, $i,j\in\{d+1,\dots,9\}$, becomes larger (in absolute value) than $\varrho  \sqrt{G^{\prime dd}}$, the number $\nB+\nF$ of KK towers such that $\M'_{L0}<\varrho M'_{(\sigma)}$ decreases. Physically, the gauge theory enters a Coulomb branch where the component  $(G'+B'+\Delta B')_{ij}$ is a flat direction of the effective potential, up to $\O\big((|\varsigma |\Ms M'_{(\sigma)})^{d\over 2}e^{-\pi{|\delta | \Ms/M'_{(\sigma)}}}\big)$ terms.


\subsection{\bm Example with $\nF$ greater, lower or equal to $\nB$ }
\label{a6}

In order to illustrate the results of Sects~(\ref{a4}) and~(\ref{a5}), we consider the $E_8\times E'_8$ heterotic string compactified on $S^1(R_d)\times T^{9-d}$, where $R_d$ is the radius of the circle, in the presence of  discrete deformations,
\begin{align}
&\Xi=\big\{ \mbox{$(4,8+8)$-tuples } (a,\dots, a; \gamma,\dots ,\gamma,\gamma',\dots,\gamma') \, ,\;  \where\; \;  a,\gamma,\gamma'\in \Z_2\big\}\; \Longrightarrow\; |\Xi|=2^3\, , \nonumber \\
&\C[^{a;\gamma,\gamma'}_{b\, ;\delta,\, \delta'}]=(-1)^{a+b+ab}\, ,\nonumber\\
&Y^L_{I\A}=\delta_{Id}\, \delta_{\A 1}\, ,\quad I\in\{d,\dots, 9\}\,,  \; \A\in\{1,\dots, 4\}\, , \nonumber \\
&Y^R_{I\J} =\delta_{Id}\eta_\J^R \quad  i.e. \quad  y^R_{I\J}=0\, , \quad  \J\in\{10,\dots, 25\}\, ,\nonumber \\
&0= (G'+B'+\Delta B')_{dJ}=\eta_J^R\quad i.e. \quad y^{\prime R}_{dJ}=0\, , \quad J\in\{d,\dots, 9\}\, .
\label{desdef}
\end{align}
As explained before, the left-moving discrete deformation implements a spontaneous breaking of  supersymmetry \via a stringy Scherk-Schwarz mechanism along the direction $X^d$. Moreover, when the supersymmetry breaking scale is low, the bosonic or fermionic nature of  the lightest states  is determined by $\vec \eta^R$, a fact that has a direct impact on the gauge symmetry. In the following, we consider in details the example where  
\be
\label{etabreak}
\eta^R_\J=\delta_{\J,10}+\delta_{\J,18}\, , \quad  \J\in\{10,\dots, 25\}\, .
\ee
We first derive the partition function of the model under the above assumptions. Then, we switch on arbitrary (but small) Wilson line deformations around such a background and derive the 1-loop effective potential at low supersymmetry breaking scale. 
 
Using Eq.~(\ref{blockB3}), the 1-loop partition function~(\ref{defi}) can be written as
\begin{align}
Z={1\over \tau_2^{d-2\over 2}}\, & {1\over \eta^{12}\, \bar \eta^{24}}\, {1\over 2} \sum_{a,b}\, (-1)^{a+b+ab}\, \theta[^a_b]^4\; {1\over 2} \sum_{\gamma,\delta}\theta[^\gamma_\delta]^8\; {1\over 2} \sum_{\gamma',\delta'}\theta[^{\gamma'}_{\delta'}]^8\nonumber \\
&\, \times  {R_d\over \sqrt{\tau_2}}\sum_{\tilde m_d,n_d}e^{-{\pi\over \tau_2}R_d^2|\tilde m_d+n_d\tau|^2}\, \Gamma_{9-d,9-d}\, (-1)^{\tilde m_d(a-\gamma-\gamma')-n_d(b-\delta-\delta')-\tilde m_dn_d}\, , 
\end{align}
where we have used the fact that the lattice of zero-modes associated with the internal torus is factorized, $\Gamma_{10-d,10-d}=\Gamma_{1,1}(R_d)\times \Gamma_{9-d,9-d}$. Defining $R_d'=2R_d$ and $\tilde m_d=2\tilde k_d+g$, $n_d=2l_d+h$, where $g,h\in\Z_2$, the above formula becomes
\begin{align}
Z= &\, {1\over \tau_2^{d-2\over 2}}\,  {1\over \eta^{8}\, \bar \eta^{8}}\, \Gamma_{9-d,9-d}\, {1\over 2}\sum_{h,g}\Gamma_{1,1}\!\left[{}^h_g\right]\!\!(R'_d)\nonumber \\
 &\, \times {1\over 2}\sum_{a,b}(-1)^{a+b+ab}\, {\theta[^a_b]^4\over \eta^4}\, (-1)^{ga-hb-hg}\; \;{1\over 2}\sum_{\gamma,\delta}{\bar \theta[^\gamma_\delta]^8\over \bar \eta^8}\, (-1)^{g\gamma-h\delta}
\; \;{1\over 2}\sum_{\gamma',\delta'}{\bar \theta[{}^{\gamma'}_{\delta'}]^8\over \bar \eta^8}\, (-1)^{g\gamma'-h\delta'}\, ,
\end{align}
where we have introduced shifted lattices, which can be considered either in Langrangian or Hamiltonian forms, 
\begin{align}
\Gamma_{1,1}\!\left[{}^h_g\right]\!\!(R_d')& ={R'_d\over \sqrt{\tau_2}}\sum_{\tilde k_d,l_d}e^{-{\pi\over \tau_2}{R'}_{\!\! d}^2\left|\tilde k_d+{g\over 2}+(l_d+{h\over2})\tau\right|^2}\nonumber \\
&\, =\sum_{k_d,l_d}e^{i\pi gk_d}\, q^{{1\over 4}\left({k_d\over R_d'}+(l_d+{h\over 2})R_d'\right)^2}\, \bar q^{{1\over 4}\left({k_d\over R_d'}-(l_d+{h\over 2})R_d'\right)^2}\, .
\end{align}
In terms of the $O(2n)$ affine characters
\begin{align}
O_{2n}&={\theta[{}^0_0]^n+\theta[{}^0_1]^n\over 2\eta^n}\, , &V_{2n}&={\theta[{}^0_0]^n-\theta[{}^0_1]^n\over 2\eta^n}\, ,\nonumber \\
S_{2n}&={\theta[{}^1_0]^n+(-i)^n\theta[{}^1_1]^n\over 2\eta^n}\, ,
&C_{2n}&={\theta[{}^1_0]^n-(-i)^n\theta[{}^1_1]^n\over 2\eta^n}\, ,
\label{charac}
\end{align}
and 
\begin{align}
\gamma_{1,1}\!\left[{}^h_g\right]\!\!(R_d')&={1\over 2}\left(\Gamma_{1,1}\!\left[{}^h_0\right]\!\!(R_d')+(-1)^g\, \Gamma_{1,1}\!\left[{}^h_1\right]\!\!(R_d')\right)\nonumber \\
&= \sum_{k'_d,l_d}q^{{1\over 4}\left({2k'_d+g\over R'_d}+(l_d+{h\over 2})R'_d\right)^2}\, \bar q^{{1\over 4}\left({2k'_d+g\over R'_d}-(l_d+{h\over 2})R'_d\right)^2}\, ,
\end{align}
we obtain the final expression
 \begin{align}
\label{Zsss}
Z={1\over \tau_2^{d-2\over 2}}\,  {1\over \eta^{8}\, \bar \eta^{8}}\, \Gamma_{9-d,9-d} \, & \Big[\; \,\gamma_{1,1}\big[{}^0_0\big]\!(R_d')\Big( V_8(\bar O_{16}  \bar O'_{16}+ \bar S_{16} \bar S'_{16} ) -S_8 (\bar O_{16}\bar S'_{16} + \bar S_{16} \bar O'_{16})\Big) \nonumber \\
& \!\!+\gamma_{1,1}\big[{}^0_1\big]\!(R_d')\Big( V_8 (\bar O_{16}\bar S'_{16} + \bar S_{16} \bar O'_{16}) -S_8  (\bar O_{16} \bar O'_{16}+ \bar S_{16}\bar S'_{16} ) \Big)\nonumber\espD\\
&\!\!+\gamma_{1,1}\big[{}^1_0\big]\!(R_d')\Big( O_{8} (\bar V_{16} \bar C'_{16}+ \bar C_{16}\bar V'_{16} ) -C_8 (\bar V_{16}\bar V'_{16} + \bar C_{16}\bar C'_{16} )\Big)\espD \nonumber\\
& \!\!+\gamma_{1,1}\big[{}^1_1\big]\!(R_d')\Big( O_8(\bar V_{16}\bar V'_{16} + \bar C_{16}\bar C'_{16}) -C_8 (\bar V_{16} \bar C'_{16}+ \bar C_{16}\bar  V'_{16} )\Big)\, \Big].
 \end{align}

To make contact with the notations of the previous subsections, we  identify  $R'_d=\sqrt{G'_{dd}}$, and recognize the momentum $\hat m_d'\equiv 2k'_d +g$ and winding number $n'_d=l_d+{h\over 2}\in\Z\cup(\Z+{1\over 2})$. When $R'_d\gg 1$, the states contributing in the third and fourth lines of Eq.~(\ref{Zsss}) are super massive. The  physical states $s_0$, which have $\hat m'_d=n_d'=0$, are massless and arise in the first line. They are $\nB=8\times [(d-2)+1+\dim \G_{9-d}+2\times 120]$ bosonic degrees of freedom, 
\be
\boldsymbol{8_{\rm v}}\otimes \big([d-2]\oplus {\rm Adj}_{U(1)\times \G_{9-d}}\oplus {\rm Adj}_{SO(16)\times SO(16)'}\big),
\label{spectB}
\ee
where $\G_{9-d}$ is the gauge symmetry induced by the $\Gamma_{9-d,9-d}$ lattice, 
and $\nF=8\times 2\times 128$ fermionic degrees of freedom, 
\be
\boldsymbol{8_{\rm s}} \otimes  \big(\mbox{Spinorial}_{SO(16)}\oplus \mbox{Spinorial}_{SO(16)'} \big).
\label{spectF}
\ee
Their superpartners, which have $\hat m'_d=1, n'_d=0$, show up in the second line and have  masses $\Ms/R_d'$. The sign of $\nF-\nB$ can be arbitrary, as can be seen for example by choosing  $\G_{9-d}=SU(2)^{9-d-s}\times U(1)^s$, which yields $\nF-\nB=16(d+s-5)$. In dimension $d=4$, this is negative for $s=0$, vanishes for $s=1$ and is positive for $s=2,3,4,5$.

To see explicitly the dependence of the 1-loop effective potential on the Wilson lines, let us consider as an example in arbitrary dimension  $d\ge 3$ the case of an initial background characterized by a maximally enhanced gauge group $\G_{9-d}=SU(2)^{9-d}$. In the notations of Eq.~(\ref{orbi'}), the $y$-deformations we introduce are given by
\begin{align}
&(G'+B'+\Delta B')_{IJ}=\left(\!\!\begin{array}{cc}G'_{dd} & \sqrt{2}\, y^{\prime R}_{dj}\\ \sqrt{2}\, y^{\prime R}_{id}&\delta_{ij}+\sqrt{2}\, y^{\prime R}_{ij}\end{array}\!\!\right)  , \;\; \vec y^{\, \prime R}_d \, , \; \vec y^{\, \prime R}_i  \, , \;\;  i,j\in\{d+1,\dots,9\}\, ,\nonumber \\
&\,  \Delta B'_{IJ} \mbox{ antisymmetric}\, , \quad \Delta B'_{dj}=y^{\prime R}_{j,10}+y^{\prime R}_{j,18}\, , \quad \Delta B'_{ij}=0\, .\esp
\label{orbi''}
\end{align}
We are going to apply Eq.~(\ref{Lf}), which is valid when $G'_{dd}\gg 1$, in the case the continuous Wilson lines are small, namely 
\be
|y^{\prime R}_{ij}|, |y^{\prime R}_{i\J}| \ll \varrho\sqrt{G^{\prime dd}}\, , \quad  |y^{\prime R}_{dj}|, |y^{\prime R}_{d\J}|, |y^{\prime R}_{id}|\ll 1 \, , \quad i,j\in\{d+1,\dots,9\}\,, \; \J\in\{10,\dots, 25\}\, .
\ee
For this purpose, we list the KK towers $s_0$, which satisfy $(\vec Q^{\prime L})^2=1$, $\ell_L=0$:

$\bullet$ For any given $j\in\{d+1,\dots,9\}$ and $\epsilon\in\{-1,1\}$, there are 8 KK towers $s_0$ associated with the root $\epsilon \sqrt{2}$ of the $SU(2)$ factor, and corresponding to momentum states along  the direction $X^j$. The quantum numbers of the KK modes are $(\gamma,\gamma')=(0,0)$, $\vec Q^{\prime R}=\vec 0$, $\ell_R=0$ and 
\be
\hat m'_d\in\Z\, , \; n'_d=0\, ,  \qquad m'_j=-n'_j=-\epsilon\, ,  \qquad m'_i=n'_i=0\, , \; i\in\{d+1,\dots,9\},\, i\neq j\, .
\ee
Using these data, we  derive 
\be
\xi_d=\epsilon\sqrt{2}\big( y^{\prime R}_{dj}+{1\over 2\sqrt{2}}\, \vec y_d^{\, \prime R}\cdot \vec y^{\, \prime R}_j\big)\, , \; P'_i=\epsilon\sqrt{2}\big( y^{\prime R}_{ij}+{1\over 2\sqrt{2}}\, \vec y_i^{\, \prime R}\cdot \vec y^{\, \prime R}_j\big)\, ,\; i\in\{d+1,\dots,9\}\, , 
\ee
and find
\be
{\zeta^d\Ms\over M'_{(\sigma)}}= y^{\prime R}_{dj}\,  \epsilon \sqrt{2}+\cdots \, , \qquad {\M'}^2_{\!\!L0}=\sum_{i=d+1}^9\big( y_{ij}^{\prime R}\,  \epsilon \sqrt{2}\big)^2+\cdots\, ,
\ee
where the ellipses stand for higher order terms in Wilson line deformations. 
The contribution of the 8 KK towers to the effective potential is then found to be
\begin{align}
\Vone^{(\sigma) j,\epsilon}= &\, 8{M'}^d_{\!\!\!(\sigma)}\Big\{\!-  v_{d}\,  2^d+  {v_{d-2}\over 4\pi} \, 2^{d-2}\nonumber \\
&\, \times \Big[(d-1)\big( y^{\prime R}_{dj}\,  \epsilon \sqrt{2}\big)^2+{1\over G^{\prime dd}}\sum_{i=d+1}^9\big(y^{\prime R}_{ij}\,  \epsilon \sqrt{2}\big)^2\Big]\Big\}+\cdots \,,
\end{align}
where we have defined
\be
v_{d}={\Gamma({d+1\over 2})\, \zeta(d+1)\over 2^{d-1}\;\pi^{3d+1\over 2}}\, \Big(1-{1\over 2^{d+1}}\Big).
\ee

$\bullet$ For any root $\vec Q^R$ of $SO(16)\times SO(16)'$, or any weight $\vec Q^R$ of the spinorial representation of $SO(16)$ or $SO(16)'$, there are 8 KK towers $s_0$. The former have $(\gamma,\gamma')=(0,0)$ and the latter $(\gamma,\gamma')=(1,0)$ or $(0,1)$. The other quantum numbers of the KK modes are $\hat m'_d\in\Z$, $n'_d=0$, $m'_i=n'_i=0$, $i\in\{d+1,\dots,9\}$, $\ell_R=0$. This leads 
\be
\xi_d=\vec y^{\, \prime R}_d\cdot \vec Q^R\, , \qquad P'_i=\vec y^{\, \prime R}_i\cdot \vec Q^R\, ,\quad  i\in\{d+1,\dots,9\}\, , 
\ee
so that
\be
{\zeta^d\Ms\over M'_{(\sigma)}}= \vec y^{\, \prime R}_d\cdot \vec Q^R+\cdots\, , \qquad {\M'}^2_{\!\!L0}=\sum_{i=d+1}^9\big(\vec y_i^{\, \prime R}\cdot Q^R\big)^2+\cdots\, .
\ee
The contribution of the 8 KK towers of charge $\vec Q^R$ to the effective potential is then 
\begin{align}
\Vone^{(\sigma)\vec Q^R}= (-1)^{\gamma+\gamma'}&\,8{M'}^d_{\!\!\!(\sigma)}\Big\{\!-  v_{d}\,  2^d+  {v_{d-2}\over 4\pi} \, 2^{d-2}\nonumber \\
&\, \times \Big[(d-1)\big(\vec y^{\, \prime R}_d\cdot \vec Q^R\big)^2+{1\over G^{\prime dd}}\sum_{i=d+1}^9\big(\vec y^{\, \prime R}_i\cdot \vec Q^R\big)^2\Big]\Big\}+\cdots\, .
\end{align}

$\bullet$ Finally, there are 8 KK towers $s_0$ for each of the 24 states at right-moving oscillator level $\ell_R=1$. Being neutral with respect to $SU(2)^{9-d}\times SO(16)^2$, the quantum numbers of the KK modes are  $\hat m'_d\in\Z$, $n'_d=0$, $m'_i=n'_i=0$, $i\in\{d+1,\dots,9\}$, $\vec Q^R=\vec 0$ and $(\gamma,\gamma')=(0,0)$. Therefore, $\xi_d=0$ and $P'_i=0$, $i\in\{d+1,\dots,9\}$, which implies
\be
{\zeta^d\Ms\over M'_{(\sigma)}}= 0\, , \qquad {\M'}^2_{\!\!L0}=0\, .
\ee
For each $e\in\{2,\dots, 25\}$, the contribution of the 8 neutral KK towers to $\Vone^{(\sigma)}$  is therefore
\be
\Vone^{(\sigma)e}= 8{M'}^d_{\!\!\!(\sigma)}\Big\{\!-  v_{d}\,  2^d\Big\}\, . \\
\ee

Combining the above results, the total 1-loop effective potential takes the form
\begin{align}
\Vone^{(\sigma)}= &\, (\nF-\nB)\,  v_{d}\,  {M'}^d_{\!\!\!(\sigma)}\, 2^d\nonumber \\
&\, +  {M'}^d_{\!\!\!(\sigma)} \, {v_{d-2}\over 2\pi} \, 2^{d-2}\Big\{\sum_{j=d+1}^9 c_{SU(2)}\Big[(d-1)( y^{\prime R}_{dj})^2+{1\over G^{\prime dd}}\sum_{i=d+1}^9(y^{\prime R}_{ij})^2\Big]\nonumber \\
&\,\;\;\quad \qquad\qquad\qquad\qquad+ c_{SO(16)}\Big[(d-1)(\vec y^{\, \prime R}_d)^2+{1\over G^{\prime dd}}\sum_{i=d+1}^9(\vec y^{\, \prime R}_i)^2\Big]\Big\}+\cdots\nonumber \\
&\, +\O\big((\Ms M'_{(\sigma)})^{d\over 2}e^{-\pi{\Ms/M'_{(\sigma)}}}\big)\, ,
\end{align}
where we have defined
\begin{align}
c_{SU(2)}&=8\, C_{[3]_{SU(2)}}=8\times 2=16\, , \nonumber \\
c_{SO(16)}&=8\big(C_{[120]_{SO(16)}}-C_{[128]_{SO(16)}}\big)=8\times (14-16)=-16\, ,
\end{align}
in terms of coefficients  $C_{\cal R}$  considered  for any representation ${\cal R}$ of a gauge group $\G$, 
\be
\label{CR}
 {1\over 2}\sum_{\underset{\mbox{\scriptsize of $\cal R$}}{\mbox{\scriptsize weights $Q$}}}\sum_{\I=1}^{{\rm rank }\,\G}A_\I Q_\I\, \sum_{\J=1}^{{\rm rank }\,\G}B_\J Q_\J=C_{\cal R}\sum_{\I=1}^{{\rm rank }\,\G}A_\I B_\I\, .
\ee
As a result, the Wilson lines of $SU(2)^{9-d}$ along $T^{10-d}$ are massive at 1-loop, while those of $SO(16)^2$ are tachyonic. Notice that $y^{\prime R}_{id}$, $i\in\{d+1,\dots,9\}$, multiplies $n_d'$ in Eq.~(\ref{momen}). Therefore, the non-exponentially suppressed contributions of $\Vone$ involve them only \via the expressions of $G^{\prime dd}, G^{\prime dj}$, $j\in\{d+1,\dots,9\}$. Expanding the cosine in Eq.~(\ref{Lf}), it happens that the $y^{\prime R}_{id}$'s appear in at least cubic interactions with other Wilson lines. In other words, they remain massless at 1-loop, but are no more flat directions of the effective potential. 

Using the dictionary~(\ref{orbi'}) and defining 
\be
(G+B+\Delta B)_{IJ}=\left(\!\!\begin{array}{cc}G_{dd} & \sqrt{2}\,  y^{R}_{dj}\\ \sqrt{2}\, y^R_{id}&\delta_{ij}+\sqrt{2}\, y^{R}_{ij}\end{array}\!\!\right)  , \quad  i,j\in\{d+1,\dots,9\}\, ,
\ee
where $\Delta B$ is given in Eq.~(\ref{Db}), we obtain the final result, 
\begin{align}
\Vone^{(\sigma)}= &\, (\nF-\nB)\,  v_{d}\,  M^d_{(\sigma)}\nonumber \\
&\, +  M^d_{(\sigma)} \, {v_{d-2}\over 2\pi} \Big\{\sum_{j=d+1}^9 c_{SU(2)}\Big[(d-1)( y^R_{dj})^2+{1\over G^{dd}}\sum_{i=d+1}^9(y^R_{ij})^2\Big]\nonumber \\
&\,\;\;\quad\qquad\qquad\qquad+ c_{SO(16)}\Big[(d-1)(\vec y^R_d)^2+{1\over G^{dd}}\sum_{i=d+1}^9(\vec y^R_i)^2\Big]\Big\}+\cdots\nonumber \\
&\, +\O\big((\Ms M_{(\sigma)})^{d\over 2}e^{-2\pi{\Ms/M_{(\sigma)}}}\big)\, ,
\label{final}
\end{align}
which is written using the redefined supersymmetry breaking scale
\be
M_{(\sigma)}=\Ms\,  \sqrt{G^{dd}}\, .
\ee
Eq.~(\ref{final}) is an example of the expression we use in the main text of the present work, Eq.(\ref{Vgen}), up to the minor change of notations consisting in omitting the  upper indices ``$R$''.



\end{document}